\documentclass[prd,aps,twocolumn,showpacs,preprintnumbers,amsmath,amssymb]{revtex4}

\usepackage{graphicx}
\usepackage{dcolumn}
\usepackage{bm}
\begin{document}

\title{Excitations of single-beauty hadrons}
\author{Tommy Burch,$^1$ 
Christian Hagen,$^2$ 
Christian B.\ Lang,$^3$ 
Markus Limmer,$^3$ and 
Andreas Sch\"afer$^{2,4}$\\
(BGR [Bern-Graz-Regensburg] Collaboration)}
\vskip1mm
\affiliation{$^1$Department of Physics, University of Utah, 
Salt Lake City, UT 84112, U.S.A.}
\affiliation{$^2$Institut f\"ur Theoretische Physik, Universit\"at 
Regensburg, D-93040 Regensburg, Germany}
\affiliation{$^3$Institut f\"ur Physik, FB Theoretische Physik, 
Universit\"at Graz, A-8010 Graz, Austria}
\affiliation{$^4$Yukawa Institute of Theoretical Physics, Kyoto University, 
Kyoto 606-8502, Japan}

\begin{abstract}
In this work we study the predominantly orbital and radial excitations of 
hadrons containing a single heavy quark. 
We present meson and baryon mass splittings and ratios of meson decay 
constants (e.g., $f_{B_s}/f_B$ and $f_{B_s'}/f_{B_s}$) 
resulting from quenched and dynamical two-flavor configurations. 
Light quarks are simulated using the chirally improved (CI) lattice Dirac 
operator at valence masses as light as $M_\pi \approx 350$ MeV. 
The heavy quark is approximated by a static propagator, appropriate for the 
$b$ quark on our lattices ($1/a \sim 1-2$ GeV). 
We also include some preliminary calculations of the $O(1/m_Q^{})$ kinetic 
corrections to the states, showing, in the process, a viable way of applying 
the variational method to three-point functions involving excited states. 
We compare our results with recent experimental findings.
\end{abstract}
\pacs{11.15.Ha, 12.38.Gc, 12.39.Hg, 14.20.Mr, 14.40.Nd}
\keywords{Lattice gauge theory, hadron spectroscopy, heavy quarks, 
excited states}
\maketitle

\section{Introduction}
\label{SectIntroduction}

\subsection{Singly beautiful hadrons: Recent progress}

With one heavy valence quark and otherwise light degrees of freedom, single 
beauty (bottom, $b$) mesons and baryons are the QCD analogs of hydrogen and 
helium atoms. 
In the past few years, there has been much progress in determining the 
properties of excited states of such hadrons. 
The D0 and CDF experiments at Fermilab have recently observed the ``orbitally 
excited'' $B_1$, $B_2^*$ \cite{B**} and $B_{s1}$, $B_{s2}^*$ mesons 
\cite{Bs**}, as well as the $\Sigma_b^{\pm}$, $\Sigma_b^{*\pm}$ \cite{Sigmab*}, 
$\Xi_b^-$ \cite{Xib-}, and $\Omega_b^-$ baryons \cite{Omegab-}. 
There have also been advances in lattice calculations of the mass spectrum of 
excited $B$ mesons 
\cite{MVR,AliKhan:1999yb,Hein:2000qu,Lewis:2000sv,Wingate:2002fh,Green:2003zz,Burch:2004aa,DDI,Foley:2007ui,Koponen:2007nx,Burch:2007xy,Jansen:2008ht} 
and $b$ baryons 
\cite{Bowler:1996ws,AliKhan:1999yb,Mathur:2002ce,Na:2007pv,Lewis:2008fu}, 
as well as calculations involving hadronic decay of the $B_0^*$ 
\cite{McNeile:2004rf}.

There are certainly daunting problems involved in the experimental efforts, 
and likewise, extracting excited hadrons is a real challenge on the lattice. 
Excited states appear as subleading, decaying exponentials in noisy 
two-point correlators and separating a given state from lower- and 
higher-lying states becomes a highly non-trivial problem since the former 
dominate at large Euclidean times, while the latter tend to be close in 
mass and therefore give similar contributions over the same region. 
There are a number of methods which have been devised to handle exactly this 
problem. 
In this paper, we focus on just one of these, the variational method 
\cite{VarMeth1,VarMeth2} (see also 
\cite{GBs,ExcAmp,Dudek:2007wv,SommerLat08}): 
From matrices of two-point correlators we determine meson and baryon mass 
splittings and meson decay constants; 
we also expand the method to handle three-point functions, in order to 
calculate ${\cal O}(1/m_b)$ kinetic corrections to our static approximation 
of the $b$ quark. 
Preliminary mass splittings from this ongoing study were reported in 
Refs.\ \cite{DDI,Burch:2007xy}, more recent results appeared in 
\cite{HagenLat08}.

This paper is laid out as follows: In Sec.\ \ref{SectMethod} we discuss the 
variational method and how we use it to extract the masses, couplings, and 
kinetic corrections of our heavy-light(-light) hadrons. 
The details behind our lattice calculation, including our basis of operators, 
are presented in Sec.\ \ref{SectLattice}. 
In Sec.\ \ref{SectAnalysis} we present our analysis of each of the items 
mentioned in \ref{SectMethod} and in Sec.\ \ref{SectDiscussion} we 
discuss our results for each of the mesons and baryons. 
We conclude and present an outlook in Sec.\ \ref{SectConclusion}.

\subsection{Operators versus states: Not just semantics}

Before moving on, we believe it necessary to point out that the words 
``orbital,'' ``radial,'' and ``gluonic excitations'' should be viewed as only 
convenient terminology since each describes only one (albeit perhaps the most 
likely) of the configurations of the constituents in a particular excited 
state. 
In QCD, mixtures of many such configurations are present in any hadronic 
state \cite{ConfigMix,ConfigMixNote} and their clear separation only begins 
to make sense in the heavy-quark limit \cite{Bali:2000gf,HeavyConfigMix}. 
The same applies to many similar terms, like ``hybrid'' and ``P-wave,'' 
and to any definition which tries to fix the number of constituents since 
higher Fock components also represent ever-present admixtures. 
What this means is that one must take care in distinguishing the properties 
of operators (or interpolators) from those of states. 
Ultimately, it is the covariant nature of the derivative (the momentum 
operator) which ensures that orbital, radial, and gluonic excitations are not 
uniquely defined.

\section{Methodology}
\label{SectMethod}

\subsection{Masses}

Since we are using a variational method, we need to choose a large enough 
basis of operators to create and annihilate the states of interest. 
We discuss the details of our choice of basis in the next section, but 
assuming for the moment that we have these, the next step is to create all 
cross-correlators between such operators: 
\begin{equation}
  C(t)_{ij} = \langle \, 0 \, | \, O_i(t) \; O_j^\dagger(0) \, | \, 0 \, \rangle 
  \; .
  \label{corrmatdef}
\end{equation}
We may write each of these correlators as a spectral sum of energy 
eigenstates 
\begin{equation}
  C(t)_{ij} = \sum_{n=1}^\infty v_i^{(n)} v_j^{(n)*} \, e^{-t \, E^{(n)}} \; ,
  \label{corrhilbert}
\end{equation}
where the $v_i^{(n)}$ represents the overlap of the $i$th operator with the 
$n$th state 
\begin{equation}
  v_i^{(n)} = \langle \, 0 \, | \, O_i \, | \, H_Q^{(n)} \, \rangle \; .
  \label{coeffs}
\end{equation}
We use the notation $|H_Q^{(n)}\rangle$ for the $n$th hadronic state 
containing a single static quark ($Q$).

Solving the generalized eigenvalue problem 
\begin{equation}
  C(t) \, \vec{\psi}^{(\alpha)} = \lambda(t,t_0)^{(\alpha)} \, C(t_0) \, \vec{\psi}^{(\alpha)} 
  \label{generalized}
\end{equation}
then leads to eigenvalues of the form \cite{VarMeth2} 
\begin{equation}
  \lambda(t,t_0)^{(\alpha)} = c^{(\alpha)} \, e^{- (t - t_0)E^{(\alpha)}} \left[1 + {\cal O}(e^{-(t - t_0) \Delta^{(\alpha)}}) 
  \right] \; ,
  \label{eigenvaluedecay}
\end{equation}
where it is the sum of all terms which equals 1 at $t=t_0$ and the quantity 
$\Delta^{(\alpha)}$ represents the minimum energy difference to an adjacent 
state (to the first state beyond the basis dimension for $t_0<t<2t_0$ 
\cite{SommerLat08}). 
So with a clever choice of basis (which can reduce the amplitudes of the 
correction terms) and for large enough values of $t$ (ideally, $t \gg t_0$), 
each eigenvalue yields the mass of a single state. 
The eigenvectors $\vec\psi^{(\alpha)}$ can be used to provide information 
about the couplings to the state $\alpha$ (see Ref.\ \cite{ExcAmp} and the 
next subsection). 
We do not indicate an explicit $t$-dependence of the eigenvectors since, 
ideally, there would not be any; however, one should realize that, due to 
higher-order corrections (see Eq.\ (\ref{ev_kronecker}) below) and 
statistical fluctuations, there is in fact one.

Following heavy quark effective theory (HQET; see, e.g., 
Ref.\ \cite{Manohar:2000dt}), the masses of hadrons containing a single 
heavy quark may be written in terms of an expansion in $1/m_Q^{}$. 
Accordingly, the heavy-light masses are 
\begin{eqnarray}
  M^{(n)} &=& m_Q^{} + \bar\Lambda^{(n)} + {\cal O}(k^{(n)2}/m_Q^{}) 
  \nonumber \\ 
  &=& m_Q^{} + E^{(n)} - E_0 + {\cal O}(k^{(n)2}/m_Q^{}) \; ,
\end{eqnarray}
where $\bar\Lambda^{(n)}$ is the static-light binding energy of the $n$th 
state and $k^{(n)}$ denotes the residual momentum of the heavy quark in the 
$n$th state \cite{subscript_note} 
(for the ground state, $k^{(1)} \sim \Lambda_{QCD}$). 
Note the shift $E_0$, indicating that we do not have {\it a priori} a 
physically meaningful, zero-point energy of the static quark. 
In fact, appropriate definitions of $m_Q^{}$ and $E_0$ must be chosen to 
avoid power divergences and renormalon ambiguities \cite{RenormsHQET} in any 
determination of $m_b^{}$ or $\bar\Lambda^{(1)}$. 
We may, however, use the lattice energy differences to determine the mass 
splittings 
\begin{eqnarray}
  E^{(n')} - E^{(n)} &=& \bar\Lambda^{(n')} - \bar\Lambda^{(n)}  \nonumber \\ 
  &=& M^{(n')} - M^{(n)} +  \nonumber \\ 
  && + \, {\cal O}(k^{(n')2}/m_Q^{} \, , \, k^{(n)2}/m_Q^{}) \; ,
\end{eqnarray}
including those between hadrons of different flavor, e.g., for 
$B_s^{(*)}-B^{(*)}$ \cite{StateNotation}, 
\begin{eqnarray}
  E_s^{(1)} - E_{ud}^{(1)} &=& \bar\Lambda_s^{(1)} - \bar\Lambda_{ud}^{(1)}  
  \nonumber \\ 
  &=& M_{B_s^{(*)}} - M_{B^{(*)}} +  \nonumber \\ 
  && + \, {\cal O}(k_s^{(1)2}/m_b^{} \, , \, k^{(1)2}/m_b^{}) \; .
\end{eqnarray}
In particular, we use this quantity to set the physical strange quark mass 
(see Sec.\ \ref{subsect_phys_ms}).

\subsection{Couplings}
\label{subsect_couplings}

If we work at large enough values of $t$, then we may trust that, up to 
higher-order corrections, each side of Eq.\ (\ref{generalized}) describes a 
single mass eigenstate. 
Working with the left side of Eq.\ (\ref{generalized}), we find 
\begin{equation}
  \sum_{j=1}^r C(t)_{ij} \psi_j^{(\alpha)} = 
  \sum_{n=1}^\infty \sum_{j=1}^r v_i^{(n)} v_j^{(n)*} \psi_j^{(\alpha)} \, e^{-t \, E^{(n)}} \; .
\end{equation}
The solution of the generalized eigenvalue problem then ensures that 
\begin{equation}
  \sum_{j=1}^r v_j^{(n)*} \psi_j^{(\alpha)} = A^{(\alpha)} \, \delta^{(n,\alpha)} 
  + {\cal O}(e^{- (t - t_0) \Delta^{(\alpha)}}) \; ,
  \label{ev_kronecker}
\end{equation}
where $A^{(\alpha)}$ is an arbitrary normalization which one can easily remove 
(see below). 
Therefore, for large $t$ values, we can write \cite{ExcAmp} 
\begin{equation}
  \sum_{j=1}^r C(t)_{ij} \psi_j^{(\alpha)} \approx 
  v_i^{(\alpha)} A^{(\alpha)} \, e^{-t \, E^{(\alpha)}} \; .
  \label{singlestate_singleop}
\end{equation}
We can also use two eigenvectors, giving 
\begin{equation}
  \sum_{i=1}^r \sum_{j=1}^r \psi_i^{(\alpha)*} C(t)_{ij} \psi_j^{(\alpha)} 
  \approx A^{(\alpha)*} A^{(\alpha)} \, e^{-t \, E^{(\alpha)}} \; .
  \label{singlestate_allops}
\end{equation}

We may then construct and fit the following ratio to determine the couplings: 
\begin{eqnarray}
  R(t)_i^{(\alpha)} &=& \frac{\left|\sum_j C(t)_{ij} \psi_j^{(\alpha)}\right|^2}
  {\sum_k \sum_l \psi_k^{(\alpha)*} C(t)_{kl} \psi_l^{(\alpha)}}  \nonumber \\
  &\approx& v_i^{(\alpha)} v_i^{(\alpha)*} \, e^{-t E^{(\alpha)}} \; .
  \label{amplitudefunction}
\end{eqnarray}
Ratios of different couplings to the same mass eigenstate \cite{ExcAmp} 
are even easier: 
\begin{equation}
  \frac{v_i^{(\alpha)}}{v_k^{(\alpha)}} \approx \frac{\sum_j C(t)_{ij} \psi_j^{(\alpha)}}{\sum_l C(t)_{kl} \psi_l^{(\alpha)}} \; .
  \label{amplituderatio}
\end{equation}

Since we are working with the static approximation for the heavy quark, the 
local vector ($O_i = \bar q \gamma_i Q$) operator overlaps are related 
to the pseudoscalar ($PS$) decay constants via \cite{Manohar:2000dt} 
\begin{equation}
  f_{PS}^{(\alpha)} = \sqrt{\frac{2}{M^{(\alpha)}}} 
  \left( v_i^{(\alpha)} + {\cal O}(k^{(\alpha)2}/m_Q^{}) \right) \; .
  \label{overlap_to_decayconst}
\end{equation}
In order to cancel renormalization factors and coefficients matching HQET to 
QCD, we deal with ratios of overlaps. 
So, for example, the ratio of meson decay constants $f_{B_s'}/f_{B_s}$ may be 
extracted from the $m_q=m_s$ point of 
\begin{eqnarray}
  \frac{f_{PS}^{(2)}}{f_{PS}^{(1)}} &=& 
  \frac{v_i^{(2)}}{v_i^{(1)}} 
  \sqrt{\frac{M_{B_s^{(*)}}}{M_{B_s^{(*)}}+(E^{(2)}-E^{(1)})}}  \nonumber \\ 
  && + \, {\cal O}\left( k_s^{(2)2}/m_b^{} \, , \, 
  k_s^{(1)2}/m_b^{} \right) \; ,
  \label{f2_over_f1}
\end{eqnarray}
where $M_{B_s^{(*)}} = 5400$ MeV and the $v_i^{(\alpha)}$ and $E^{(\alpha)}$ come 
from fits to Eq.\ (\ref{amplitudefunction}) for $\alpha = 1$ and 2. 
Similarly, we can find $f_{B_s}/f_B$ from 
\begin{eqnarray}
  \frac{\left. f_{PS}^{(1)} \right|_{m_q=m_s}}
       {\left. f_{PS}^{(1)} \right|_{m_q=m_{ud}}} &=& 
  \frac{\left. v_i^{(1)} \right|_{m_q=m_s}}{\left. v_i^{(1)} \right|_{m_q=m_{ud}}} 
  \sqrt{\frac{M_{B^{(*)}}}{M_{B_s^{(*)}}}}  \nonumber \\ 
  && + \, {\cal O}\left( k_s^{(1)2}/m_b^{} \, , \, 
  k^{(1)2}/m_b^{} \right) \; ,
  \label{fBs_over_fB}
\end{eqnarray}
where $M_{B^{(*)}} = 5314$ MeV.

\subsection{Kinetic corrections}
\label{subsect_kin_corr}

In order to go beyond the static approximation for the heavy quark, we begin 
calculations of the ${\cal O}(1/m_Q^{})$ kinetic corrections to the 
static-light states. 
The new term in the Lagrangian is just the kinetic energy of the heavy quark 
and the corresponding operator is the lattice-discretized Laplacian. 
The relevant matrix elements are then 
\begin{equation}
  \varepsilon^{(n,n')} = - 
  \langle \, H_Q^{(n)} \, | \, \bar Q \vec D^2 Q \, | \, H_Q^{(n')} \, \rangle 
  \; .
\end{equation}
What we call $\varepsilon^{(1,1)}$ corresponds to the negative of the usual 
HQET parameter $\lambda_{1}$. 
From the $\varepsilon^{(n,n')}$ values, the energy shifts ($n = n'$) and 
static-light state mixings ($n \not= n'$) can be calculated: 
\begin{equation}
  \delta M^{(n)} = \frac{1}{2m_Q^{}} Z (\varepsilon^{(n,n)} - \varepsilon_0) 
  \; ,
  \label{kin_corr_mass}
\end{equation}
\begin{equation}
  \delta v_i^{(n)} = \frac{1}{2m_Q^{}} \sum_{n'\not=n} 
  \frac{Z(\varepsilon^{(n',n)}-\varepsilon_0)}{E^{(n)} - E^{(n')}} \, 
  v_i^{(n')} \; .
  \label{kin_corr_coupling}
\end{equation}
The subtraction of $\varepsilon_0$ represents the need to remove the 
ultraviolet (power-law) divergence in the lattice version of the 
$\bar Q \vec D^2 Q$ matrix element and the $Z$ denotes the corresponding 
renormalization factor for this operator \cite{RenormsHQET}. 
For a more complete, future study we will need to calculate these quantities; 
however, our focus here is the technique we use to find the bare corrections 
for $n,n'>1$. 
So for now we report differences of the $\varepsilon^{(n,n')}$ to cancel 
the first quantity and use a tadpole-improved \cite{Lepage:1992xa} version of 
$\vec D^2$ (dividing the spatial links used in the operator by $u_0$) to 
try to tame the second.

To extract the bare corrections, the quantities of interest are the lattice 
three-point functions 
\begin{equation}
  T(t,t')_{ij} = - 
  \langle \, 0 \, | \, O_i(t) \; \bar Q \vec D^2 Q(t') \; O_j^\dagger(0) \, | \, 0 \, \rangle \; .
  \label{3pt_function}
\end{equation}
The trouble with applying the variational method directly to this new matrix 
of correlators is that all entries involve a valence quark participating in 
the current insertion $\bar Q \vec D^2 Q$. 
This interaction therefore excites its own tower of states in the correlators 
$T_{ij}$, much like the source ($O_j^\dagger$) and sink operators ($O_i$) do.

One possible solution (yet an approximate one; see discussion below) is to 
view the three-point functions as taking part in two variational problems: 
\begin{eqnarray}
  C(t-t')_{ij} &=& \langle \, 0 \, | \, O_i(t) \; O_j^\dagger(t') \, | \, 0 \, \rangle  \nonumber \\ 
  C(t')_{ij} &=& \langle \, 0 \, | \, O_i(t') \; O_j^\dagger(0) \, | \, 0 \, \rangle \; .
  \label{2_2pt_functions}
\end{eqnarray}
Solving the two corresponding generalized eigenvalue equations, 
\begin{eqnarray}
  C(t-t') \, \vec{\psi}^{(\alpha)} &=& \lambda(t-t',t_0'-t')^{(\alpha)} \, C(t_0'-t') \, \vec{\psi}^{(\alpha)} \nonumber \\
  C(t') \, \vec{\phi}^{(\beta)} &=& \lambda(t',t_0)^{(\beta)} \, C(t_0) \, \vec{\phi}^{(\beta)} \; ,
  \label{two_generalized}
\end{eqnarray}
then gives sets of eigenvectors which one can use to project the states of 
interest on either side of the interaction: 
\begin{eqnarray}
  && \sum_{i=1}^r \sum_{j=1}^r \psi_i^{(\alpha)*} T(t,t')_{ij} \phi_j^{(\beta)}  
  \nonumber \\
  && \;\;\;\; \approx A^{(\alpha)*} \varepsilon^{(\alpha,\beta)} A^{(\beta)} \, 
  e^{-(t-t') E^{(\alpha)}} e^{-t' E^{(\beta)}} \; .
\end{eqnarray}
With a large ensemble and all-to-all quark propagators, translational 
invariance should strictly apply and one can use just one set of 
eigenvectors $\vec\psi^{(\alpha)}=\vec\phi^{(\alpha)}$, referenced by just the 
time difference $\Delta t=t'$ or $t-t'$ (this we do.) 
One may then form ratios to cancel the exponentials and some of the overlap 
factors: 
\begin{eqnarray}
  {\cal R}_{kl}^{(\alpha,\beta)} &=& 
  \frac{ \sum_i \sum_j \psi_i^{(\alpha)*} T(t,t')_{ij} \phi_j^{(\beta)} }
  { \sum_a \psi_a^{(\alpha)*} C(t-t')_{ak} \cdot 
    \sum_b C(t')_{lb} \phi_b^{(\beta)} }  \nonumber \\ 
  &\approx& \frac{\varepsilon^{(\alpha,\beta)}}{v_k^{(\alpha)*} v_l^{(\beta)}} \; .
  \label{generalized_3pt}
\end{eqnarray}
Similarly, it is possible to contract the open $k,l$ indices with the 
corresponding eigenvector components and end up with $A^{(\alpha)}A^{(\beta)*}$ 
in the denominator. 
Combining this with earlier ratios for determining the couplings, we obtain 
(assuming real overlaps) 
\begin{eqnarray}
  && {\cal R}_{ij}^{(\alpha,\beta)} \sqrt{R(t-t')_i^{(\alpha)}R(t')_j^{(\beta)}} 
  \nonumber \\ 
  && \;\;\;\; \approx \varepsilon^{(\alpha,\beta)} \, 
  e^{-[(t-t') E^{(\alpha)} + t' E^{(\beta)}]/2} \; .
\end{eqnarray}
We can fit such a quantity directly for the bare kinetic corrections 
$\varepsilon^{(\alpha,\beta)}$. 
(This technique and first results were presented in \cite{HagenLat08}.)

This shows how the variational method may be used to extract matrix elements 
and form factors (including transitions) involving excited hadronic states. 
The fact that we are dealing with ${\cal O}(1/m_Q^{})$ corrections to HQET 
states is irrelevant. 
The procedure can be easily generalized to, say, an electroweak current 
applied to full QCD states.

The salient feature of our analysis is that the three-point functions are 
treated in the context of two variational problems: one before the 
interaction and one after. 
It is conceivable, however, that a chosen basis of source and sink operators 
may not be robust enough to well approximate the intrusion of the probing 
current. 
In other words, the current itself must be viewed as a linearly independent 
operator exposing the higher dimensionality of the problem at hand 
(with lower lying states suddenly ``reappearing'' in a given eigenvalue 
channel $\alpha$, perhaps first in the form of seemingly additional noise). 
Indeed, in the limit of large statistics, this should be seen with {\it any 
finite basis in which the interaction is not included}. 
To be completely accurate, the current, together with the evolution to the 
sink or from the source, has to be added to the basis as an additional 
operator. 
Such computations may be prohibitively expensive since the new diagonal 
elements of such an expanded correlator matrix would be four-point functions.

One may also try a more simplified version of Eq.\ (\ref{generalized_3pt}) 
for the case of $\alpha=\beta$: 
\begin{eqnarray}
  {\cal R}^{(\alpha)} &=& 
  \frac{ \sum_i \sum_j \psi_i^{(\alpha)*} T(t,t')_{ij} \psi_j^{(\alpha)} }
  { \sum_k \sum_l \psi_k^{(\alpha)*} C(t)_{kl} \psi_l^{(\alpha)} }  \nonumber \\ 
  &\approx& \varepsilon^{(\alpha,\alpha)} \; ,
  \label{simp_generalized_3pt}
\end{eqnarray}
where $\vec\psi^{(\alpha)}$ is now the eigenvector from the solution of the 
generalized eigenvalue problem for $C(t)$, Eq.\ (\ref{generalized}). 
As long as $t'$ and $t-t'$ are large enough, using these eigenvectors or the 
ones from Eqs.\ (\ref{two_generalized}) should give consistent results.

\section{Lattice calculation details}
\label{SectLattice}

\subsection{Configuration details}

We perform runs on a combination of both quenched ($N_f=0$) and dynamical 
($N_f=2$) ensembles \cite{DynCI}. 
The gauge action is one-loop improved L\"uscher-Weisz \cite{LWaction} and 
we use (for both sea and valence quarks) chirally improved (CI) fermions 
\cite{CIaction}, which have been shown to have good chiral properties 
\cite{BGRlarge}. 
We set the scale of the lattice spacing via the static-quark potential 
\cite{scale}, using the value $r_0=0.49(1)$ fm \cite{r0_error}. 
Except where noted otherwise, the links of the quenched configurations are 
Hyp smeared \cite{Hyp} and the dynamical ones are Stout smeared 
\cite{Stout}. 
The details of the configurations can be found in Table \ref{latticetable}.

\begin{table}[!b]
\caption{
Relevant lattice parameters. The scales are set using $r_0=0.49(1)$ fm 
\cite{r0_error}.
}
\label{latticetable}
\begin{center}
\begin{tabular}{ccccccc}
\hline \hline
$\beta$ & $N_s^3 \times N_t$ & $N_{conf}$ & $a^{-1}$ (MeV) & 
$M_{\pi,\mbox{sea}}$ (MeV) & $L$ (fm) \\
\hline
8.15 & $20^3 \times 40$ & 100 & 1682(34) & $\infty$ & 2.34 \\
7.90 & $16^3 \times 32$ & 100 & 1360(28) & $\infty$ & 2.32 \\
7.57 & $12^3 \times 24$ & 200 & 1010(21) & $\infty$ & 2.34 \\
4.65 & $16^3 \times 32$ & 100 & 1263(26) & $461(6)(9)$ & 2.50 \\
5.20 & $12^3 \times 24$ & 74 & 1750(36) & $525(22)(11)$ & 1.35 \\
\hline \hline
\end{tabular}
\end{center}
\end{table}

In Table \ref{scaletable} we outline the basic results of a simple test of 
our choice of a physical value for $r_0$. 
Using tadpole-improved, $O(v^4)$-accurate lattice NRQCD (with unsmeared gauge 
links; see Ref.\ \cite{ExcAmp} for details of the action) we create heavy 
quarkonium correlators at zero and finite momentum. 
We use the spin-averaged $1P-\Upsilon$ splitting (440 MeV) to set the lattice 
spacing for these runs and the kinetic $\Upsilon$ mass to set the physical 
$b$ quark mass. 
The results on the $\beta=4.65$ dynamical lattice give $1/a = 1231(7)$ MeV, 
which, together with the value of $r_0/a$ from the static-quark potential, 
gives $r_0=0.5027(29)$ fm. 
This value is quite precise, but incorporates partial quenching effects since 
the sea quarks are still quite heavy ($M_{\pi,\mbox{sea}} \approx 460$ MeV). 
However, we are encouraged by its consistency with our ``working value'' of 
0.49(1) fm, which we have chosen to be slightly lower than our previous 
value (0.50 fm in Refs.\ \cite{DDI,Burch:2007xy}) due to a seeming average 
trend in the results of the larger lattice community (see, e.g., 
Ref.\ \cite{McNeile:2007fu}).

\begin{table}[!b]
\caption{
Scale setting on the $\beta=4.65$ ($r_0/a=3.1414(21)$) dynamical lattice 
using ${\cal O}(v^4)$ NRQCD $\bar bb$ correlators, $\Delta E(1P-\Upsilon)=440$ 
MeV (to set $a^{-1}$), and $M_\Upsilon=9460$ MeV (to set $am_b$). 
Result: $r_0=0.5027(29)$ fm, consistent with our more conservative use of 
$r_0=0.49(1)$ fm throughout this work.
}
\label{scaletable}
\begin{center}
\begin{tabular}{cccccc}
\hline \hline
$am_b$ & state & $t/a$ & $aE$ & $a^{-1}$ & $M_{1S,\mbox{kin}}$ \\
\hline
3.2 & $1S$ & 20-25 & 0.44338(29) & & \\
 & $1P$, spin- & & & & \\
 & averaged & 3-8 & 0.8004(21) & 1232(7) & \\
 & $1S$ @ & & & MeV & \\
 & $p=2\pi/aN_s$ & 20-25 & 0.45369(30) & & 9214 MeV \\
\hline
3.6 & $1S$ & 20-25 & 0.42580(26) & & \\
 & $1P$, spin- & & & & \\
 & averaged & 3-8 & 0.7851(21) & 1225(7) & \\
 & $1S$ @ & & & & \\
 & $p=2\pi/aN_s$ & 20-25 & 0.43486(27) & & 10426 \\
\hline
3.28 & & & & 1231(7) & 9460 \\
\hline \hline
\end{tabular}
\end{center}
\end{table}

\subsection{Light-quark propagators}

In this subsection we briefly cover the improved estimates of half-to-half 
propagators which we use for the light quarks. 
Details of this so-called domain-decomposition improvement (DDI) can be found 
in Ref.\ \cite{DDI}. 
The technique is inspired by earlier work for pseudofermion estimators 
\cite{MVR}, but represents a more tractable method for improved lattice 
fermions.

If we decompose the lattice into two distinct domains, the lattice Dirac 
operator (${\cal M}$) may be written in terms of submatrices 
\begin{equation}
  {\cal M} = \left( 
    \begin{array}{cc}
      {\cal M}_{11} & {\cal M}_{12} \\
      {\cal M}_{21} & {\cal M}_{22} 
    \end{array} \right) \, ,
\end{equation}
where ${\cal M}_{11}$ and ${\cal M}_{22}$ connect sites within a region and 
${\cal M}_{12}$ and ${\cal M}_{21}$ connect sites from the different regions. 
We may also write the propagator (${\cal P}$) in this form: 
\begin{equation}
  {\cal M}^{-1} = {\cal P} = \left( 
    \begin{array}{cc}
      {\cal P}_{11} & {\cal P}_{12} \\
      {\cal P}_{21} & {\cal P}_{22} 
    \end{array} \right) \, .
\end{equation}
The propagator between regions 1 and 2 can then be estimated using $N$ random 
sources ($\chi^i$, $i=1,..,N$) \cite{DDI}: 
\begin{eqnarray}
  {\cal P}_{12} &=& - {\cal M}^{-1}_{11} {\cal M}_{12} {\cal P}_{22}  \nonumber \\ 
  &\approx& - {\cal M}^{-1}_{11} {\cal M}_{12} \frac1N \sum_i \chi_2^{i} \chi_2^{i\dagger} {\cal P}  
  \nonumber \\ 
  &\approx& -\frac1N \sum_i \left( {\cal M}^{-1}_{11} {\cal M}_{12} \chi_2^{i} \right) \left( \gamma_5 {\cal P} \gamma_5 \chi^{i}_2 \right)^{\dagger} \equiv  \nonumber \\
  && \;\;\;\;\; -\frac1N \sum_i \Psi^{i}_1 \Phi^{i\dagger}_2 \; .
\end{eqnarray}
The $2N$ solution vectors ($\Psi^{i}$,$\Phi^{j}$) are calculated and saved to 
later reconstruct the estimated light-quark propagator from anywhere in 
region 1 to anywhere in region 2. 
In this form we see that no sources are needed in region 1 and those in 
region 2 should reach region 1 with one application of ${\cal M}$. 
Since ${\cal M}$ is usually a sparse matrix (even for the CI operator), this 
greatly reduces the number of lattice sites which the random sources need to 
cover and therefore greatly improves the signal-to-noise ratio for 
long-distance correlators.

We stick with the implementation as originally detailed in \cite{DDI}. 
In particular, each domain contains half of the timeslices, the sources 
appear in both regions at the boundary edges (in order to obtain two 
estimates of ${\cal P}_{12}$), and we divide our random sources (originally, 
$N=12$) into a factor four greater in number by making the spin summation 
explicit (spin ``dilution'').

\subsection{Operator basis and correlator construction}
\label{subsect_basis_corrs}

The local versions of the operators we use can be found in Table 
\ref{operatortable}. 
This is just the start of our basis and we expand it further by 
gauge-covariantly smearing \cite{Jacobi} the light-quark wavefunctions.

\begin{table}[!b]
\caption{
Heavy hadron operators. 
Shown are the total light-quark (plus glue) angular momentum $j$ (lowest 
value in the continuum), the possible quantum numbers $J^P$ when including 
the heavy-quark spin, and the spin-averaged heavy-chiral (-strange) states 
\cite{StateNotation}. 
The last column displays the local version of the operator; color ($a,b,c$) 
and Lorentz indices are explicit, while spatial and spin indices 
are implicit.
}
\label{operatortable}
\begin{center}
\begin{tabular}{ccccc}
\hline \hline
label & $j$ & $J^P$ & states & operator \\
\hline
\vspace*{1mm}
$S$ & $\frac12$ & $(0,1)^-$ & $B^{(*)}$ ($B_s^{(*)}$) & $\bar Q_a \, \gamma_{5,i} \, q_a$ \\
\hline
\vspace*{1mm}
$P_-$ & $\frac12$ & $(0,1)^+$ & $B_{0,1}^*$ ($B_{s0,1}^*$) & $\bar Q_a \, \gamma_i (D_i)_{ab} \, q_b$ \\
\hline
\vspace*{1mm}
$P_+$ & $\frac32$ & $(1,2)^+$ & $B_{1(2)}^{(*)}$ ($B_{s1(2)}^{(*)}$) & $\bar Q_a \, (\gamma_1 D_1 - \gamma_2 D_2)_{ab} \, q_b$ \\
\hline
\vspace*{1mm}
$D_\pm$ & $(\frac32,\frac52)$ & $(1,2,3)^-$ & $B_{2(3)}^{(*',*)}$ ($B_{s2(3)}^{(*',*)}$) & $\bar Q_a \, \gamma_5 (D_1^2 - D_2^2)_{ab} \, q_b$ \\
\hline
\vspace*{1mm}
$\Lambda_Q$ & 0 & $\frac12^+$ & $\Lambda_b$ (-) & $\epsilon_{abc} \, Q_a \, q_b \, C \gamma_5 \, q_c$ \\
 & & & & $i \epsilon_{abc} \, Q_a \, q_b \, C \gamma_4 \gamma_5 \, q_c$ \\
\hline
\vspace*{1mm}
$\Sigma_Q^{(*)}$ & 1 & $(\frac12,\frac32)^+$ & $\Sigma_b^{(*)}$ ($\Omega_b^{(*)}$) & $\epsilon_{abc} \, Q_a \, q_b \, C \gamma_i \, q_c$ \\
 & & & & $i \epsilon_{abc} \, Q_a \, q_b \, C \gamma_4 \gamma_i \, q_c$ \\
\hline \hline
\end{tabular}
\end{center}
\end{table}

In choosing how to smear the light-quark sources (and sinks), we work with a 
different philosophy than we employed previously (see, e.g., 
Ref.\ \cite{Burch:2004he}). 
Here, rather than designing the sources, we simply make choices such that the 
resulting operators are drastically different from each other (having, e.g., 
relatively small overlap in the free case). 
As long as we do not make the sources too large or too small, this should 
ensure that the states of interest (the lower-lying ones) have various 
amounts of overlap with the different operators, hopefully making the states 
easier to resolve.

Accordingly, for each set of meson quantum numbers, in addition to the local 
operator, we construct three smeared versions: one left as is, one with a 
subsequent gauge-covariant Laplacian applied, and the last with two such 
applications. 
For the baryons, we have two different local operators and we use smeared 
versions as well (so far, only smeared on the $\beta=8.15$ lattice). 
This allows us to use up to a $4 \times 4$ basis for each desired $J^P$ 
(reduced to $3 \times 3$ for the mesons on the dynamical lattices due to 
lower statistics). 
The parameters chosen for the smearings are shown in Table 
\ref{smear_params}.

\begin{table}[!b]
\caption{
Parameters for the quark source smearings. 
Listed are the number of smearing steps $N_{sm,i}$, the hopping weight factor 
$\kappa$, and the number of Laplacians applied $l_i$ for each set of 
configurations. 
Except where noted otherwise, the link smearing is also in place for the 
propagation of both the CI light quark and the static quark.
}
\label{smear_params}
\begin{center}
\begin{tabular}{ccccc} \hline \hline
$\beta$ & link smearing & $l_i$ & $N_{sm,i}$ & $\kappa$ \\ \hline
8.15  & Hyp\cite{Hyp} & (0,0,1,2) & (0,16,24,32) & 0.2 \\
7.90  & Hyp & (0,0,1,2) & (0,12,18,24) & 0.2 \\
7.57  & Hyp & (0,0,1,2) & (0,8,12,16) & 0.2 \\
4.65  & Stout\cite{Stout} & (0,0,1,2) & (0,12,18,24) & 0.2 \\
5.20  & Stout & (0,0,1,2) & (0,12,18,24) & 0.2 \\ \hline \hline
\end{tabular}
\end{center}
\end{table}

\begin{figure*}[!t]
\begin{center}
\includegraphics*[width=5.7cm]{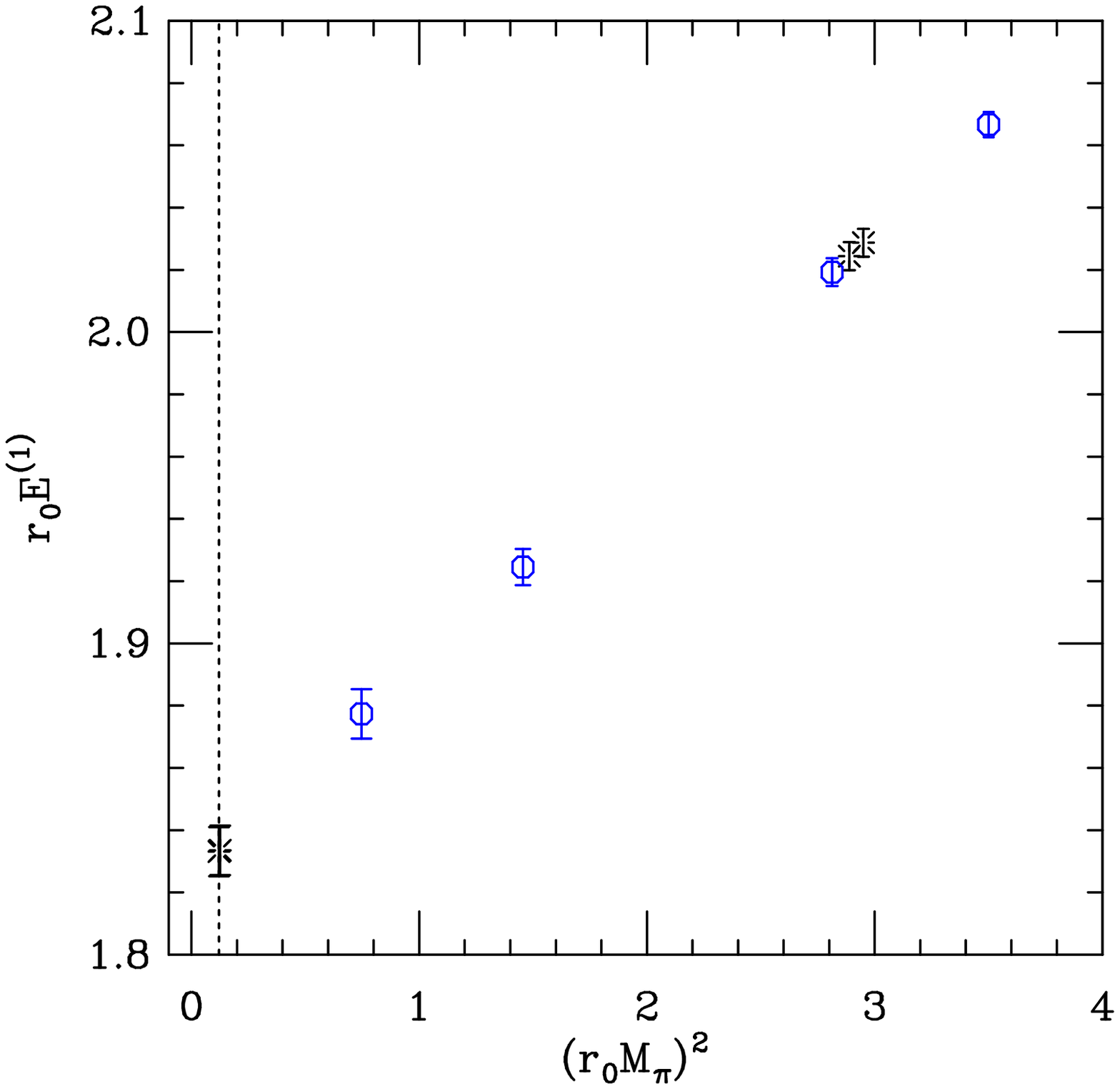}
\includegraphics*[width=5.7cm]{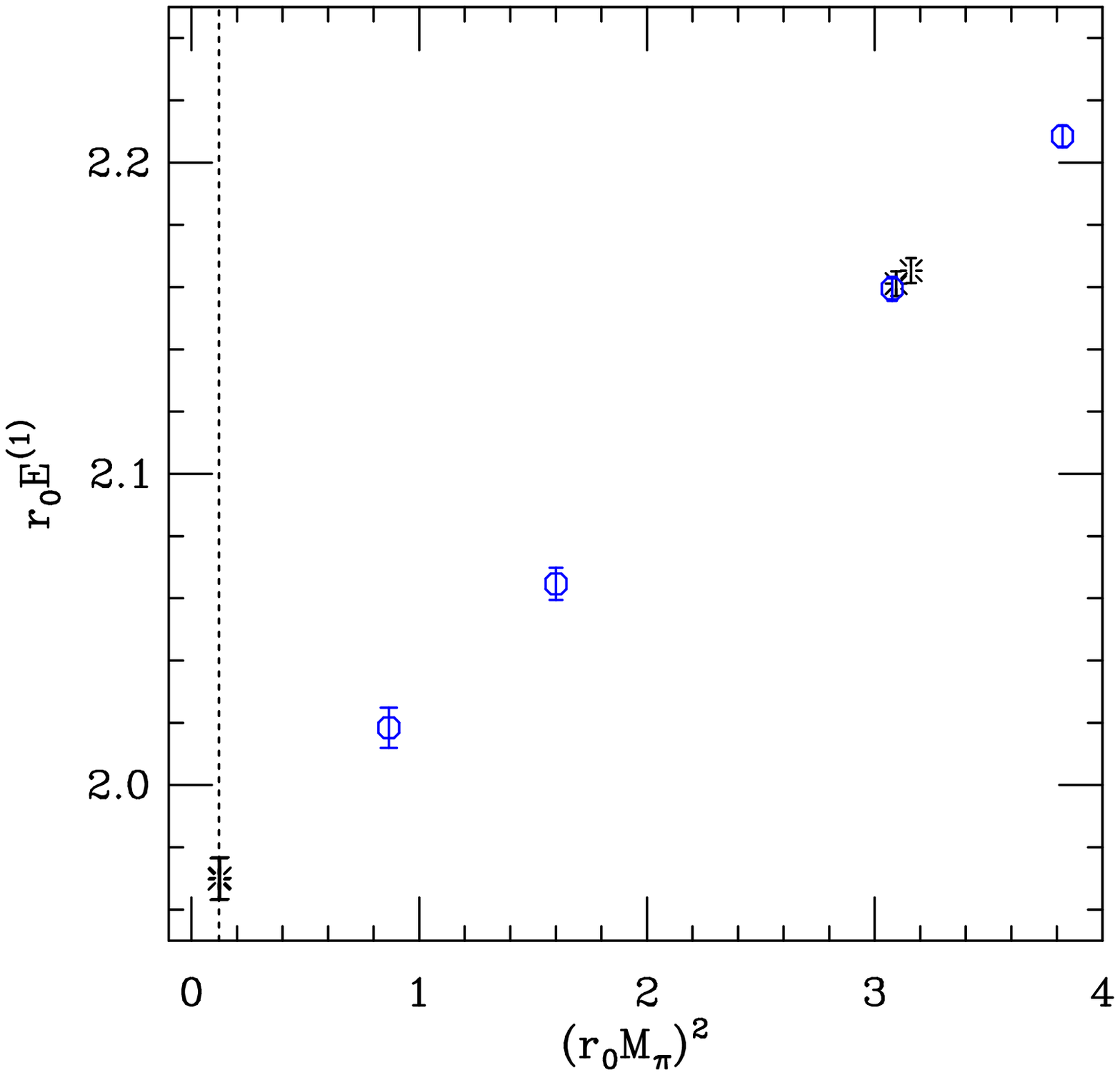}
\includegraphics*[width=5.7cm]{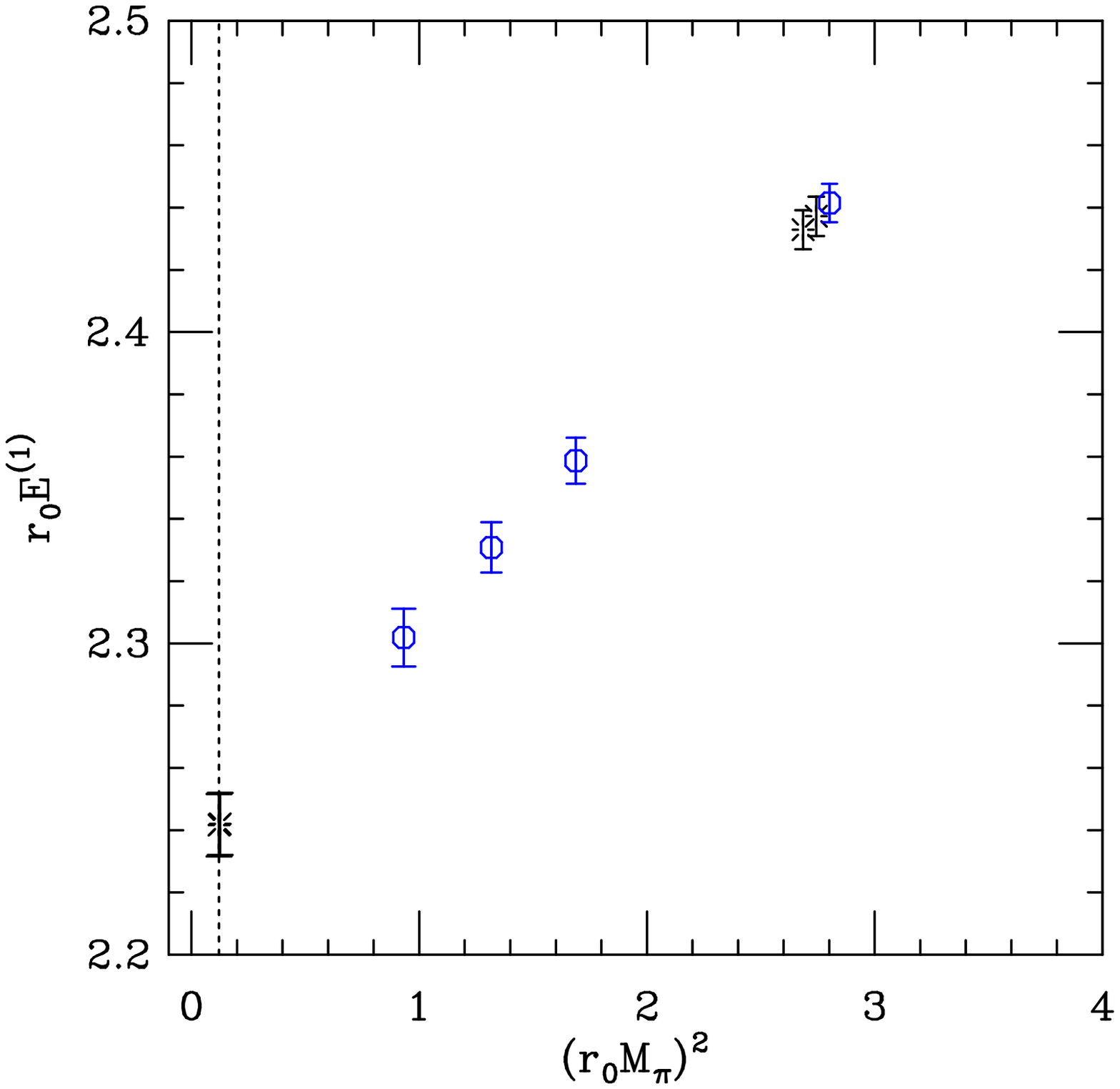}
\end{center}
\caption{
The setting of the strange quark mass on the $\beta=7.90$ and $\beta=8.15$ 
quenched lattices and the $\beta=4.65$ dynamical lattice (from left to right, 
respectively). 
Plotted are the $1S$ lattice energies $r_0E^{(1)}$ versus $(r_0M_\pi)^2$. 
The $E_s^{(1)}-E_{ud}^{(1)}$ difference is set to the 
static-strange--static-light mass difference of 76.9 MeV, which is the 
$1/M_{H^{(*)}} \rightarrow 0$ linear extrapolation of the experimental values 
$M_{B_s^{(*)}}-M_{B^{(*)}}=86.8$ MeV and $M_{D_s^{(*)}}-M_{D^{(*)}}=103.5$ MeV. 
The burst at the left shows the linear chiral extrapolation ($E_{ud}^{(1)}$) 
and the two bursts at the right show the linear strange-mass interpolations 
($E_s^{(1)}$) using $r_0=0.49$ and 0.50 fm.
}
\label{r0E1_vs_r0Mpi2}
\end{figure*}

Given the operators of the basis, the correlator matrix can then be 
constructed. 
For example, for the static-light mesons we have 
\begin{eqnarray}
  && \!\!\!\!\!\! C(t)_{jk} = \langle \, 0 \, | \, O_j(t) \; O_k^\dagger(0) \, | \, 0 \, \rangle =  \nonumber \\
  && \!\!\!\!\! \left\langle \sum_x \mbox{Tr} \left[ 
    \frac{1+\gamma_4}{2}\prod_{i=0}^{t-1} U_4^\dagger(x+i\hat{4}) \, 
    o_j {\cal P}_{x+t\hat{4},x} o_k^\dagger \right] \right\rangle \, ,
\end{eqnarray}
where the product of the (smeared) links in the time direction corresponds to 
the evolution of the (spinless) static quark, the lower-case operators $o_j$ 
correspond to what remains of the $O_j$ after integration over the fermion 
fields ($Q$, $\bar Q$, $q$, $\bar q$), 
and ${\cal P}_{x+t\hat{4},x}$ is the estimated light-quark propagator 
(discussed in the previous subsection). 
So the lattice points $x$ and $x+t\hat{4}$ must be within different domains 
and the ensemble average implied by the angled brackets also includes a 
division by the number of correlators of extent $t$ which cross the boundary: 
$min(t,T-t)$. 
For the static-light-light baryons, there is a similar expression, but 
without the $(1+\gamma_4)/2$ projection and where the (now, two) estimated 
light quarks are constructed without using the same random sources in the 
correlator contributions (i.e., 
${\cal P}_{xy}{\cal P}_{xy}'\propto\sum_i\sum_j\Psi_x^i\Phi_y^{i\dagger}
\Psi_x^j\Phi_y^{j\dagger}$ has the restriction that $i\not=j$).

Before proceeding to solve the generalized eigenvalue problem, we check that 
the correlator matrices are real and symmetric (within errors). 
We then explicitly symmetrize them and use a Cholesky decomposition of 
$C(t_0)$ to solve Eqs.\ (\ref{generalized}) and (\ref{two_generalized}).

\subsection{First look at the ground state: Physical strange quark mass}
\label{subsect_phys_ms}

To determine the physical strange quark mass, we consider spin-averaged 
experimental heavy-light meson masses. 
Up to ${\cal O}(1/m_Q^{})$ in HQET, the average mass of the heavy-light 
ground-state spin multiplet is given by (using our notation) 
\begin{equation}
  M_{H_q^{(*)}} = m_Q^{} + E_q^{(1)} - E_0 + 
  \frac{Z(\varepsilon_q^{(1,1)}-\varepsilon_0)}{2m_Q^{}} \; .
\end{equation}
Since we want to know the strange quark mass, we then consider the following 
difference 
\begin{equation}
  E_s^{(1)} - E_{ud}^{(1)} = M_{H_s^{(*)}} - M_{H^{(*)}} - 
  \frac{Z(\varepsilon_s^{(1,1)}-\varepsilon_{ud}^{(1,1)})}{2m_Q^{}} \; .
\end{equation}
To minimize the effect of the $1/m_Q^{}$ term on the right hand side, we 
linearly extrapolate the experimental numbers \cite{PDG} 
$M_{D_s^{(*)}}-M_{D^{(*)}}=103.5$ MeV at $1/M_{D^{(*)}}$ and 
$M_{B_s^{(*)}}-M_{B^{(*)}}=86.8$ MeV at $1/M_{B^{(*)}}$ to 
$1/M_{H^{(*)}} \rightarrow 0$. 
This gives $M_{H_s^{(*)}}-M_{H^{(*)}}=76.9$ MeV and we use this number to set the 
value of $(r_0M_\pi)^2$ corresponding to the physical strange quark mass.

Figure \ref{r0E1_vs_r0Mpi2} shows the ground-state static-light $S$-wave 
masses $E^{(1)}$ versus the corresponding light-quark masses 
($\propto M_\pi^2$) on three of the ensembles we use (see the next section 
for details surrounding the fits). 
The data points are consistent with a solely linear dependence over the region 
of simulated quark masses and we therefore perform only linear extrapolations 
(to $m_{ud}$) and interpolations (to $m_s$).

Given the quality of these results and the completely linear behavior they 
display, one can imagine using them as input in a determination of 
$\overline{m_s}$(2 GeV). 
We plan to pursue this in the near future.

\section{Analysis}
\label{SectAnalysis}

\begin{figure*}[!t]
\begin{center}
\includegraphics*[width=4.3cm]{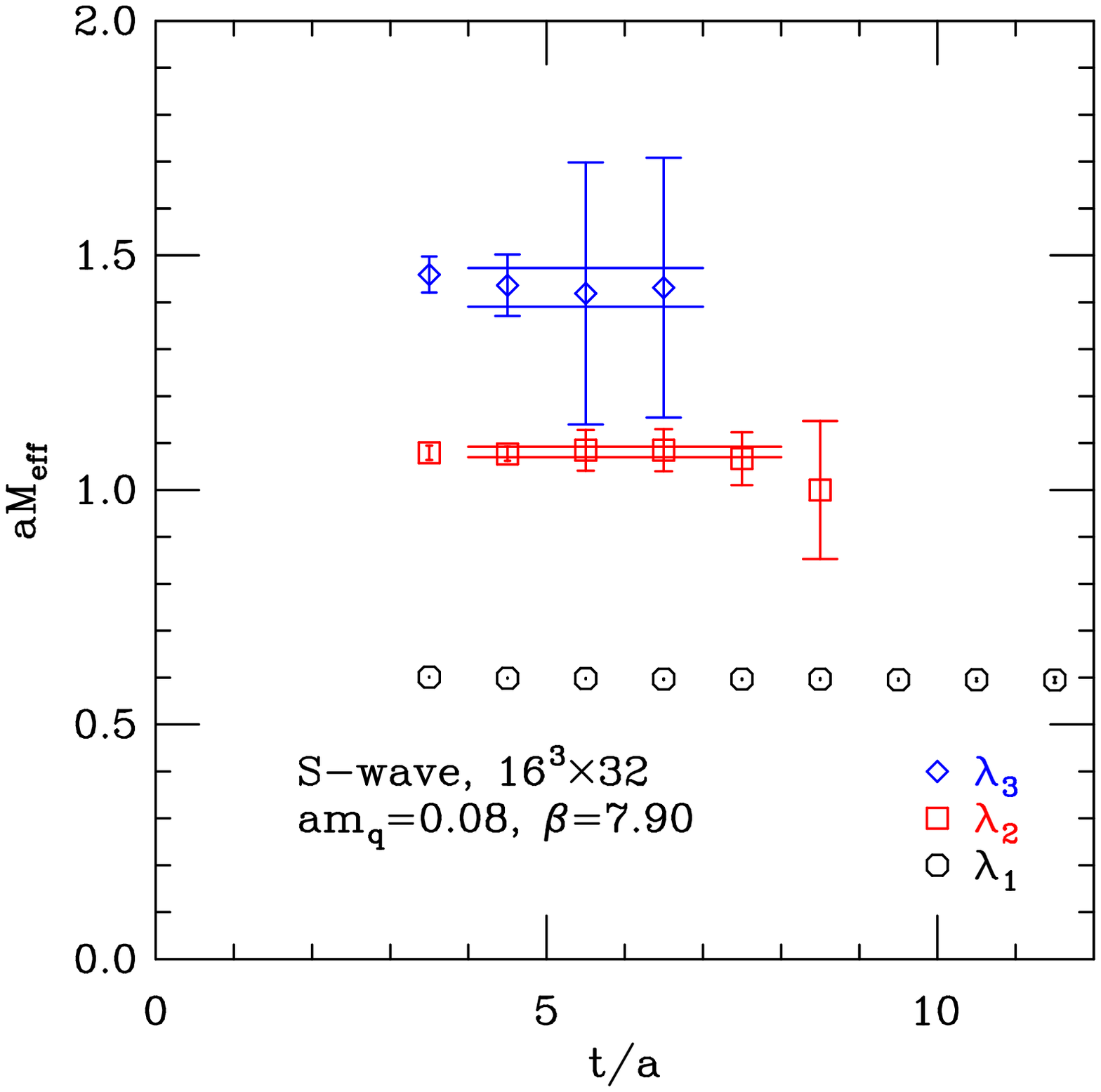}
\includegraphics*[width=4.3cm]{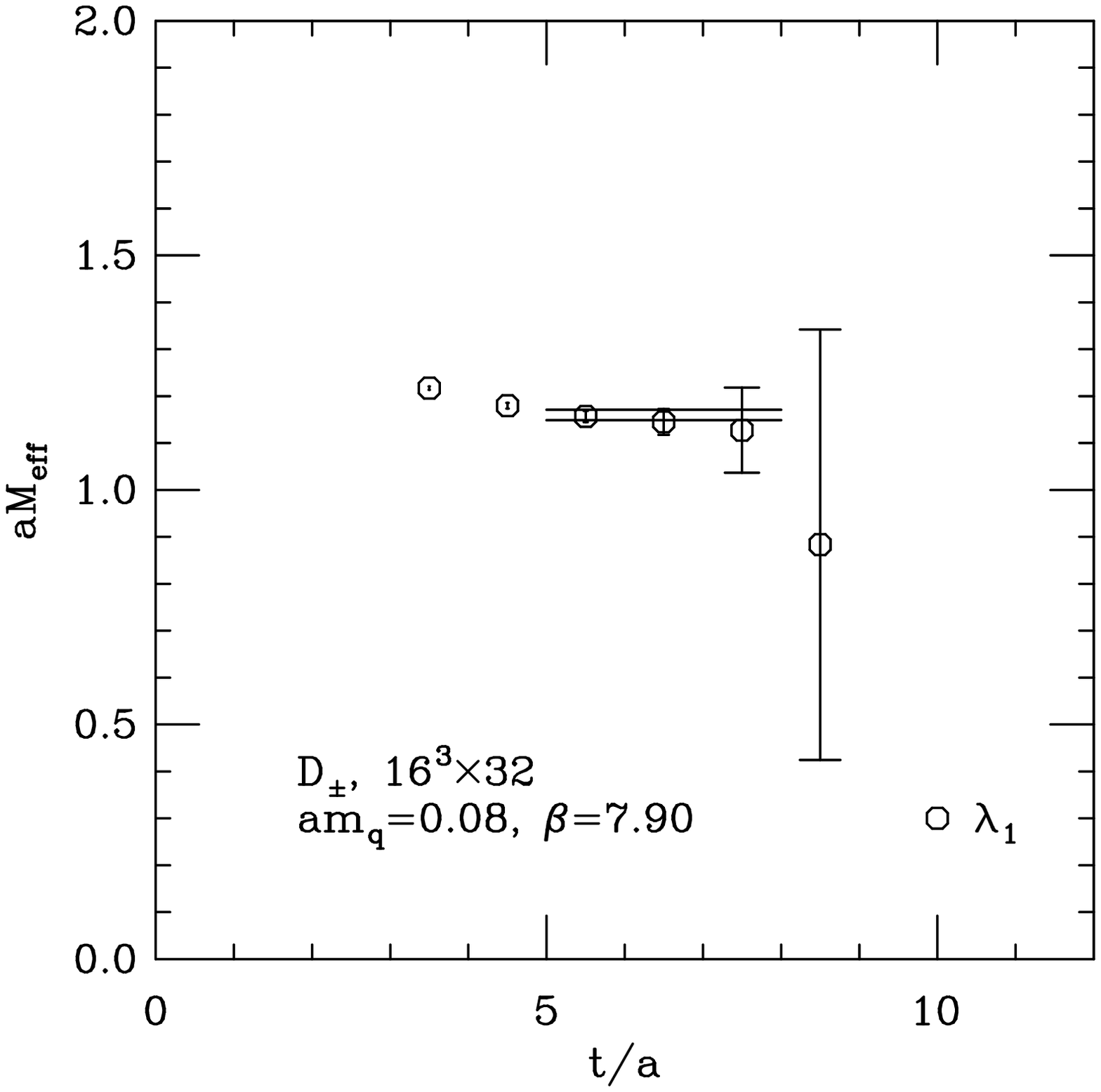}
\includegraphics*[width=4.3cm]{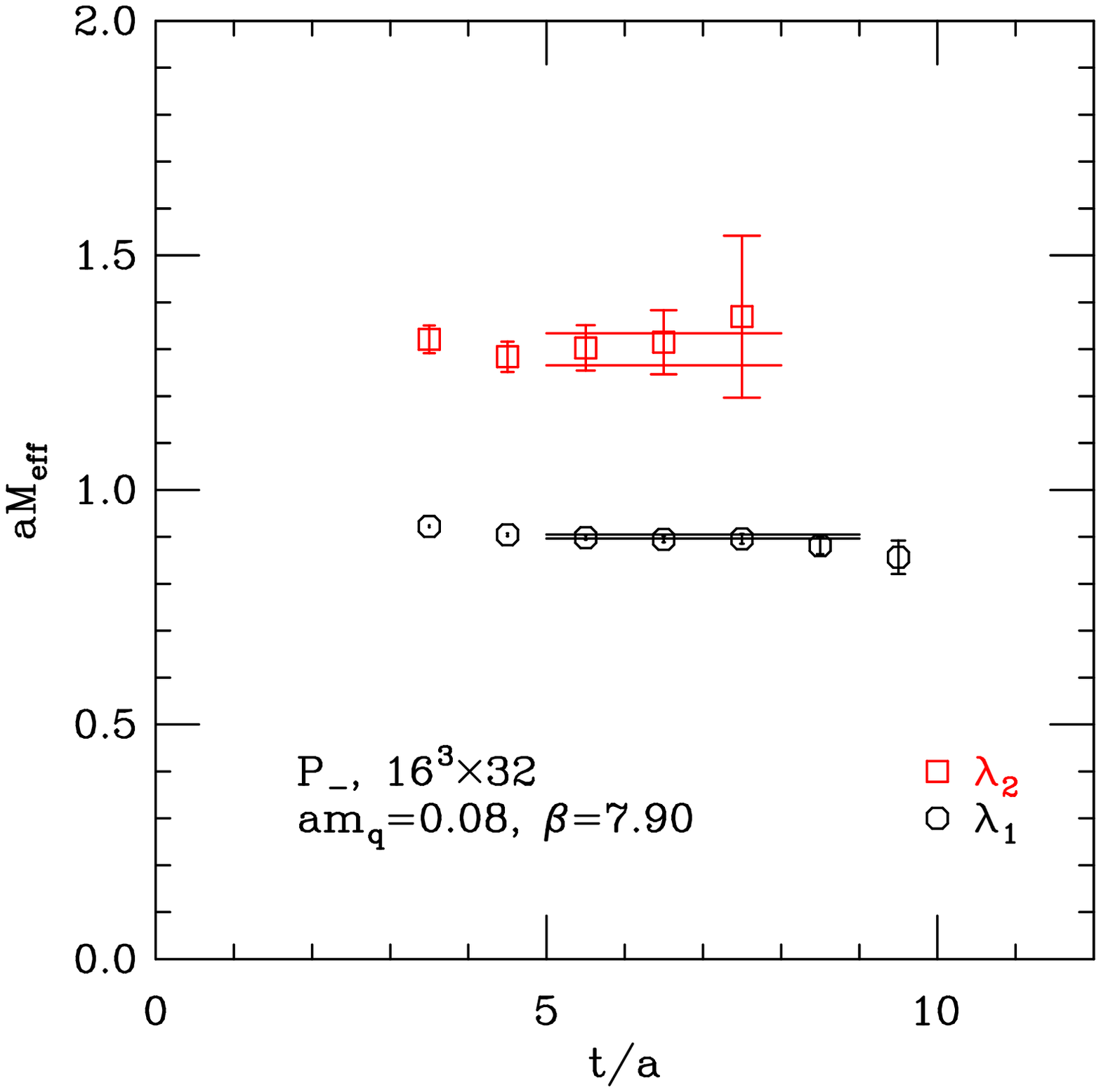}
\includegraphics*[width=4.3cm]{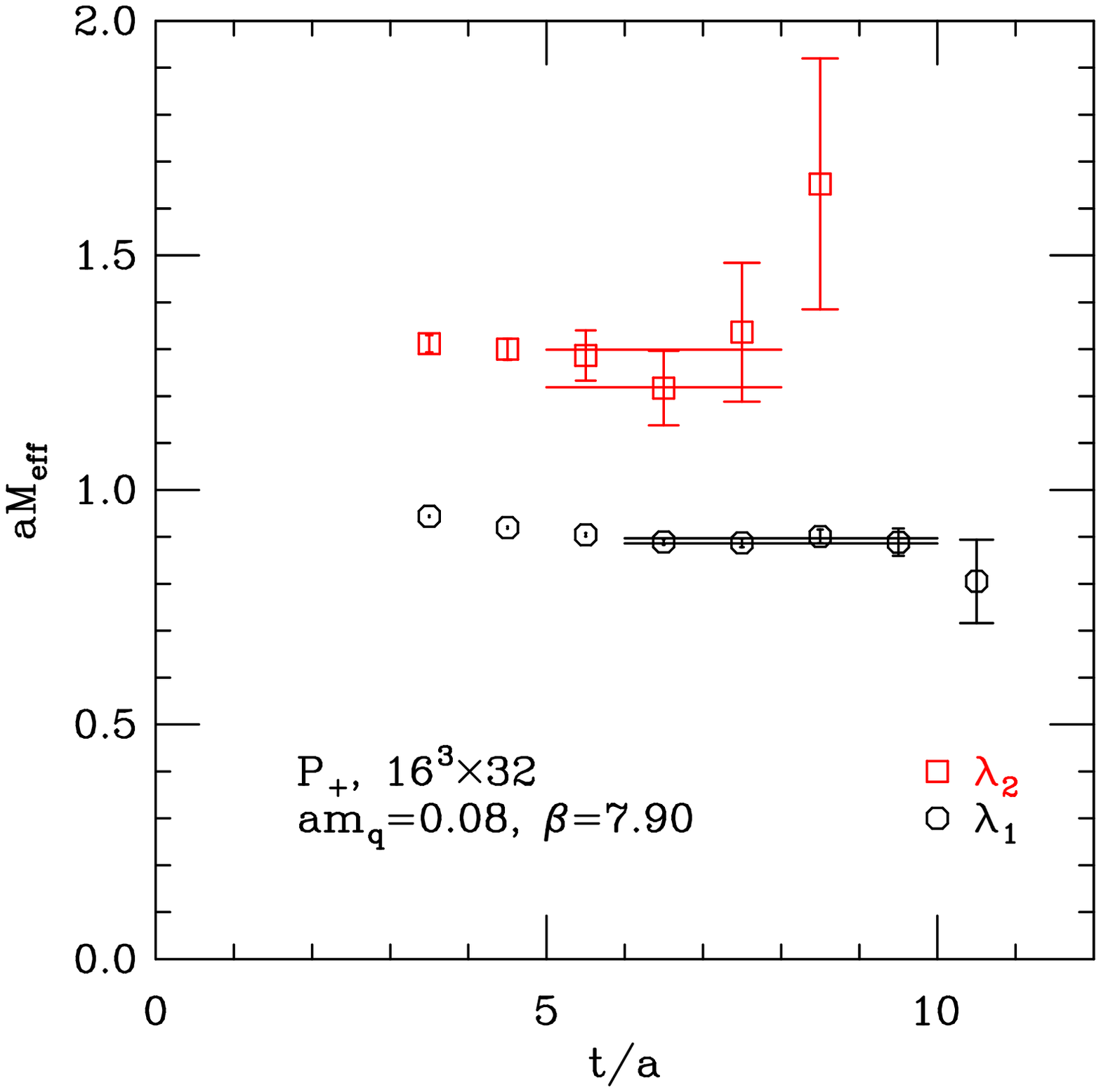}
\end{center}
\caption{
Sample of meson effective masses from the $16^3\times32$, $\beta=7.90$ 
quenched lattice ($am_q=0.08$). 
From left to right are the $S$, $D_\pm$, $P_-$, and $P_+$. 
The horizontal lines denote the $aE^{(n)} \pm a\sigma^{(n)}$ correlated fit 
values and the corresponding time ranges.
}
\label{effmass_b790}
\end{figure*}

\begin{figure*}[!t]
\begin{center}
\includegraphics*[width=4.3cm]{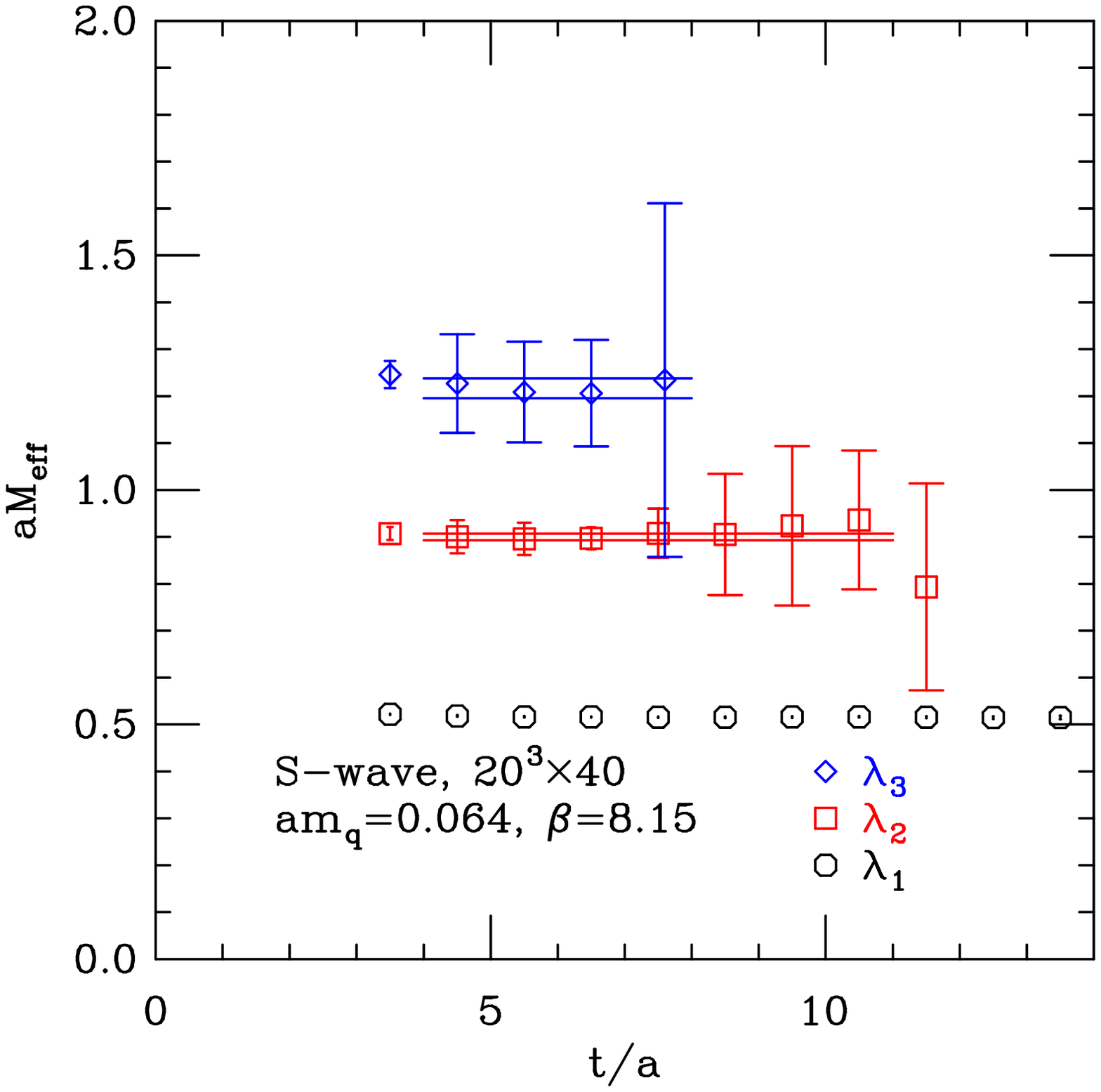}
\includegraphics*[width=4.3cm]{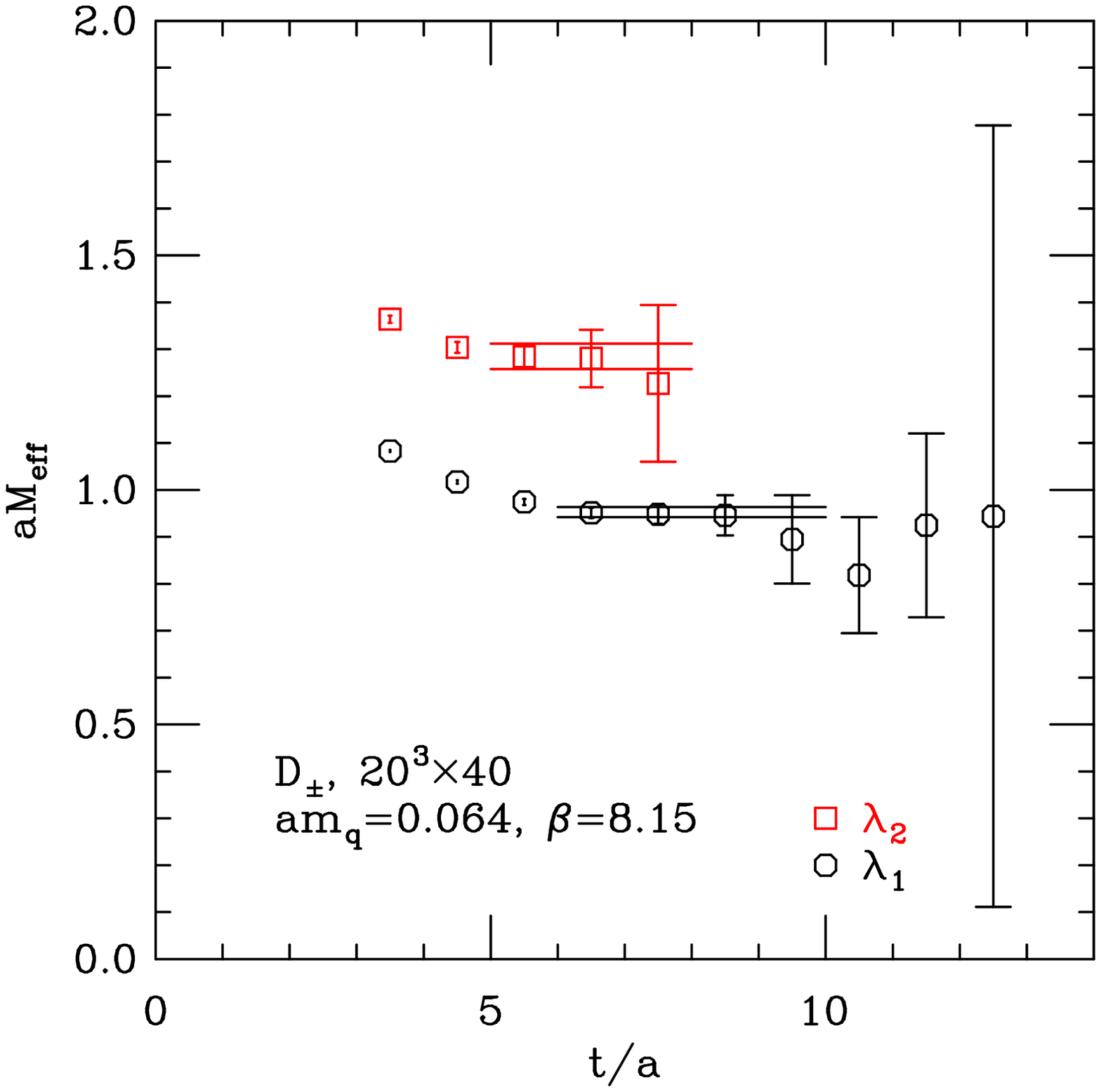}
\includegraphics*[width=4.3cm]{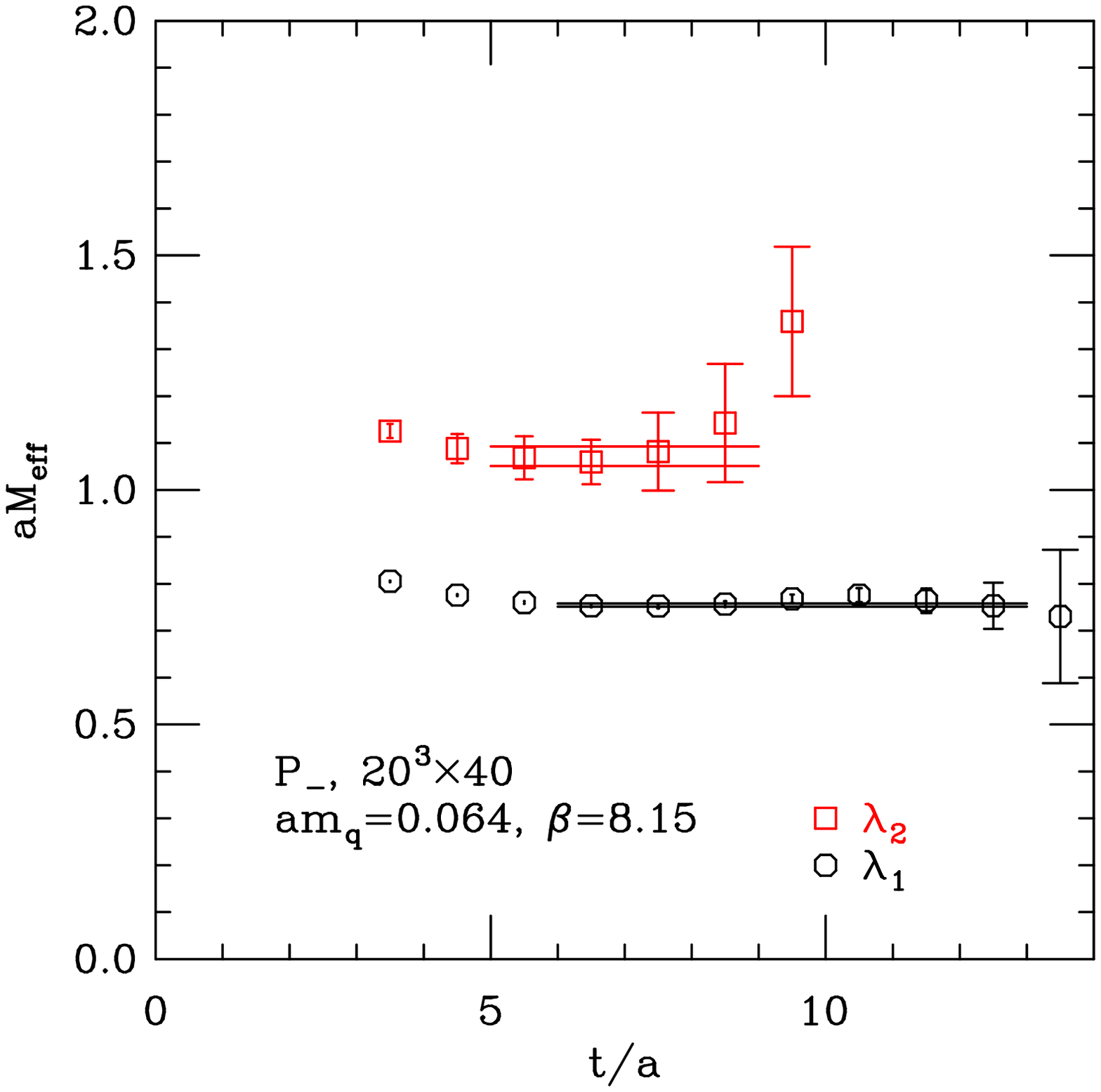}
\includegraphics*[width=4.3cm]{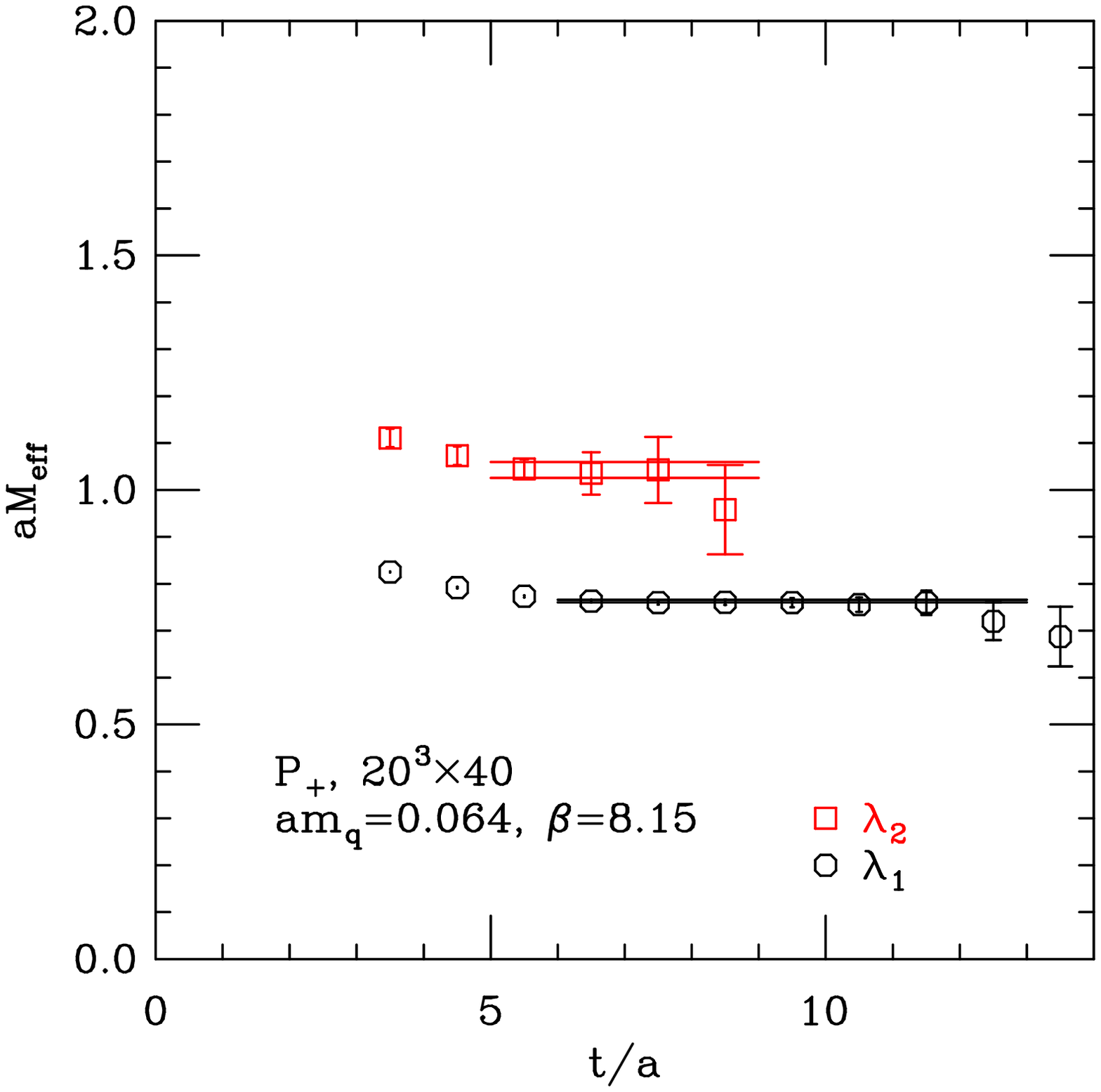}
\end{center}
\caption{
Sample of meson effective masses from the $20^3\times40$, $\beta=8.15$ 
quenched lattice ($am_q=0.064$). 
From left to right are the $S$, $D_\pm$, $P_-$, and $P_+$. 
The horizontal lines denote the $aE^{(n)} \pm a\sigma^{(n)}$ correlated fit 
values and the corresponding time ranges.
}
\label{effmass_b815}
\end{figure*}

\begin{figure*}[!t]
\begin{center}
\includegraphics*[width=4.3cm]{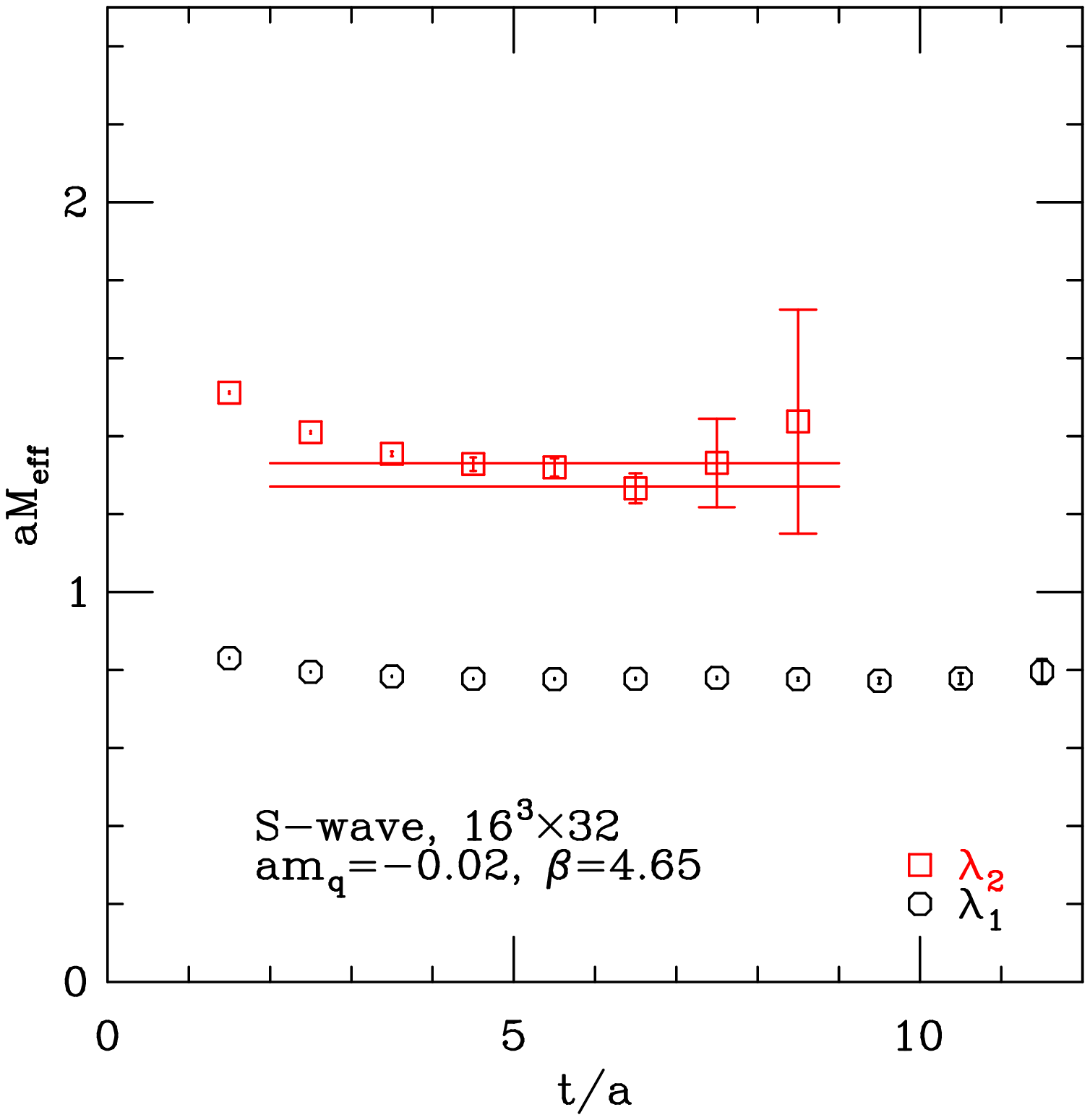}
\includegraphics*[width=4.3cm]{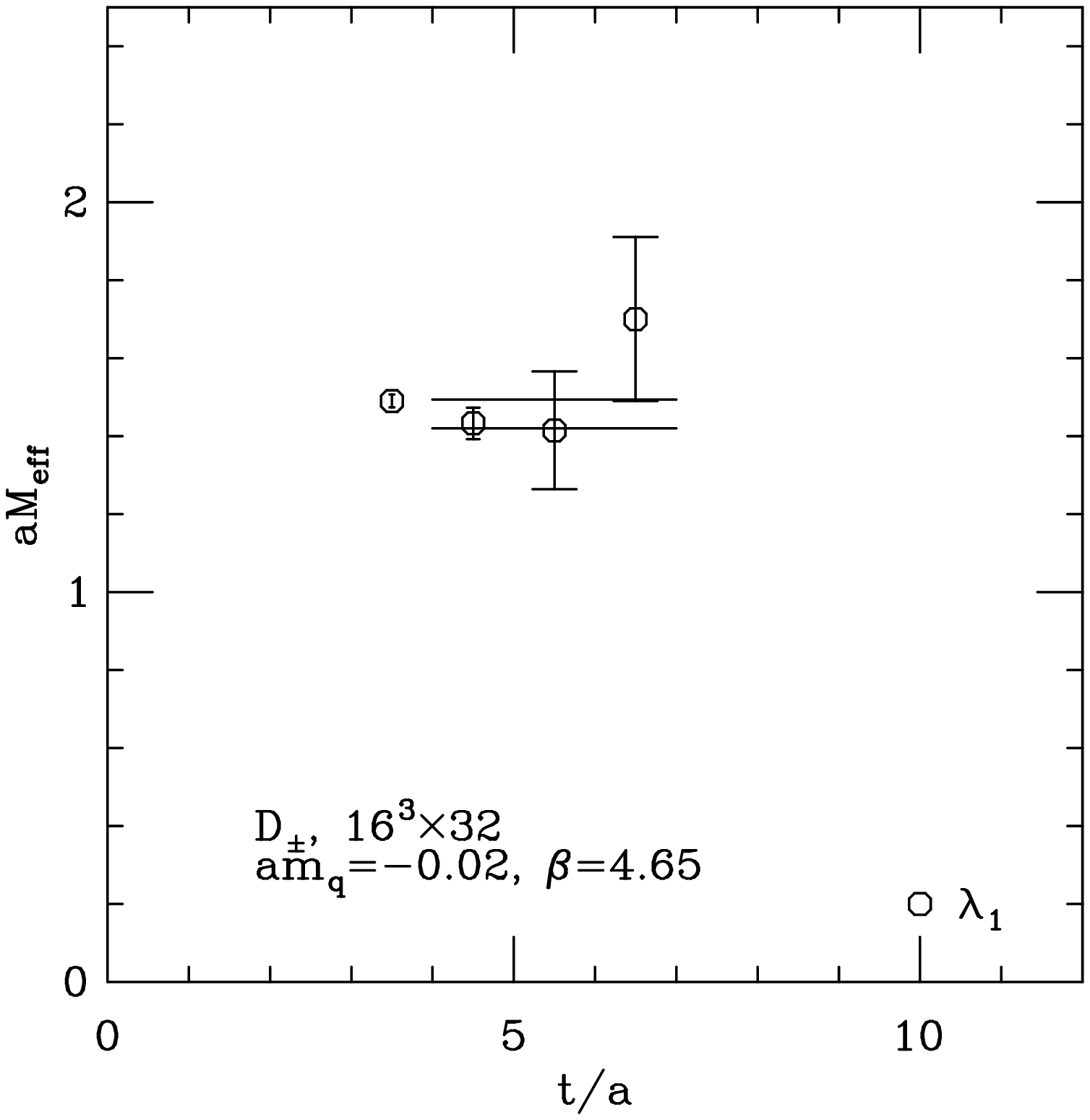}
\includegraphics*[width=4.3cm]{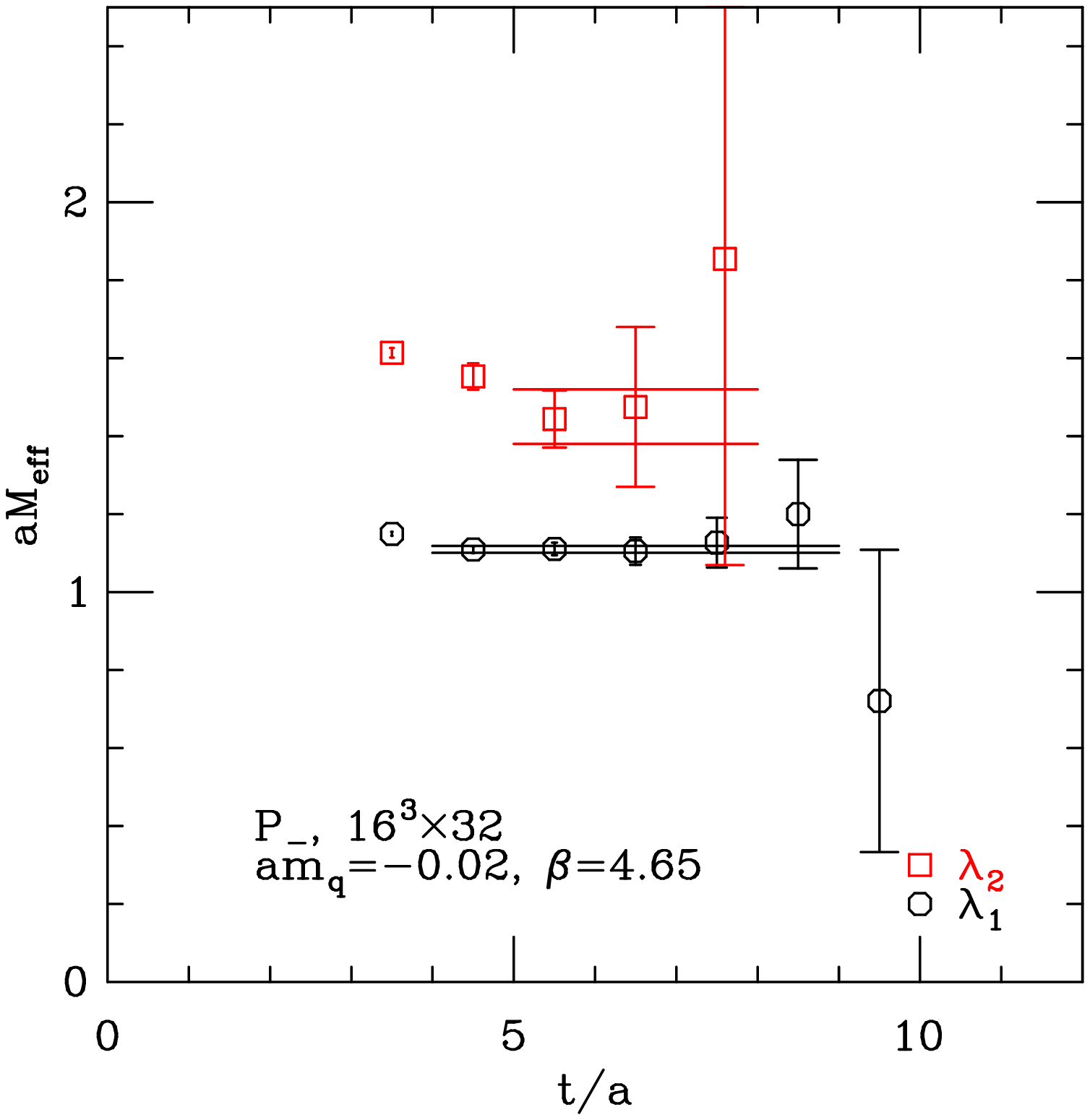}
\includegraphics*[width=4.3cm]{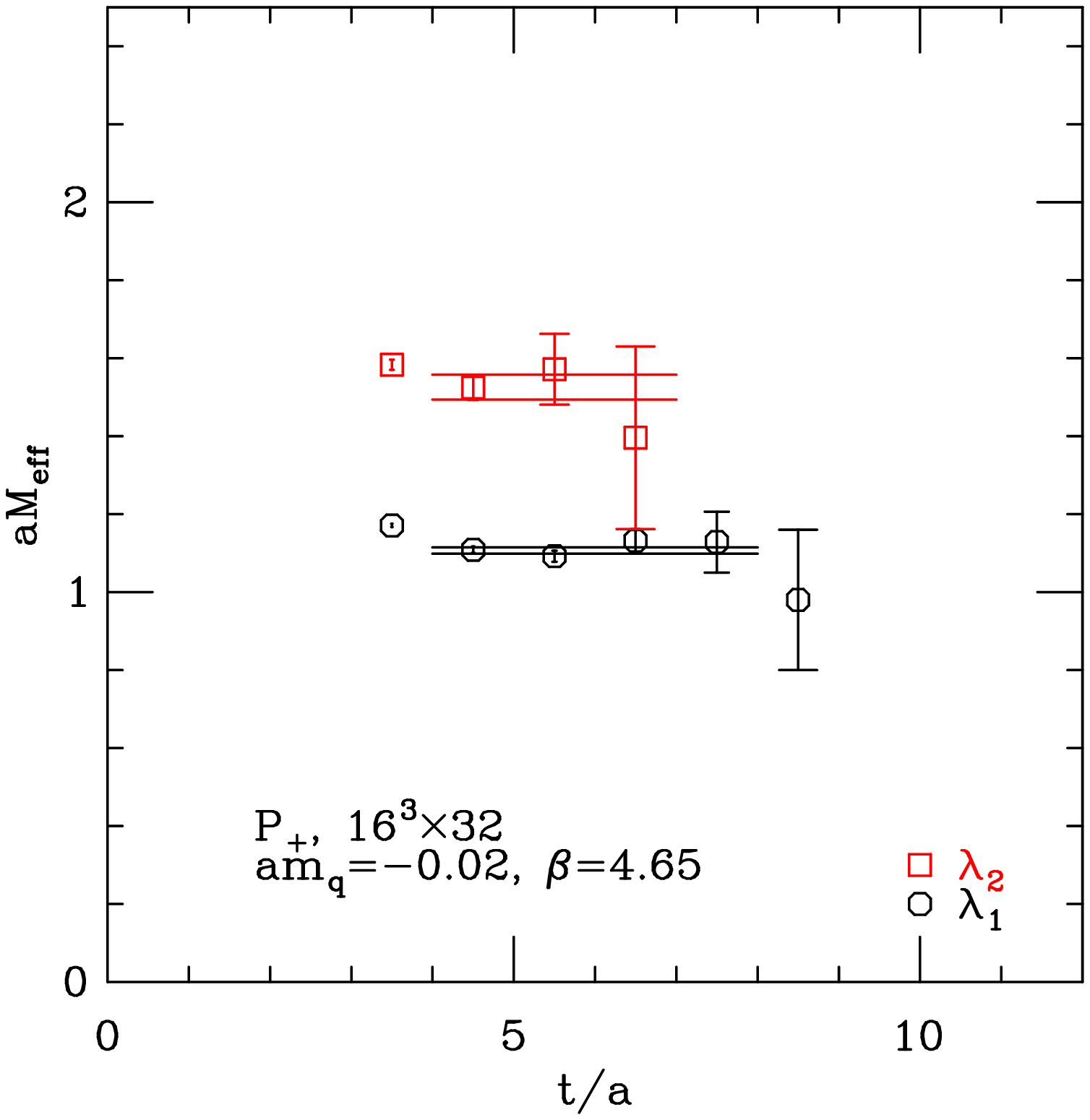}
\end{center}
\caption{
Sample of meson effective masses from the $16^3\times32$, $\beta=4.65$ 
dynamical lattice ($am_q=-0.02$). 
From left to right are the $S$, $D_\pm$, $P_-$, and $P_+$. 
The horizontal lines denote the $aE^{(n)} \pm a\sigma^{(n)}$ correlated fit 
values and the corresponding time ranges. 
The $2S$ states are extracted from two-mass fits to the eigenvalues 
$\lambda^{(2)}$ with $t_0/a=1$.
}
\label{effmass_b465}
\end{figure*}

In this section we present our analysis of the static-light(-light) lattice 
energies and couplings. 
These lead to the corresponding results for the meson and baryon mass 
splittings and ratios of meson decay constants. 
Lastly, we show our first results for the bare kinetic corrections 
$\varepsilon^{(n,n')}$ to the $nS$ and $1P_+$ meson states on a subset of the 
quenched $a \approx 0.15$ fm ($\beta=7.90$) configurations.

\subsection{Masses}

\subsubsection{Diagonalization, effective masses, and fits}

Constructing the correlator matrices and solving (or diagonalizing) the 
corresponding generalized eigenvalue problem as outlined in 
Sec.\ \ref{subsect_basis_corrs}, we then have a set of eigenvalues 
$\lambda(t,t_0)^{(n)}$ and eigenvectors $\vec\psi^{(n)}$ (up to $n=4$ for the 
$4 \times 4$ basis) for each $J^P$ channel.

\begin{figure*}[!t]
\begin{center}
\includegraphics*[width=5.7cm]{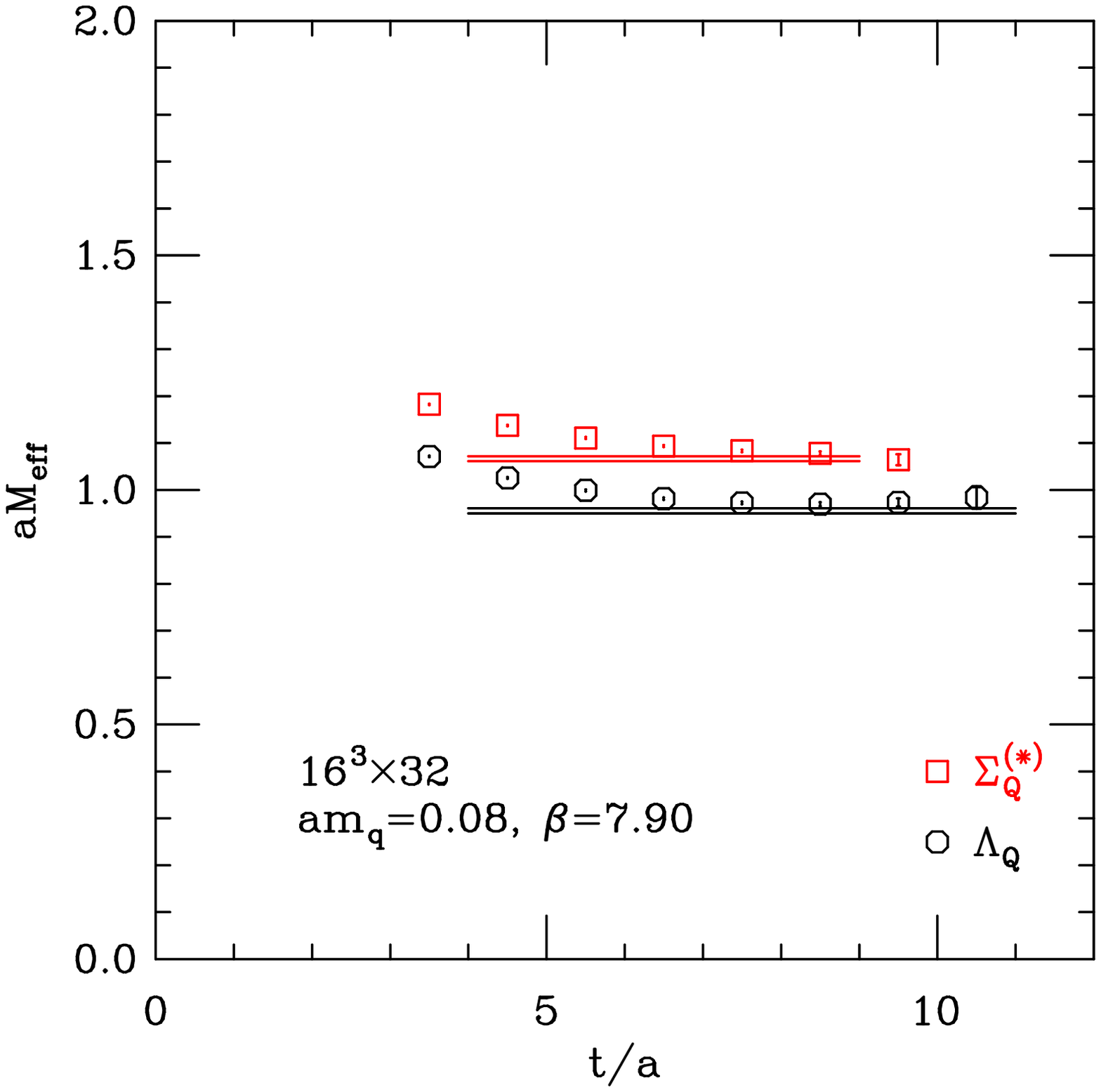}
\includegraphics*[width=5.7cm]{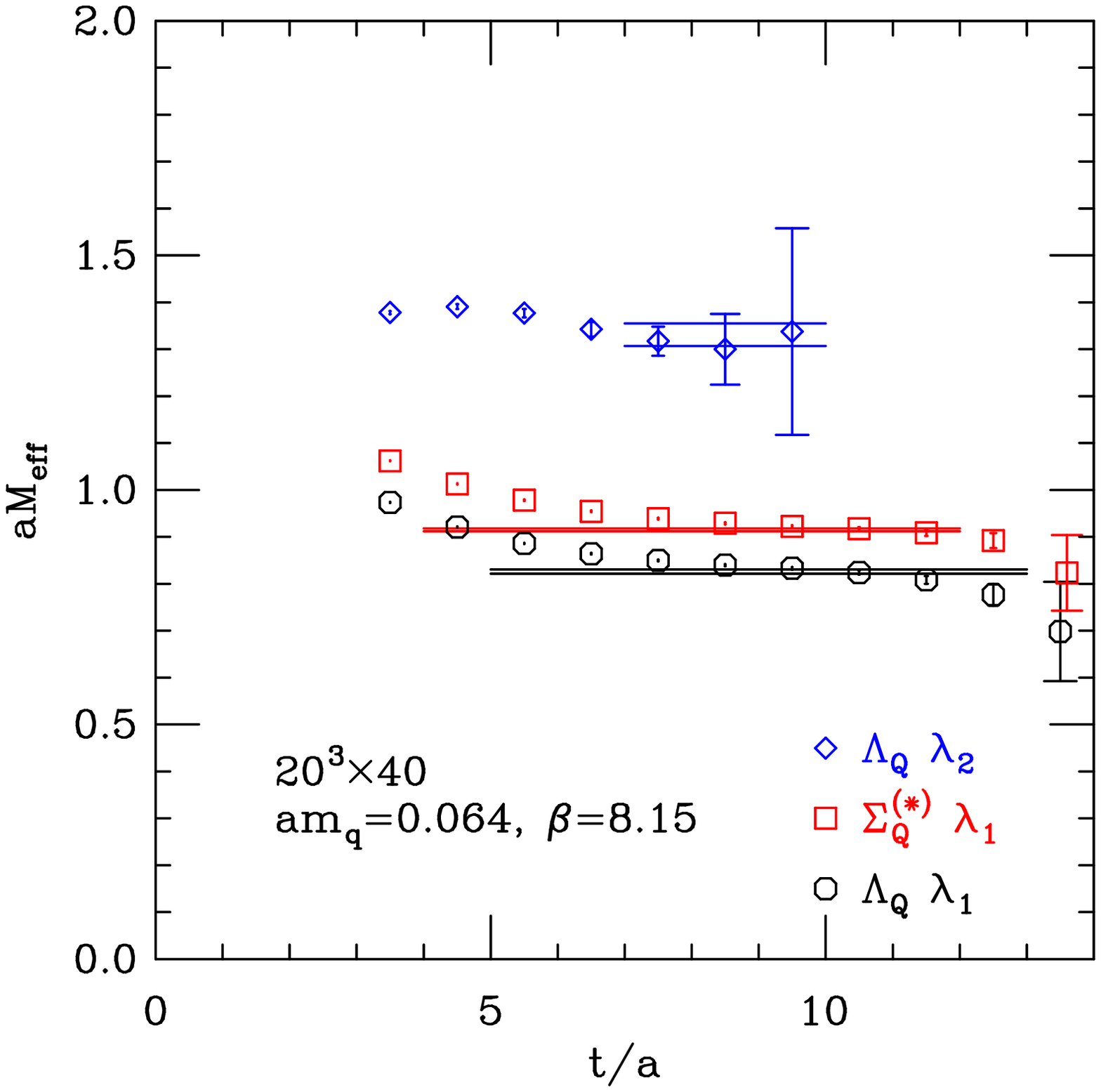}
\includegraphics*[width=5.7cm]{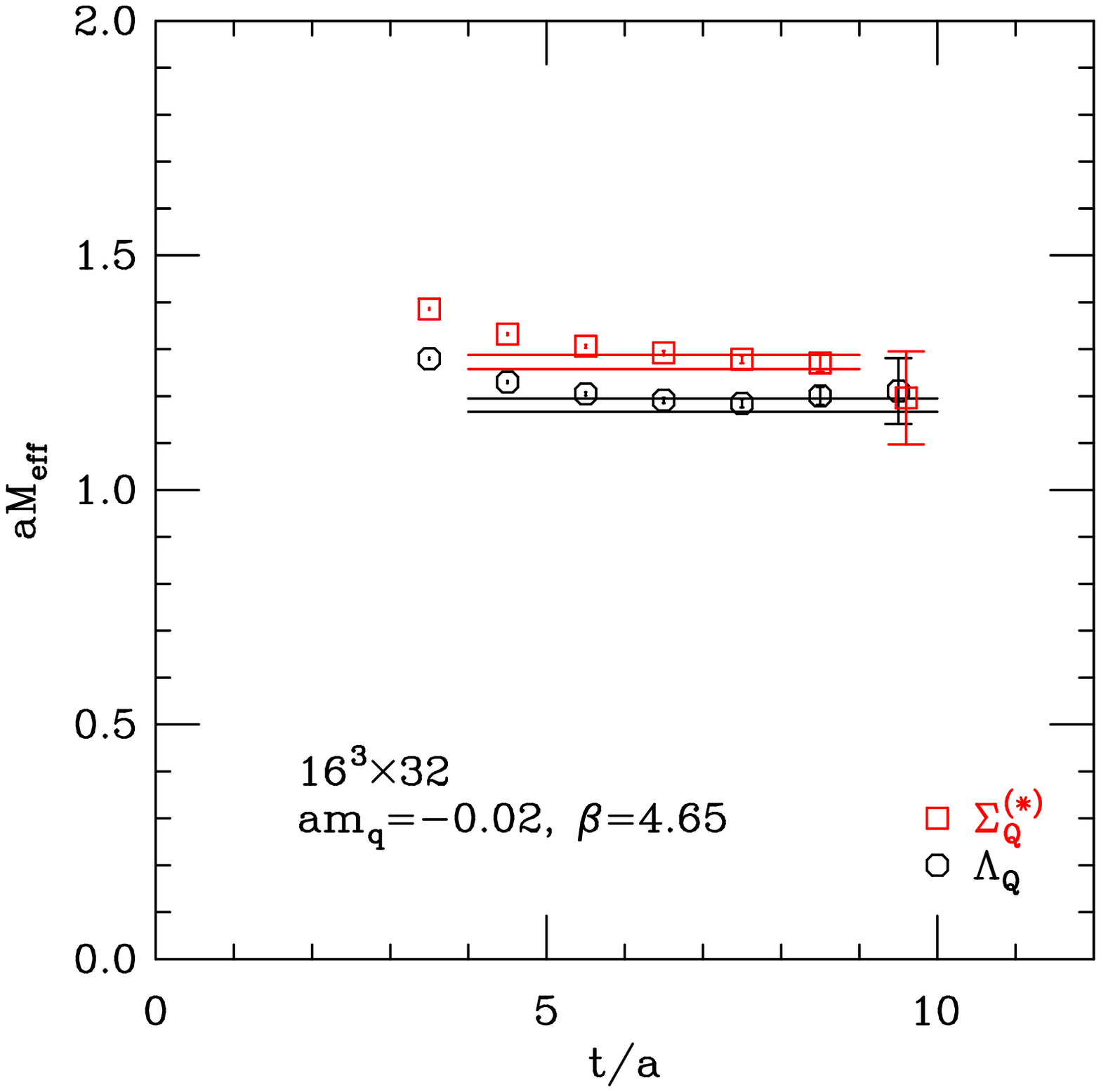}
\end{center}
\caption{
Sample of baryon effective masses from the $\beta=7.90$ and $\beta=8.15$ 
quenched lattices and the $\beta=4.65$ dynamical lattice (from left to right, 
respectively). 
The horizontal lines denote the $aE^{(n)} \pm a\sigma^{(n)}$ correlated fit 
values and the corresponding time ranges. 
The ground states are extracted from two-mass fits to the eigenvalues 
$\lambda^{(1)}$.
}
\label{effmass_Baryons}
\end{figure*}

From the eigenvalues, we create effective masses 
\begin{equation}
  aM_{eff}^{(n)}(t+a/2) = \ln\left( 
  \frac{\lambda(t,t_0)^{(n)}}{\lambda(t+a,t_0)^{(n)}} \right) \; .
\end{equation}
Plotting these versus $t$, we look for plateaus of at least three points, 
within the single-elimination jackknife errors, when searching for a good 
fitting range in $t$. 
We also look for stable eigenvectors over the same region (the so-called 
``fingerprint'' of the state; these are discussed in more detail in the 
next subsection, in the context of the couplings).

Figures \ref{effmass_b790} -- \ref{effmass_b465} show samples of meson 
effective-mass plots from the two finer quenched ensembles and the larger 
dynamical lattice. 
The horizontal lines show the $aE^{(n)} \pm a\sigma^{(n)}$ values from 
correlated fits (as before, we construct the covariance matrix with a 
single-elimination jackknife procedure) to the corresponding time ranges. 
Figure \ref{effmass_Baryons} shows similar plots for the baryons.

We try to be more conservative with our choice of $t_0$ here (previously we 
used $t_0/a=1$) and work with $t_0/a=3$ where we can \cite{new_results_note}. 
The notable exceptions are the $2S$ states on the $\beta=4.65$ dynamical 
lattice, where we use $t_0/a=1$, but there we fit $\lambda^{(2)}$ with a 
two-mass ansatz, trusting only the lower mass. 
Similar two-mass fits are also used for the $2S$ states on the smaller, 
finer, dynamical lattice, as well as for all of the ground-state baryons. 
For the remainder of the results, the bases seem to work well, suppressing 
the higher-order corrections (see Eqs.\ (\ref{eigenvaluedecay}) and 
(\ref{ev_kronecker})), and we can perform many single-mass fits.

As mentioned previously (Sec.\ \ref{subsect_basis_corrs}), most of the 
results come from the full $4 \times 4$ basis. 
The exceptions are: the mesons on the dynamical lattices, where we find it 
beneficial to prune the basis to the first 3 operators (with the full basis, 
the solution of the eigenvalue problem becomes unstable since, within the 
statistically limited ensemble, the fourth operator adds more noise, but 
seemingly little new information); 
and the baryons on the $\beta=8.15$ quenched lattice, where we only have 
results from the smeared light-quark sources thus far ($2 \times 2$ basis).

In Tables \ref{fittable_b815} -- \ref{fittable_b520} we show details of our 
best fits for the $1S$, $2S$, $1P_-$, $1P_+$, $1D_\pm$, $\Lambda_Q$, and 
$\Sigma_Q^{(*)}$ states at all values of the simulated valence quark masses on 
each ensemble. 
(In the interest of some brevity, we do not present such detail for all 
states; there are, however, some $3S$, $2P$, $2D$, and $\Lambda_Q'$ results 
which appear in the effective-mass and chiral-extrapolation plots and later 
tables.)

\begin{table}[!b]
\caption{
Results from correlated mass fits on the $\beta=8.15$ quenched lattice. 
The $\dagger$ symbols indicate where two-mass fits are used.
}
\label{fittable_b815}
\begin{center}
\begin{tabular}{ccccc}
\hline \hline
state & $am_q$ & $t_0/a$ , $t/a$ & $aE$ & $\chi^2/dof$ \\
\hline
$1S$ & 0.016 & 3 , 6--15 & 0.4824(15) & 2.85/8 \\
 & 0.032 & & 0.4935(12) & 1.51/8 \\
 & 0.064 & & 0.51612(93) & 1.97/8 \\
 & 0.080 & & 0.52783(85) & 2.31/8 \\
\hline
$2S$ & 0.016 & 3 , 4--11 & 0.8837(99) & 0.67/6 \\
 & 0.032 & & 0.8881(86) & 0.56/6 \\
 & 0.064 & & 0.9000(72) & 0.50/6 \\
 & 0.080 & & 0.9075(65) & 0.35/6 \\
\hline
$1P_-$ & 0.016 & 3 , 6--13 & 0.7302(53) & 3.70/6 \\
 & 0.032 & & 0.7371(41) & 3.26/6 \\
 & 0.064 & & 0.7549(34) & 4.55/6 \\
 & 0.080 & & 0.7649(31) & 5.56/6 \\
\hline
$1P_+$ & 0.016 & 3 , 6--13 & 0.7425(49) & 2.58/6 \\
 & 0.032 & & 0.7472(37) & 4.07/6 \\
 & 0.064 & & 0.7634(28) & 6.26/6 \\
 & 0.080 & & 0.7727(26) & 7.08/6 \\
\hline
$1D_\pm$ & 0.016 & 3 , 6--10 & 0.930(19) & 0.77/3 \\
 & 0.032 & & 0.937(15) & 0.62/3 \\
 & 0.064 & & 0.953(11) & 0.63/3 \\
 & 0.080 & & 0.961(10) & 0.66/3 \\
\hline
$\Lambda_Q$ & 0.016 & 3 , 4--12 & 0.7329(79) & 4.16/5$^\dagger$ \\
 & 0.032 & & 0.7644(47) & 6.08/5$^\dagger$ \\
 & 0.064 & 3 , 5--13 & 0.8262(44) & 13.4/5$^\dagger$ \\
 & 0.080 & & 0.8586(38) & 15.7/5$^\dagger$ \\
\hline
$\Sigma_Q^{(*)}$ & 0.016 & 3 , 4--12 & 0.8521(66) & 4.05/5$^\dagger$ \\
 & 0.032 & & 0.8711(43) & 2.42/5$^\dagger$ \\
 & 0.064 & & 0.9152(30) & 3.70/5$^\dagger$ \\
 & 0.080 & & 0.9390(27) & 4.54/5$^\dagger$ \\
\hline \hline
\end{tabular}
\end{center}
\end{table}

\begin{table}[!b]
\caption{
Results from correlated mass fits on the $\beta=7.90$ quenched lattice. 
The $\dagger$ symbols indicate where two-mass fits are used.
}
\label{fittable_b790}
\begin{center}
\begin{tabular}{ccccc}
\hline \hline
state & $am_q$ & $t_0/a$ , $t/a$ & $aE$ & $\chi^2/dof$ \\
\hline
$1S$ & 0.020 & 3 , 6--11 & 0.5548(24) & 0.56/4 \\
 & 0.040 & & 0.5688(17) & 0.65/4 \\
 & 0.080 & & 0.5968(13) & 0.91/4 \\
 & 0.100 & & 0.6108(12) & 0.90/4 \\
\hline
$2S$ & 0.020 & 3 , 4--8 & 1.065(17) & 0.15/3 \\
 & 0.040 & & 1.064(14) & 0.06/3 \\
 & 0.080 & & 1.081(11) & 0.22/3 \\
 & 0.100 & & 1.090(10) & 0.10/3 \\
\hline
$1P_-$ & 0.020 & 3 , 5--9 & 0.8738(77) & 0.98/3 \\
 & 0.040 & & 0.8830(57) & 2.88/3 \\
 & 0.080 & & 0.9009(43) & 4.83/3 \\
 & 0.100 & & 0.9113(39) & 5.54/3 \\
\hline
$1P_+$ & 0.020 & 3 , 6--10 & 0.875(12) & 5.16/3 \\
 & 0.040 & & 0.8737(80) & 5.30/3 \\
 & 0.080 & & 0.8919(56) & 4.34/3 \\
 & 0.100 & & 0.9035(50) & 3.69/3 \\
\hline
$1D_\pm$ & 0.020 & 3 , 5--8 & 1.148(20) & 0.24/2 \\
 & 0.040 & & 1.149(15) & 0.26/2 \\
 & 0.080 & & 1.160(11) & 0.31/2 \\
 & 0.100 & & 1.167(10) & 0.52/2 \\
\hline
$\Lambda_Q$ & 0.020 & 3 , 4--11 & 0.848(13) & 8.0/4$^\dagger$ \\
 & 0.040 & & 0.8787(92) & 7.2/4$^\dagger$ \\
 & 0.080 & & 0.9557(57) & 3.0/4$^\dagger$ \\
 & 0.100 & & 0.9952(48) & 2.1/4$^\dagger$ \\
\hline
$\Sigma_Q^{(*)}$ & 0.020 & 3 , 4--9 & 0.954(24) & 2.6/2$^\dagger$ \\
 & 0.040 & & 1.0063(98) & 2.3/2$^\dagger$ \\
 & 0.080 & & 1.0667(52) & 0.78/2$^\dagger$ \\
 & 0.100 & & 1.0959(43) & 0.59/2$^\dagger$ \\
\hline \hline
\end{tabular}
\end{center}
\end{table}

\begin{table}[!b]
\caption{
Results from correlated mass fits on the $\beta=7.57$ quenched lattice. 
The $\dagger$ symbols indicate where two-mass fits are used.
}
\label{fittable_b757}
\begin{center}
\begin{tabular}{ccccc}
\hline \hline
state & $am_q$ & $t_0/a$ , $t/a$ & $aE$ & $\chi^2/dof$ \\
\hline
$1S$ & 0.025 & 3 , 5--10 & 0.7073(18) & 3.21/4 \\
 & 0.050 & & 0.7227(15) & 2.77/4 \\
 & 0.100 & & 0.7551(12) & 2.02/4 \\
 & 0.125 & & 0.7715(11) & 1.62/4 \\
\hline
$2S$ & 0.025 & 3 , 4--8 & 1.482(31) & 4.67/3 \\
 & 0.050 & & 1.467(30) & 4.97/3 \\
 & 0.100 & & 1.461(29) & 0.20/3 \\
 & 0.125 & & 1.468(30) & 0.07/3 \\
\hline
$1P_-$ & 0.025 & 3 , 5--9 & 1.107(11) & 1.05/3 \\
 & 0.050 & & 1.1184(86) & 0.65/3 \\
 & 0.100 & & 1.1462(66) & 0.63/3 \\
 & 0.125 & & 1.1605(60) & 0.70/3 \\
\hline
$1P_+$ & 0.025 & 3 , 4--8 & 1.1569(46) & 6.18/3 \\
 & 0.050 & & 1.1638(39) & 6.92/3 \\
 & 0.100 & & 1.1843(31) & 6.63/3 \\
 & 0.125 & & 1.1958(29) & 6.38/3 \\
\hline
$1D_\pm$ & 0.025 & 3 , 4--8 & 1.462(17) & 0.51/3 \\
 & 0.050 & & 1.472(14) & 0.84/3 \\
 & 0.100 & & 1.498(13) & 1.37/3 \\
 & 0.125 & & 1.508(12) & 1.34/3 \\
\hline
$\Lambda_Q$ & 0.025 & 3 , 4--9 & 1.105(23) & 3.01/2$^\dagger$ \\
 & 0.050 & & 1.163(12) & 5.34/2$^\dagger$ \\
 & 0.100 & & 1.2609(80) & 8.16/2$^\dagger$ \\
 & 0.125 & & 1.3061(71) & 8.65/2$^\dagger$ \\
\hline
$\Sigma_Q^{(*)}$ & 0.025 & 3 , 4--9 & 1.3027(53) & 1.51/2$^\dagger$ \\
 & 0.050 & & 1.3337(39) & 1.42/2$^\dagger$ \\
 & 0.100 & & 1.3936(66) & 0.25/2$^\dagger$ \\
 & 0.125 & & 1.4250(63) & 0.05/2$^\dagger$ \\
\hline \hline
\end{tabular}
\end{center}
\end{table}

\begin{table}[!b]
\caption{
Results from correlated mass fits on the $\beta=4.65$ dynamical lattice. 
The $\dagger$ symbols indicate where two-mass fits are used.
}
\label{fittable_b465}
\begin{center}
\begin{tabular}{ccccc}
\hline \hline
state & $am_q$ & $t_0/a$ , $t/a$ & $aE$ & $\chi^2/dof$ \\
\hline
$1S$ & -0.07 & 3 , 5--12 & 0.7328(30) & 7.81/6 \\
 & -0.06 & & 0.7420(26) & 6.57/6 \\
 & -0.05 & & 0.7508(24) & 5.79/6 \\
 & -0.02 & & 0.7772(20) & 3.47/6 \\
\hline
$2S$ & -0.07 & 1 , 2--9 & 1.261(47) & 4.17/4$^\dagger$ \\
 & -0.06 & & 1.271(42) & 3.72/4$^\dagger$ \\
 & -0.05 & & 1.281(38) & 3.19/4$^\dagger$ \\
 & -0.02 & & 1.301(30) & 1.84/4$^\dagger$ \\
\hline
$1P_-$ & -0.07 & 3 , 4--9 & 1.087(13) & 1.98/4 \\
 & -0.06 & & 1.087(12) & 1.33/4 \\
 & -0.05 & & 1.091(11) & 1.00/4 \\
 & -0.02 & & 1.1096(88) & 0.53/4 \\
\hline
$1P_+$ & -0.07 & 3 , 4--8 & 1.081(12) & 1.57/3 \\
 & -0.06 & & 1.083(11) & 1.80/3 \\
 & -0.05 & & 1.0870(95) & 1.99/3 \\
 & -0.02 & & 1.1065(78) & 2.61/3 \\
\hline
$1D_\pm$ & -0.07 & 3 , 4--7 & 1.430(56) & 1.14/2 \\
 & -0.06 & & 1.439(49) & 1.35/2 \\
 & -0.05 & & 1.445(45) & 1.53/2 \\
 & -0.02 & & 1.457(37) & 1.77/2 \\
\hline
$\Lambda_Q$ & -0.07 & 3 , 4--10 & 1.054(39) & 0.34/3$^\dagger$ \\
 & -0.06 & & 1.078(30) & 0.25/3$^\dagger$ \\
 & -0.05 & & 1.103(23) & 0.29/3$^\dagger$ \\
 & -0.02 & & 1.181(14) & 1.40/3$^\dagger$ \\
\hline
$\Sigma_Q^{(*)}$ & -0.07 & 3 , 4--9 & 1.191(32) & 1.24/2$^\dagger$ \\
 & -0.06 & & 1.205(27) & 1.31/2$^\dagger$ \\
 & -0.05 & & 1.221(22) & 1.19/2$^\dagger$ \\
 & -0.02 & & 1.273(15) & 0.88/2$^\dagger$ \\
\hline \hline
\end{tabular}
\end{center}
\end{table}

\begin{table}[!b]
\caption{
Results from correlated mass fits on the $\beta=5.20$ dynamical lattice. 
The $\dagger$ symbols indicate where two-mass fits are used.
}
\label{fittable_b520}
\begin{center}
\begin{tabular}{ccccc}
\hline \hline
state & $am_q$ & $t_0/a$ , $t/a$ & $aE$ & $\chi^2/dof$ \\
\hline
$1S$ & 0.02 & 3 , 6--12 & 0.6223(57) & 6.18/5 \\
 & 0.04 & & 0.6390(45) & 4.39/5 \\
 & 0.08 & & 0.6729(35) & 4.30/5 \\
 & 0.10 & & 0.6904(32) & 4.90/5 \\
\hline
$2S$ & 0.02 & 3 , 4--9 & 1.024(82) & 0.96/2$^\dagger$ \\
 & 0.04 & & 1.038(66) & 0.56/2$^\dagger$ \\
 & 0.08 & & 1.060(64) & 0.28/2$^\dagger$ \\
 & 0.10 & & 1.074(64) & 0.24/2$^\dagger$ \\
\hline
$1P_-$ & 0.02 & 3 , 6--10 & 0.852(42) & 2.42/3 \\
 & 0.04 & & 0.855(33) & 2.63/3 \\
 & 0.08 & & 0.885(25) & 3.50/3 \\
 & 0.10 & & 0.901(23) & 3.61/3 \\
\hline
$1P_+$ & 0.02 & 3 , 6--10 & 0.852(29) & 2.52/3 \\
 & 0.04 & & 0.866(22) & 2.62/3 \\
 & 0.08 & & 0.897(17) & 1.35/3 \\
 & 0.10 & & 0.913(15) & 0.93/3 \\
\hline
$1D_\pm$ & 0.02 & 3 , 4--7 & 1.114(62) & 0.26/2 \\
 & 0.04 & & 1.125(53) & 0.20/2 \\
 & 0.08 & & 1.157(44) & 0.37/2 \\
 & 0.10 & & 1.173(41) & 0.46/2 \\
\hline
$\Lambda_Q$ & 0.02 & 3 , 4--9 & 0.783(103) & 0.21/2$^\dagger$ \\
 & 0.04 & & 0.893(50) & 0.61/2$^\dagger$ \\
 & 0.08 & & 1.024(27) & 1.23/2$^\dagger$ \\
 & 0.10 & & 1.076(22) & 1.21/2$^\dagger$ \\
\hline
$\Sigma_Q^{(*)}$ & 0.02 & 3 , 4--9 & 1.001(32) & 1.05/2$^\dagger$ \\
 & 0.04 & & 1.027(23) & 1.39/2$^\dagger$ \\
 & 0.08 & & 1.101(16) & 2.80/2$^\dagger$ \\
 & 0.10 & & 1.139(15) & 3.15/2$^\dagger$ \\
\hline \hline
\end{tabular}
\end{center}
\end{table}

\subsubsection{Extrapolations and interpolations}
\label{subsubsect_extraps}

\begin{figure}[!t]
\begin{center}
\includegraphics*[width=8cm]{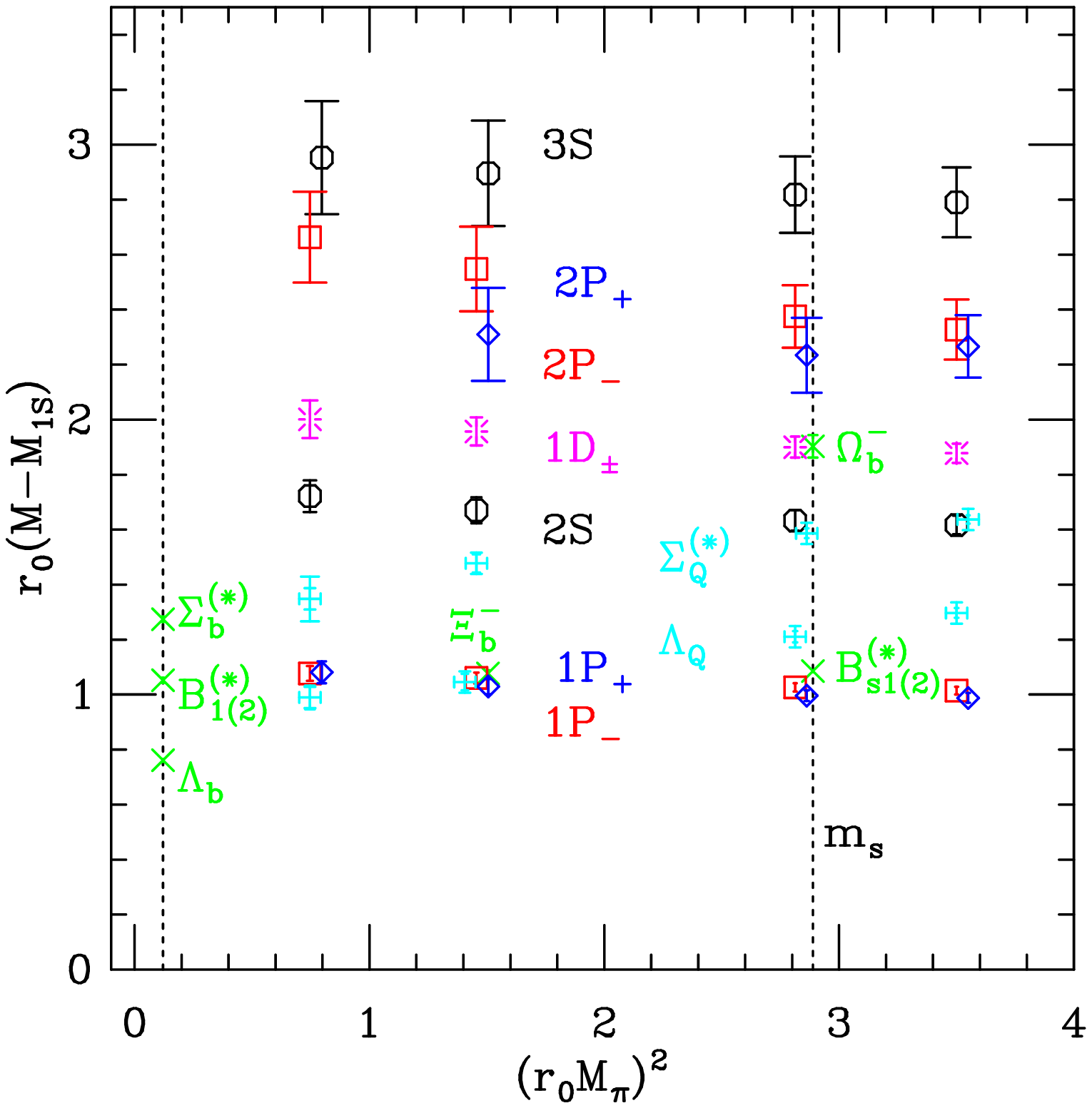}
\end{center}
\caption{
Meson and baryon masses, relative to the meson ground state 
$M_{1S}=M(B_q^{(*)})$, as a function of $M_\pi^2$ ($\propto m_q$) on the 
$16^3\times32$, $\beta=7.90$ quenched lattice. 
All masses result from fits to the eigenvalues of the complete $4\times4$ 
basis. 
Circles represent the $S$ states; squares, the $P_-$ states; diamonds, the 
$P_+$ states; bursts, the $1D_\pm$; plus signs, the baryons $\Lambda_Q$ and 
$\Sigma_Q^{(*)}$; and crosses, the experimental results 
\cite{B**,Bs**,Sigmab*,Xib-,Omegab-,PDG} (using $r_0=0.49$ fm).
}
\label{Mdiff_b790}
\end{figure}

\begin{figure}[!t]
\begin{center}
\includegraphics*[width=8cm]{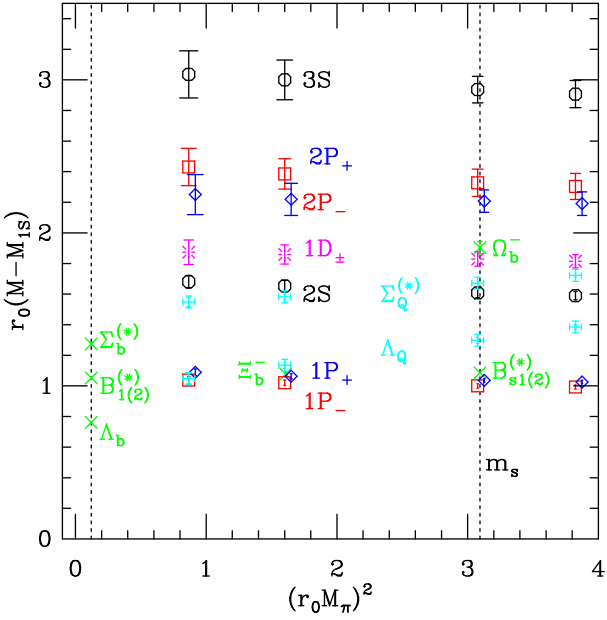}
\end{center}
\caption{
Meson and baryon masses, relative to the meson ground state 
$M_{1S}=M(B_q^{(*)})$, as a function of $M_\pi^2$ on the $20^3\times40$, 
$\beta=8.15$ quenched lattice. 
All meson masses result from fits to the eigenvalues of the complete 
$4\times4$ basis. 
The baryons masses are from the smaller $2 \times 2$ basis (smeared only). 
The symbols have the same meaning as in Fig.\ \ref{Mdiff_b790}.
}
\label{Mdiff_b815}
\end{figure}

\begin{figure}[!t]
\begin{center}
\includegraphics*[width=8cm]{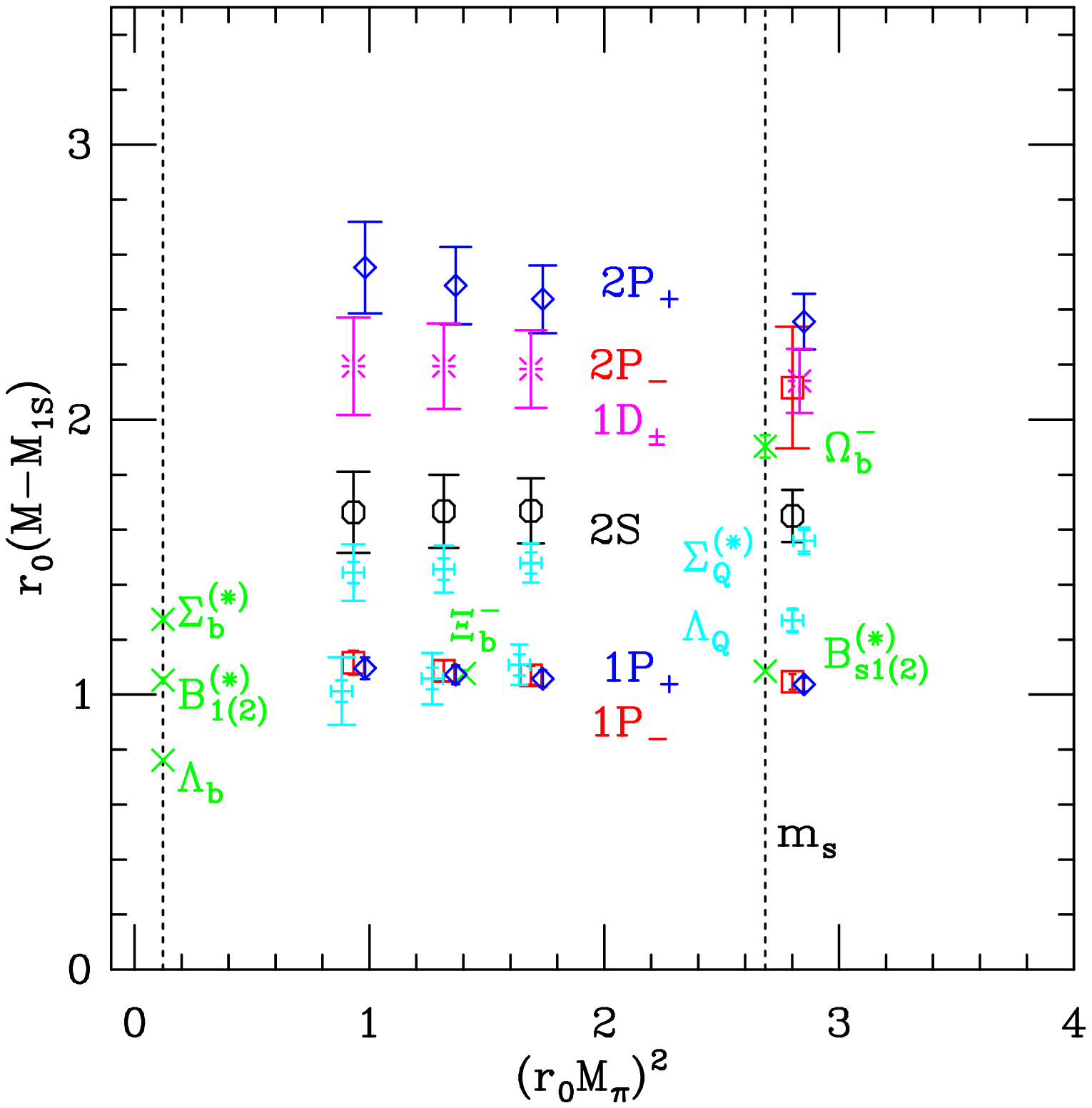}
\end{center}
\caption{
Meson and baryon masses, relative to the meson ground state 
$M_{1S}=M(B_q^{(*)})$, as a function of $M_\pi^2$ on the $16^3\times32$, 
$\beta=4.65$ dynamical lattice. 
All meson masses result from fits to the eigenvalues of a $3\times3$ basis. 
The baryons masses are from the complete $4 \times 4$ basis. 
The symbols have the same meaning as in Fig.\ \ref{Mdiff_b790}.
}
\label{Mdiff_b465}
\end{figure}

Having the lattice energies, we may now reference each of them with respect 
to, say, the ground state: $E-E_{1S}=M-M_{1S}+{\cal O}(1/m_b^{})$. 
Figures \ref{Mdiff_b790} -- \ref{Mdiff_b465} show these mass differences (in 
units of $r_0$) as functions of the light-quark mass on the two finer, 
quenched lattices and the larger, dynamical lattice. 
(The pion masses associated with the light-quark valence masses were computed 
from light-light correlators using standard source techniques 
\cite{DynCI,BGRlarge,Burch:2006dg}.) 
The experimental mass differences \cite{B**,Bs**,Sigmab*,Xib-,Omegab-,PDG} 
for $B_{1(2)}^{(*)}-B^{(*)}$, $B_{s1(2)}^{(*)}-B_s^{(*)}$, $\Lambda_b-B^{(*)}$, 
$\Sigma_b^{(*)}-B^{(*)}$, $\Omega_b^--B_s^{(*)}$, and $\Xi_b^--B_{(s)}^{(*)}$ 
are plotted as the green crosses ($\times$) using $r_0=0.49$ fm 
(note that, for the last difference, we use the average mass of the $B^{(*)}$ 
and $B_s^{(*)}$ plotted at $(m_{ud}+m_s)/2$, enabling an easier comparison 
with the $M_{\Lambda_Q}-M_{1S}$ results at that $m_q$ value; see 
Sec.\ \ref{subsect_baryons} for further discussion of this).

For the most part, the plots display rather slight dependences of $M-M_{1S}$ 
on the light-quark mass and we extrapolate (interpolate) to $m_{ud}$ ($m_s$) 
by fitting linear functions of $M_\pi^2$. 
For a few of the chiral (strange) extrapolations (interpolations), we leave 
out some of the points. 
These include the lightest quark mass for the $2S$ state on the $\beta=7.57$ 
lattice (due to a questionable rise there; see Table \ref{fittable_b757}) and 
the heaviest quark mass for the $\Lambda_Q$ on the $\beta=7.57$ and 
$\beta=8.15$ lattices (due to rather large values for $\chi^2$; see Tables 
\ref{fittable_b815} and \ref{fittable_b757}). 
We point out, however, that the corresponding extrapolation results change by 
an amount within the final error bars when these points are left out.

The results of the light-quark chiral extrapolations of the $M-M_{1S}$ 
($\rightarrow M-M_{B^{(*)}}$) values for the lower-lying meson excitations and 
baryons ($2S$, $1P$, $1D$, $\Lambda_b$, $\Sigma_b^{(*)}$) are displayed in 
Table \ref{chiral_contin_table}. 
Also included are continuum extrapolations of the quenched results, using 
linear fits in $a^2$ (see further discussion of this below). 
Since some of the fits for the continuum extrapolations, or some of the 
points which are used in them (labeled with ?), are rather poor (e.g., large 
$\chi^2$'s), we present alternative fits for these in an attempt to check for 
strong systematics. 
For example, each of the eigenvalue fits for the $1P_+$ states on the 
$\beta=7.57$ lattice have $\chi^2/dof \gtrsim 2$; so we also try a constant 
fit in $a$ to just the two finer quenched lattices. 
The result, however, is only $\approx 1.2\sigma$ higher and both the old and 
new continuum-fit $\chi^2$ values are $<1$ with one degree of freedom.

Contrary to our expectations and the results on all of the other lattices, 
the result of the chiral extrapolation for the $2S$ on the coarse quenched 
lattice lands above that for the $1D_\pm$. 
For this reason we try a constant continuum fit to the two finer lattices 
and find a result $\approx 2.4\sigma$ higher; however, both continuum 
extrapolations have decent $\chi^2$ values.

For the baryons on the finest lattice, we unfortunately have the smallest 
basis (thus far only $2 \times 2$) and are therefore left with a suspicion 
that we may not have well isolated the ground-state mass. 
We thus try constant fits in $a$ to the two coarser lattices and the results 
end up $\approx 2.6\sigma$ and $\approx 1.9\sigma$ lower for the $\Lambda_b$ 
and $\Sigma_b^{(*)}$, respectively, with the continuum-fit $\chi^2/dof$ value 
for the latter greatly improving (from 4.4 to 1.8). 
This consistency of the mass splittings at the two larger values of $a$, 
together with the formerly poor $\Sigma_b^{(*)}$ extrapolation, suggests that 
we may indeed need to enlarge the baryon basis on the $\beta=8.15$ lattice to 
get a more reliable (perhaps lower) value of $M_{\Sigma_b^{(*)}}-M_{B^{(*)}}$ (and 
maybe $M_{\Lambda_b}-M_{B^{(*)}}$ also).

We also check the baryon masses on the coarser quenched lattices with a 
similar two-operator basis (only smeared). 
We find that the results change, but only slightly (i.e., well within the 
errors). 
This suggests that the higher-state terms on the fine lattice may also be 
well isolated, even with the small basis; but since the smeared operators are 
not quite the same on all lattices, we cannot be sure of this.

\begin{table}[!b]
\caption{
Static-light mass differences and their quenched continuum values (linear 
fits in $a^2$). 
Values are given using $r_0=0.49(1)$ fm \cite{r0_error}. 
The $**$ symbols indicate constant fits in $a$ using only the two finer quenched 
lattices (thereby avoiding some possibly problematic fits on the $\beta=7.57$ 
lattice). 
The $\dagger$ symbols indicate constant fits using only the two coarser quenched 
lattices (these are the only ones which use the full $4\times4$ basis for the 
baryons).
}
\label{chiral_contin_table}
\begin{center}
\begin{tabular}{cccc}
\hline \hline
operator & states & $\beta$ & $M-M_{B^{(*)}}$ (MeV) \\
\hline
$2S$ & $B^{(*)'}$ & 7.57 & 800(29)(?) \\  
 & & 7.90 & 695(24) \\  
 & & 8.15 & 683(18) \\  
 & & $\infty$ & 612(31)(13) ($\chi^2=0.67$) \\  
 & & $\infty$(?) & 687(14)(14) ($\chi^2=0.14$) $**$ \\  
 & & 4.65 & 674(66)(14) \\
 & & 5.20 & 712(139)(15) \\
\hline
$1P_-$ & $B_{0,1}^*$ & 7.57 & 408(12) \\  
 & & 7.90 & 438(11) \\  
 & & 8.15 & 420(9) \\  
 & & $\infty$ & 435(15)(9) ($\chi^2=2.81$) \\
 & & 4.65 & 454(19)(9) \\
 & & 5.20 & 398(78)(8) \\
\hline
$1P_+$ & $B_{1(2)}^{(*)}$ & 7.57 & 465(6)(?) \\  
 & & 7.90 & 433(16) \\  
 & & 8.15 & 441(9) \\  
 & & $\infty$ & 423(13)(9) ($\chi^2=0.81$) \\
 & & $\infty$(?) & 439(8)(9) ($\chi^2=0.20$) $**$ \\
 & & 4.65 & 446(17)(9) \\
 & & 5.20 & 402(54)(8) \\
\hline
$1D_\pm$ & $B_{2(3)}^{(*',*)}$ & 7.57 & 771(18) \\  
 & & 7.90 & 811(27) \\  
 & & 8.15 & 759(33) \\  
 & & $\infty$ & 794(42)(16) ($\chi^2=1.80$) \\
 & & 4.65 & 896(78)(18) \\
 & & 5.20 & 858(115)(18) \\
\hline
$\Lambda_Q$ & $\Lambda_b$ & 7.57 & 340(25) \\  
 & & 7.90 & 361(17) \\  
 & & 8.15 & 389(12)(?) \\  
 & & $\infty$ & 415(23)(8) ($\chi^2=0.41$) \\
 & & $\infty$(?) & 355(14)(7) ($\chi^2=0.49$) $\dagger$ \\
 & & 4.65 & 358(55)(7) \\
 & & 5.20 & 260(142)(5) \\
\hline
$\Sigma_Q^{(*)}$ & $\Sigma_b^{(*)}$ & 7.57 & 574(4) \\  
 & & 7.90 & 545(21) \\  
 & & 8.15 & 601(11)(?) \\  
 & & $\infty$ & 604(16)(12) ($\chi^2=4.41$) \\
 & & $\infty$(?) & 573(4)(12) ($\chi^2=1.79$) $\dagger$ \\
 & & 4.65 & 555(47)(11) \\
 & & 5.20 & 611(58)(12) \\
\hline \hline
\end{tabular}
\end{center}
\end{table}

Table \ref{strange_contin_table} displays the strange-quark interpolations of 
the $M-M_{1S}$ ($\rightarrow M-M_{B_s^{(*)}}$) values for the lower-lying 
states. 
Quenched continuum results are also present, again including some alternative 
fits for a few questionable cases: 
A similar $1P_+$ constant continuum fit to the two smaller $a$ values is 
displayed, resulting in a marked improvement in the fit 
($\chi^2=5.4 \rightarrow 2.5$), but an increase in the splitting of only 
$\approx 1.6\sigma$. 
Although the coarse static-strange $2S$ falls below the $1D_\pm$, we 
nevertheless show the same constant continuum fit as for the static-light 
case and get a number $\approx 1.8\sigma$ higher, but again, with reasonable 
$\chi^2$ values for both types of fits. 
As with the $\Sigma_b^{(*)}$, the $\Omega_b^{(*)}$ continuum $a^2$ 
extrapolation is poor ($\chi^2=7.9$ !) and the two results at larger $a$ 
agree quite well, resulting in a significantly lower 
$\Omega_b^{(*)}-B_s^{(*)}$ splitting (by $\approx 4.7\sigma$) with a constant 
fit to just these two.

\begin{table}[!b]
\caption{
Static-strange mass differences and their quenched continuum values (linear 
fits in $a^2$). 
Values are given using $r_0=0.49(1)$ fm \cite{r0_error}. 
The $**$ symbols indicate constant fits in $a$ using only the two finer quenched 
lattices (thereby avoiding some possibly problematic fits on the $\beta=7.57$ 
lattice). 
The $\dagger$ symbol indicates a constant fit using only the two coarser quenched 
lattices (these are the only ones which use the full $4\times4$ basis for the 
baryons).
}
\label{strange_contin_table}
\begin{center}
\begin{tabular}{cccc}
\hline \hline
operator & states & $\beta$ & $M-M_{B_s^{(*)}}$ (MeV) \\
\hline
$2S$ & $B_s^{(*)'}$ & 7.57 & 722(30) \\  
 & & 7.90 & 657(15) \\  
 & & 8.15 & 647(12) \\  
 & & $\infty$ & 604(26)(12) ($\chi^2=0.42$) \\
 & & $\infty$(?) & 651(10)(12) ($\chi^2=0.24$) $**$ \\  
 & & 4.65 & 664(39)(13) \\
 & & 5.20 & 691(125)(14) \\
\hline
$1P_-$ & $B_{s0,1}^*$ & 7.57 & 396(7) \\  
 & & 7.90 & 413(6) \\  
 & & 8.15 & 403(6) \\  
 & & $\infty$ & 412(10)(8) ($\chi^2=2.29$) \\
 & & 4.65 & 421(12)(9) \\
 & & 5.20 & 382(59)(8) \\
\hline
$1P_+$ & $B_{s1(2)}^{(*)}$ & 7.57 & 435(4)(?) \\  
 & & 7.90 & 402(8) \\  
 & & 8.15 & 417(5) \\  
 & & $\infty$ & 400(8)(8) ($\chi^2=5.39$) \\
 & & $\infty$(?) & 413(4)(8) ($\chi^2=2.54$) $**$ \\  
 & & 4.65 & 417(10)(9) \\
 & & 5.20 & 396(40)(8) \\
\hline
$1D_\pm$ & $B_{s2(3)}^{(*',*)}$ & 7.57 & 750(13) \\  
 & & 7.90 & 764(16) \\  
 & & 8.15 & 735(20) \\  
 & & $\infty$ & 749(26)(15) ($\chi^2=1.33$) \\
 & & 4.65 & 863(47)(18) \\
 & & 5.20 & 850(93)(17) \\
\hline
$\Sigma_Q^{(*)}$ & $\Omega_b^{(*)}$ & 7.57 & 642(6) \\  
 & & 7.90 & 639(8) \\  
 & & 8.15 & 674(6)(?) \\  
 & & $\infty$ & 683(9)(14) ($\chi^2=7.87$) \\
 & & $\infty$(?) & 641(5)(13) ($\chi^2=0.13$) $\dagger$ \\
 & & 4.65 & 624(21)(13) \\
 & & 5.20 & 694(42)(14) \\
\hline \hline
\end{tabular}
\end{center}
\end{table}

\begin{figure}[!t]
\begin{center}
\includegraphics*[width=8cm]{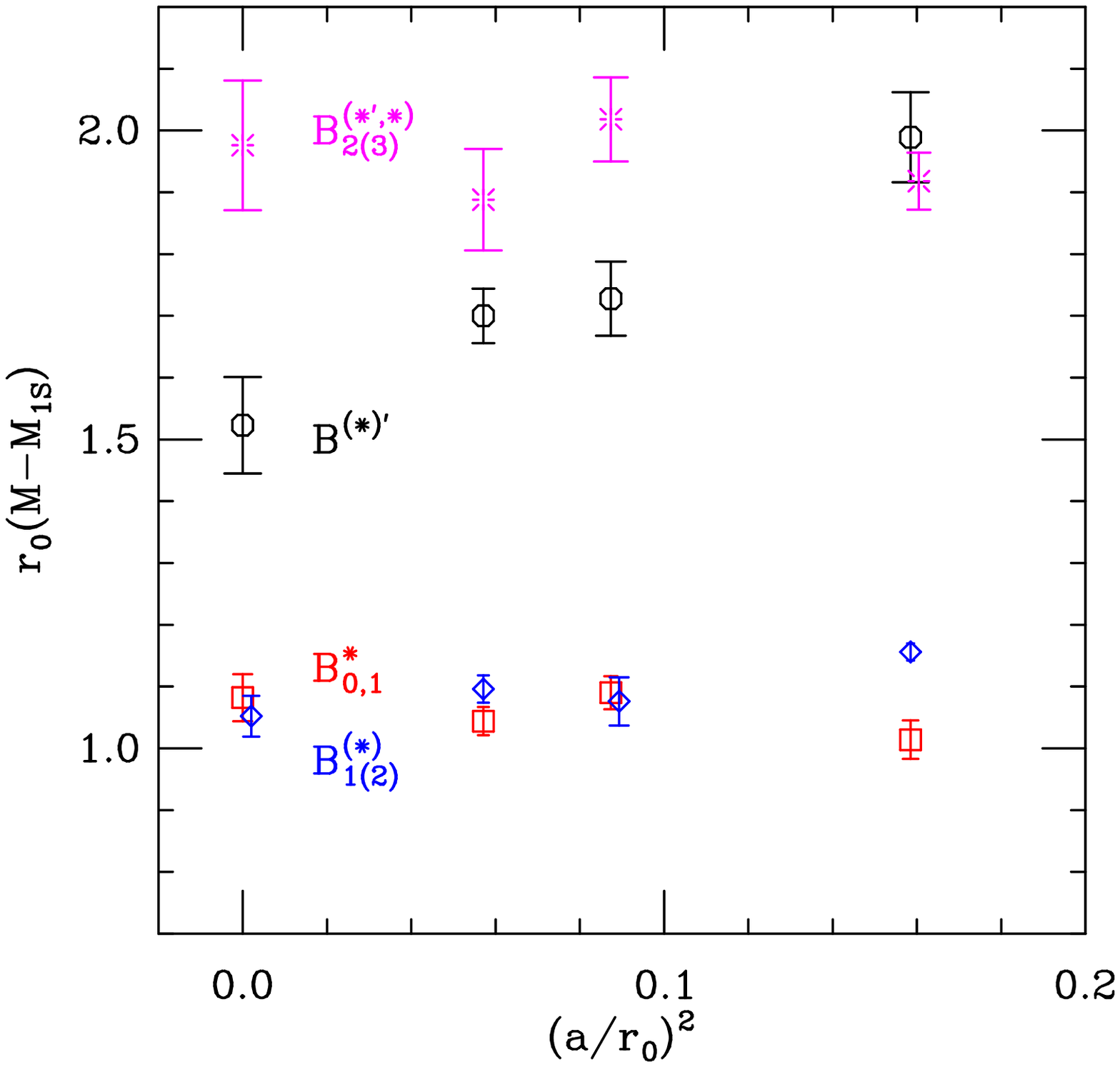}
\end{center}
\caption{
Quenched continuum extrapolations of excited $B$ meson masses, relative to 
$B^{(*)}$.
}
\label{contin_B}
\end{figure}

\begin{figure}[!t]
\begin{center}
\includegraphics*[width=8cm]{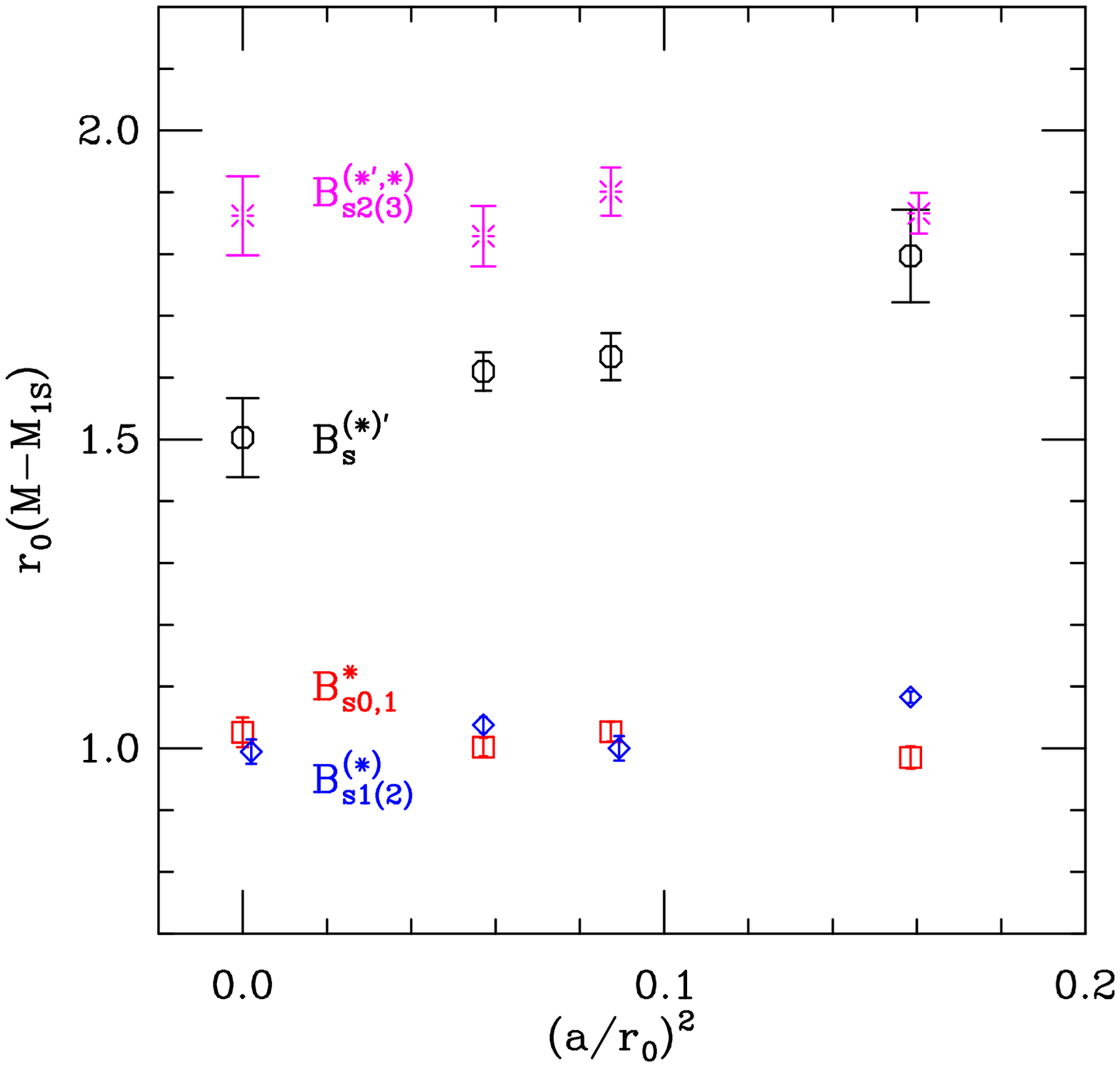}
\end{center}
\caption{
Quenched continuum extrapolations of excited $B_s$ meson masses, relative to 
$B_s^{(*)}$.
}
\label{contin_Bs}
\end{figure}

\begin{figure}[!t]
\begin{center}
\includegraphics*[width=8cm]{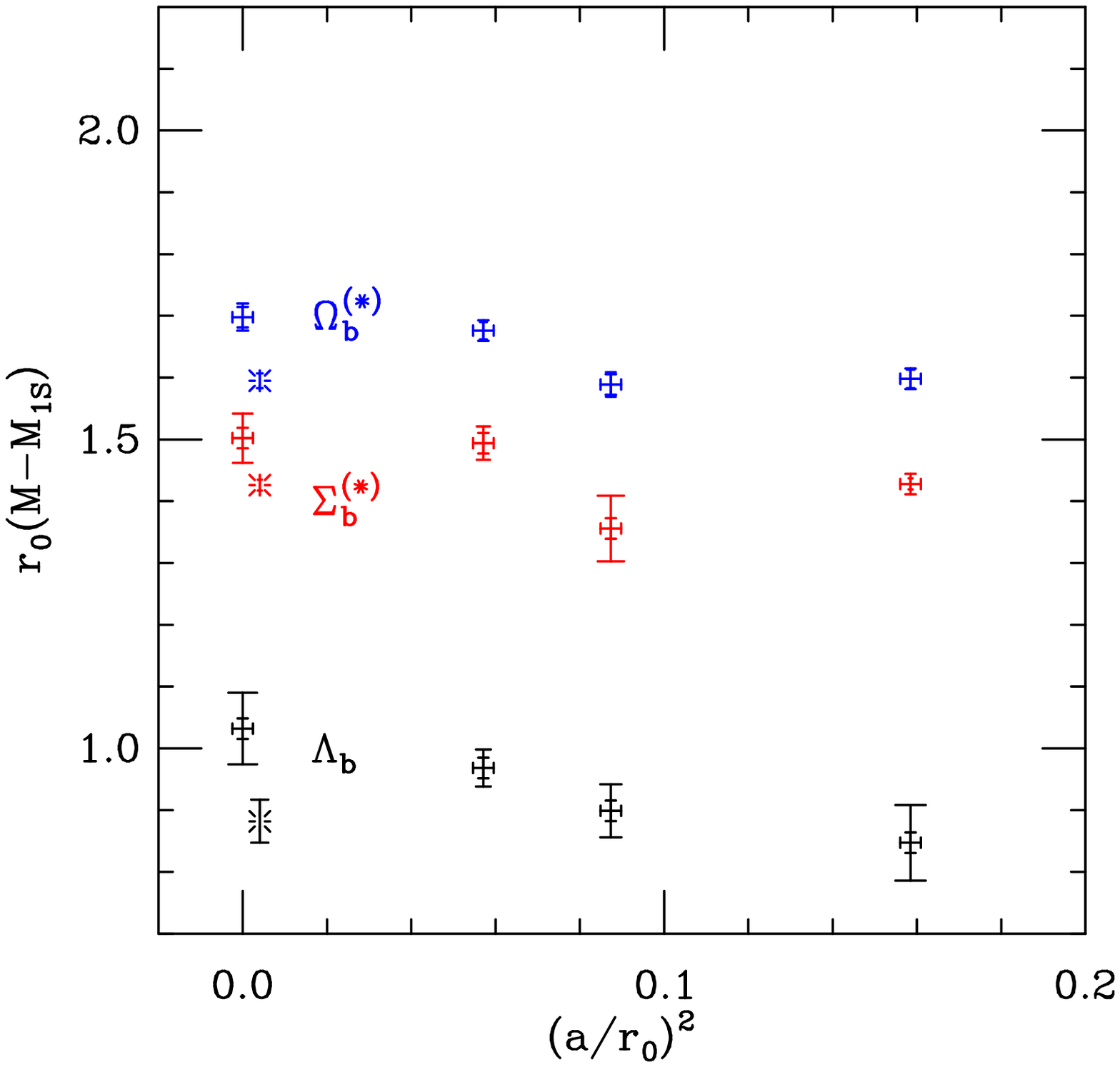}
\end{center}
\caption{
Quenched continuum extrapolations of baryon masses, relative to $B^{(*)}$ 
(for $\Lambda_b$ and $\Sigma_b^{(*)}$) and $B_s^{(*)}$ (for $\Omega_b^{(*)}$). 
The bursts denote constant extrapolations using only the two coarser 
lattices (the ones where we have results for baryons from the full 
$4\times4$ bases).
}
\label{contin_Baryon}
\end{figure}

Figures \ref{contin_B} and \ref{contin_Bs} show the static-light and 
static-strange mass splittings as functions of the squared lattice spacing. 
The results at $a^2 \rightarrow 0$ correspond to the linear extrapolations 
using all three $(a/r_0)^2$ values. 
In Fig.\ \ref{contin_Baryon} the baryon-meson mass differences are plotted 
versus $(a/r_0)^2$. 
Shown here are both the linear-$a^2$ extrapolations and the constant fits to 
the two coarser lattices, the differences between the two sets being clear.

\begin{figure}[!t]
\begin{center}
\includegraphics*[width=8cm]{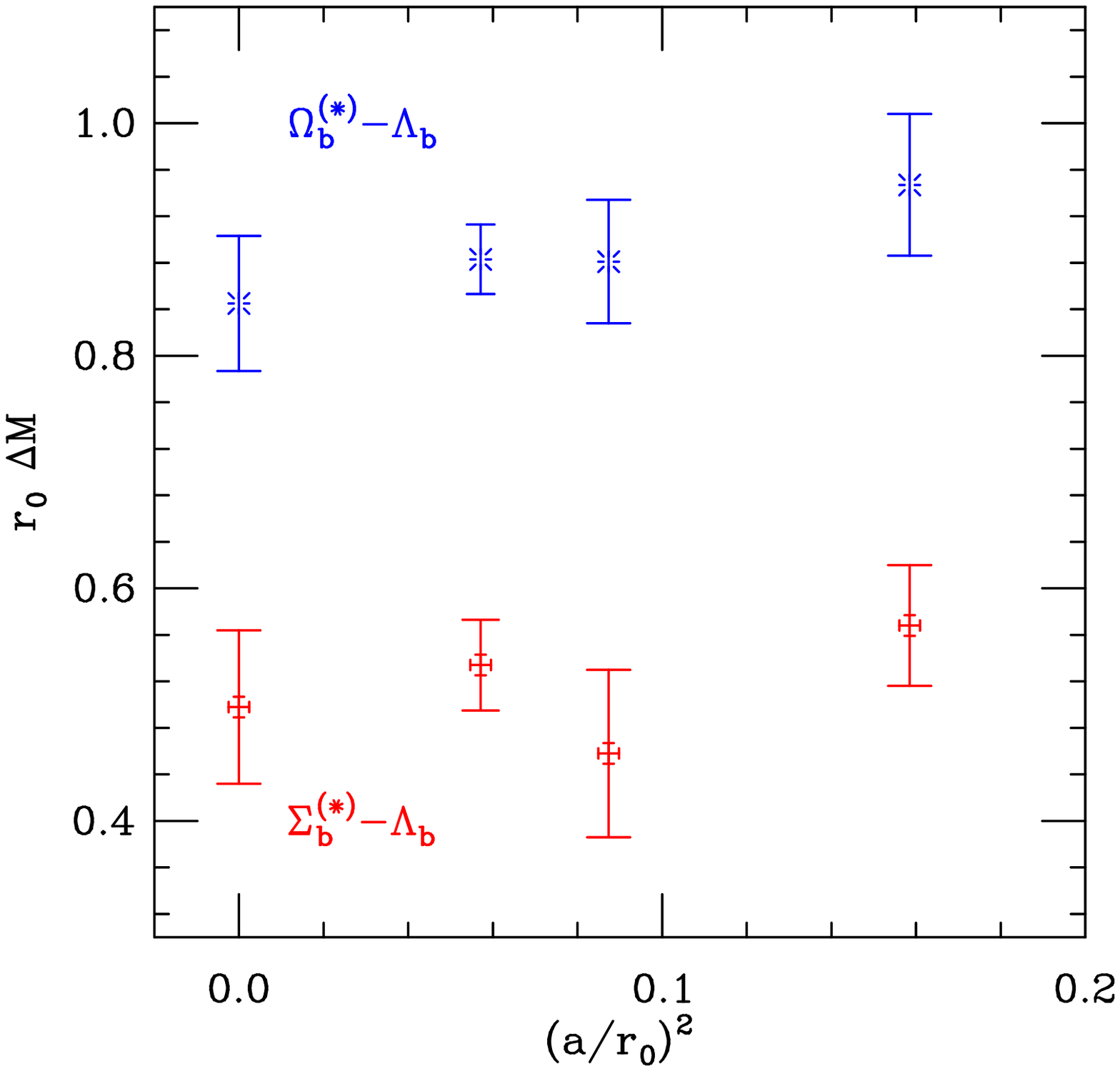}
\end{center}
\caption{
Quenched continuum extrapolations of the $\Sigma_b^{(*)}-\Lambda_b$ and 
$\Omega_b^{(*)}-\Lambda_b$ baryon mass differences.
}
\label{contin_Baryons}
\end{figure}

As mentioned above, the baryon-meson mass differences on the fine lattice 
are significantly larger than on the coarser lattices, and by so much that 
the $a^2$ continuum extrapolations for our $M_{\Sigma_b^{(*)}}-M_{B^{(*)}}$ and 
$M_{\Omega_b^{(*)}}-M_{B_s^{(*)}}$ values appear to be problematic. 
If we believe there are still some significant higher-order corrections in 
the fine baryon masses, we might try mitigating some of these effects by 
considering differences of just baryon masses. 
Accordingly, in Table \ref{chiral_contin_Baryon_table} we present results of 
chiral (strange) extrapolations (interpolations) of $M-M_{\Lambda_b}$ values, 
along with quenched continuum extrapolations. 
An associated plot of $M-M_{\Lambda_b}$ versus $(a/r_0)^2$ appears in 
Fig.\ \ref{contin_Baryons}. 
These extrapolations appear much improved and suggest that we have to work 
harder to get reliable baryon-meson mass differences on the fine lattice.

\begin{table}[!b]
\caption{
Chirally extrapolated and strange-interpolated baryon mass differences and 
their quenched continuum values (where possible). 
Values are given using $r_0=0.49(1)$ fm \cite{r0_error}.
}
\label{chiral_contin_Baryon_table}
\begin{center}
\begin{tabular}{cccc}
\hline \hline
operator & states & $\beta$ & $M-M_{\Lambda_b}$ (MeV) \\
\hline
$\Sigma_Q^{(*)}$ & $\Sigma_b^{(*)}$ & 7.57 & 228(21) \\  
 & & 7.90 & 184(29) \\  
 & & 8.15 & 228(16) \\  
 & & $\infty$ & 200(27)(4) ($\chi^2=1.20$) \\
 & & 4.65 & 195(72)(4) \\
 & & 5.20 & 353(156)(7) \\
\hline
$\Lambda_Q'$ & $\Lambda_b'$ & 8.15 & 900(68)(18) \\
\hline
$\Sigma_Q^{(*)}$ & $\Omega_b^{(*)}$ & 7.57 & 391(25) \\  
 & & 7.90 & 364(21) \\  
 & & 8.15 & 365(12) \\  
 & & $\infty$ & 350(23)(7) ($\chi^2=0.12$) \\
 & & 4.65 & 352(55)(7) \\
 & & 5.20 & 390(142)(8) \\
\hline \hline
\end{tabular}
\end{center}
\end{table}

In Tables \ref{dubious_chiral_contin_table} and 
\ref{dubious_strange_contin_table} we present results of chiral 
extrapolations and strange interpolations, respectively, of $M-M_{1S}$ 
mass differences involving higher-lying states: $3S$, $2P$, $2D$, 
$\Lambda_b'$. 
(Also included are quenched continuum extrapolations for the $2P$ 
states.) 
We consider these masses to be somewhat more dubious since they arise 
from fits over rather short $t$ ranges (see Figs.\ \ref{effmass_b790} -- 
\ref{effmass_Baryons}).

\begin{table}[!b]
\caption{
Static-light mass differences, and their quenched continuum values (where 
possible), involving states whose successful isolation is more dubious. 
Values are given using $r_0=0.49(1)$ fm \cite{r0_error}.
}
\label{dubious_chiral_contin_table}
\begin{center}
\begin{tabular}{cccc}
\hline \hline
operator & states & $\beta$ & $M-M_{B^{(*)}}$ (MeV) \\
\hline
$3S$ & $B^{(*)''}$ & 7.90 & 1198(91)(24) \\
 & & 8.15 & 1232(65)(25) \\
\hline
$2P_-$ & $B_{0,1}^{*'}$ & 7.57 & 1105(57) \\  
 & & 7.90 & 1093(72) \\  
 & & 8.15 & 986(49) \\  
 & & $\infty$ & 941(78)(19) ($\chi^2=0.78$) \\
 & & 5.20 & 1195(67)(24) \\
\hline
$2P_+$ & $B_{1(2)}^{(*)'}$ & 7.57 & 1062(35) \\  
 & & 7.90 & 933(83) \\  
 & & 8.15 & 906(53) \\  
 & & $\infty$ & 811(79)(17) ($\chi^2=0.05$) \\
 & & 4.65 & 1047(72)(21) \\
\hline
$2D_\pm$ & $B_{2(3)}^{(*',*)'}$ & 8.15 & 1401(64)(29) \\
\hline
$\Lambda_Q'$ & $\Lambda_b'$ & 8.15 & 1289(68)(26) \\
\hline \hline
\end{tabular}
\end{center}
\end{table}

\begin{table}[!b]
\caption{
Static-strange mass differences, and their quenched continuum values (where 
possible), involving states whose successful isolation is more dubious. 
Values are given using $r_0=0.49(1)$ fm \cite{r0_error}.
}
\label{dubious_strange_contin_table}
\begin{center}
\begin{tabular}{cccc}
\hline \hline
operator & states & $\beta$ & $M-M_{B_s^{(*)}}$ (MeV) \\
\hline
$3S$ & $B_s^{(*)''}$ & 7.90 & 1134(57)(23) \\
 & & 8.15 & 1181(39)(24) \\
\hline
$2P_-$ & $B_{s0,1}^{*'}$ & 7.57 & 1016(37) \\  
 & & 7.90 & 959(47) \\  
 & & 8.15 & 937(36) \\  
 & & $\infty$ & 891(56)(18) ($\chi^2=0.001$) \\
 & & 4.65 & 851(89)(17) \\
 & & 5.20 & 1126(64)(23) \\
\hline
$2P_+$ & $B_{s1(2)}^{(*)'}$ & 7.57 & 1011(26) \\  
 & & 7.90 & 910(53) \\  
 & & 8.15 & 886(33) \\  
 & & $\infty$ & 811(50)(17) ($\chi^2=0.05$) \\
 & & 4.65 & 949(41)(19) \\
\hline
$2D_\pm$ & $B_{s2(3)}^{(*',*)'}$ & 8.15 & 1301(44)(27) \\
\hline \hline
\end{tabular}
\end{center}
\end{table}

\subsection{Decay constant ratios}
\label{subsect_decay_const_ratios}

In this subsection we turn to the analysis of the state couplings to the 
local operators, which provide us with the decay constants.

Using the eigenvectors from the solutions of the generalized eigenvalue 
problem, Eq.\ (\ref{generalized}), we create the appropriate ratios of 
projected correlator matrices, $R(t)_i^{(n)}$ (see 
Eq.\ (\ref{amplitudefunction})), for the $S$-wave states and the case where 
the open index $i$ refers to the local operator. 
From these altered correlators, we find very similar effective-mass 
behaviors as for the original eigenvalues and we perform fits to the 
$R(t)_i^{(n)}$ over the same $t$ ranges as before. 
From the amplitudes to these fits we achieve the quantity 
$v_i^{(n)*}v_i^{(n)}$. 
Assuming real overlaps, we arrive at the desired couplings (see 
Eq.\ (\ref{overlap_to_decayconst})) and we can then form the appropriate 
ratios for, say, $f_{B_s'}/f_{B_s}$ (see Eq.\ (\ref{f2_over_f1})) or 
$f_{B_s}/f_B$ (see Eq.\ (\ref{fBs_over_fB})).

\begin{figure}[!t]
\begin{center}
\includegraphics*[width=8cm]{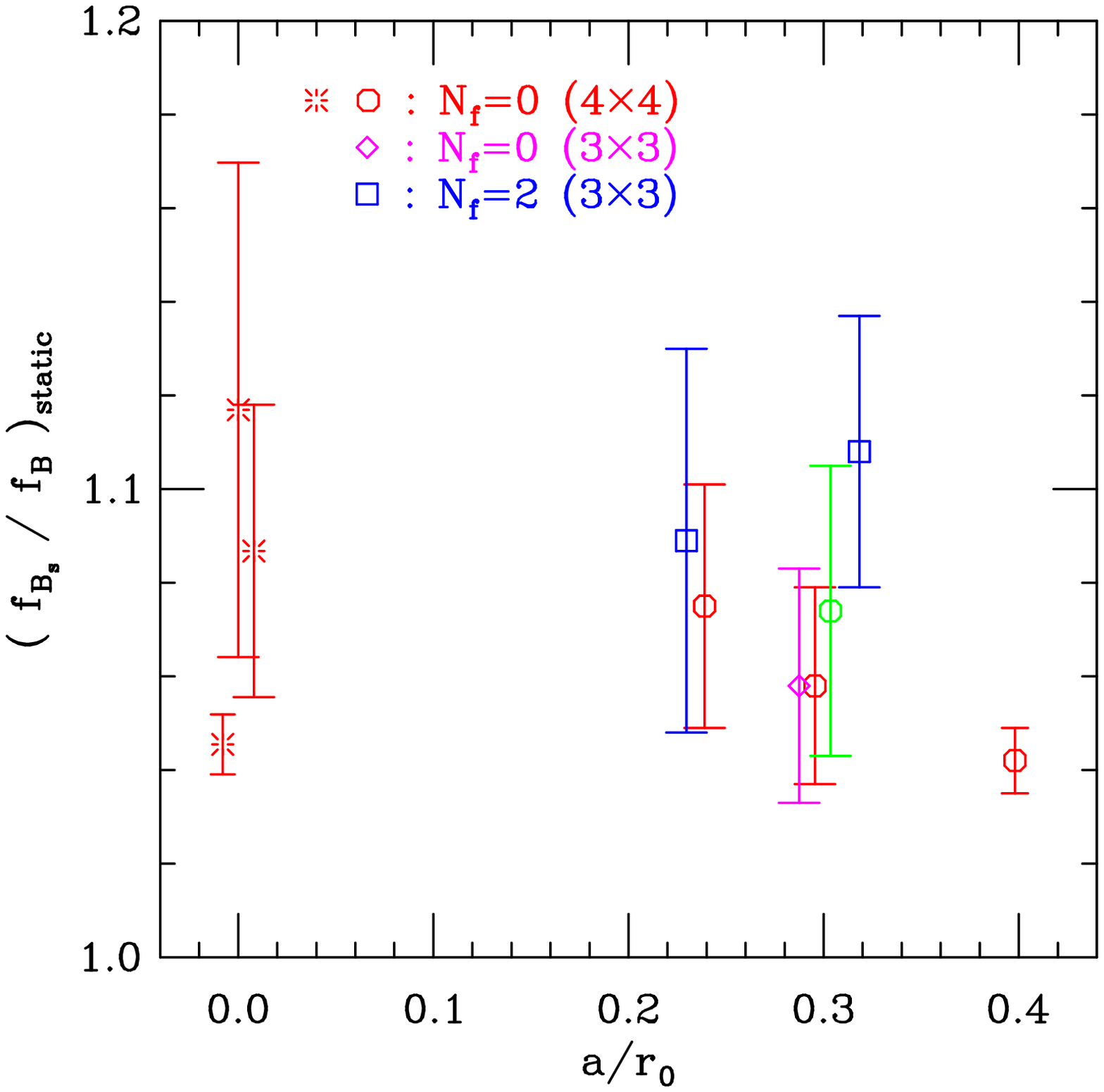}
\end{center}
\caption{
Ratio of meson decay constants $(f_{B_s}/f_B)_{static}$ as a function of 
lattice spacing. 
The three quenched, continuum extrapolations of the $4 \times 4$-basis 
results correspond to (from left to right): a constant fit, a fit linear in 
$a$, a fit linear in $a^2$. 
The green circle at $a/r_0 \approx 0.3$ (shifted to the right for clarity) 
corresponds to the adjacent quenched result, but with naive links for the 
static quark.
}
\label{f_Bs_over_f_B_vs_a}
\end{figure}

Figure \ref{f_Bs_over_f_B_vs_a} shows our resulting values for the 
static-light $f_{B_s}/f_B$ versus the lattice spacing, $a/r_0$. 
Both quenched and dynamical points are plotted, as well as results from the 
$\beta=7.90$ lattice where we use the ``naive'' (or non-Hyp-smeared) links 
for the static quark (the green circle) and a smaller $3 \times 3$ basis 
(the magenta diamond). 
One can easily see that each of these additional points agrees with the 
original result using the Hyp-smeared static quark and the $4 \times 4$ 
basis.

As we are using improved gauge and fermion actions, we were originally 
expecting to use linear extrapolations in $a^2$ for these decay constants. 
However, since we do not improve the local static-light currents, we check 
these results for significant ${\cal O}(a)$ effects by extrapolating the 
quenched points (where we have three $a$ values) using three different 
functional forms: constant, linear-$a$, and linear-$a^2$ fits (the resulting 
extrapolations go from left to right, respectively, in the plot). 
Not surprisingly, the constant fit (which is dominated by the largest $a$ 
value) is inconsistent with the other two extrapolations, showing that there 
are likely lattice-spacing effects in this ratio. 
The difference between the other two central values is almost as large as 
that between the constant and linear-$a^2$ fits, but they do overlap, given 
the rather large error for the linear-$a$ fit. 
Therefore, we do not believe that we are able to conclusively determine 
whether the leading-order effects in $a$ are negligible. 
However, since the linear-$a$ and linear-$a^2$ results are consistent (even 
more so for the other ratios; see below), we choose to report our final 
value(s) from the linear-$a^2$ extrapolation(s) (see 
Table \ref{decay_constant_table}).

\begin{table}[!b]
\caption{
Static-light decay constant ratios and the quenched continuum values 
(linear-$a^2$ fits). 
Values are given using $r_0=0.49(1)$ fm \cite{r0_error} to set the physical 
$m_s$ point (see Sec.\ \ref{subsect_phys_ms}).
}
\label{decay_constant_table}
\begin{center}
\begin{tabular}{cccc}
\hline \hline
$\beta$ & $(f_{B_s}/f_B)_{static}$ & $(f_{B_s'}/f_{B_s})_{static}$ & $(f_{B_s'}/f_{B'})_{static}$ \\
\hline
7.57 & 1.042(7) & 1.309(75) & 0.976(142) \\
7.90 & 1.058(21) & 1.237(15) & 0.996(59) \\
8.15 & 1.075(26) & 1.259(25) & 0.972(61) \\
$\infty$ & 1.087(31) & 1.240(58) & 0.972(123) \\
 & ($\chi^2=0.07$) & ($\chi^2=1.34$) & ($\chi^2=0.07$) \\
4.65 & 1.108(29) & 1.356(142) & 1.089(259) \\
5.20 & 1.089(41) & 1.453(168) & 1.026(128) \\
\hline \hline
\end{tabular}
\end{center}
\end{table}

The result from the larger dynamical lattice is significantly higher than the 
quenched result at similar lattice spacing. 
The finer dynamical result is consistent with the quenched, but the volume is 
rather small (1.35 fm; perhaps this enhances $f_B$ more than $f_{B_s}$, 
reducing the ratio).

\begin{figure*}
\begin{center}
\includegraphics*[width=5.7cm]{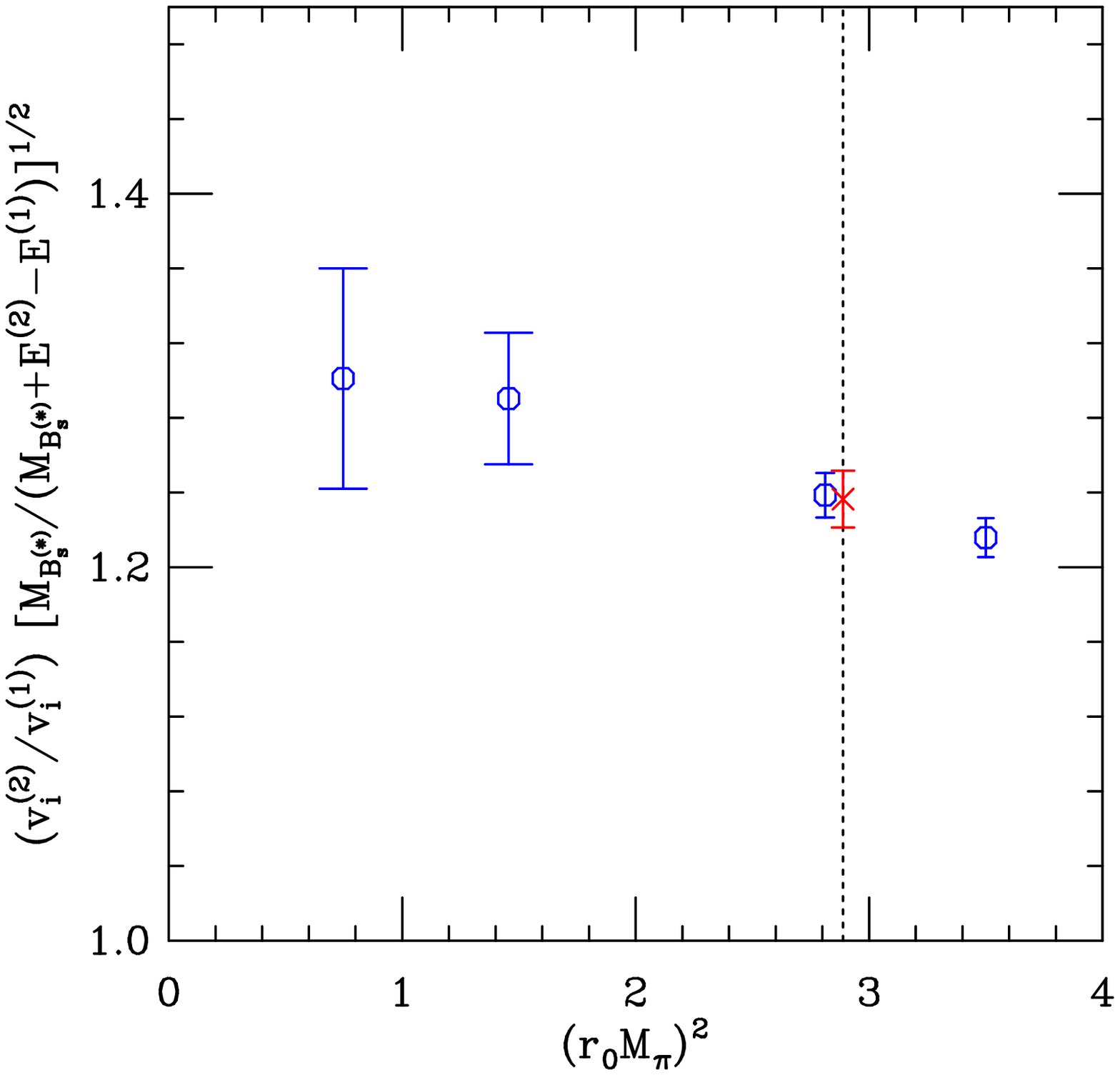}
\includegraphics*[width=5.7cm]{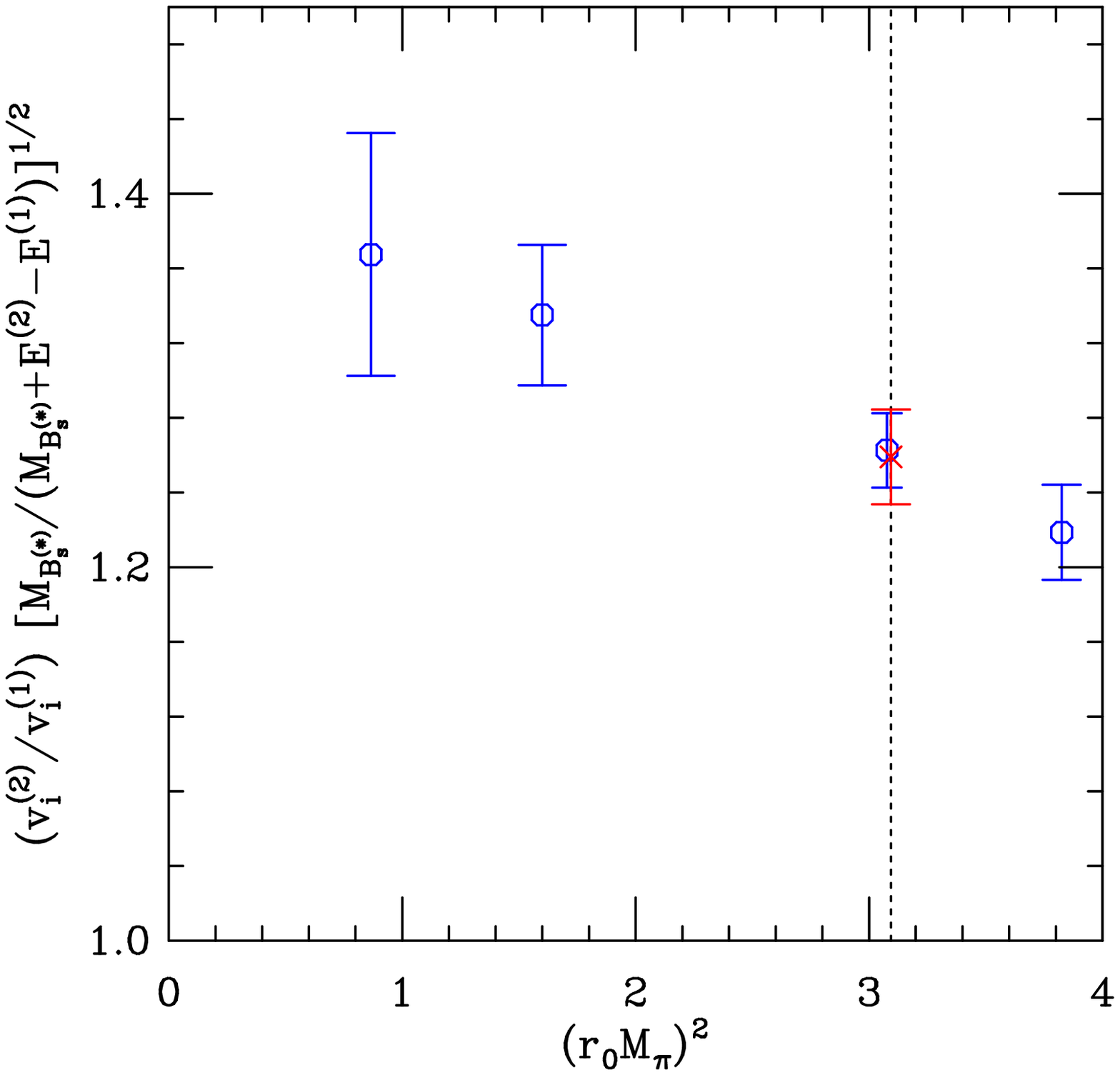}
\includegraphics*[width=5.7cm]{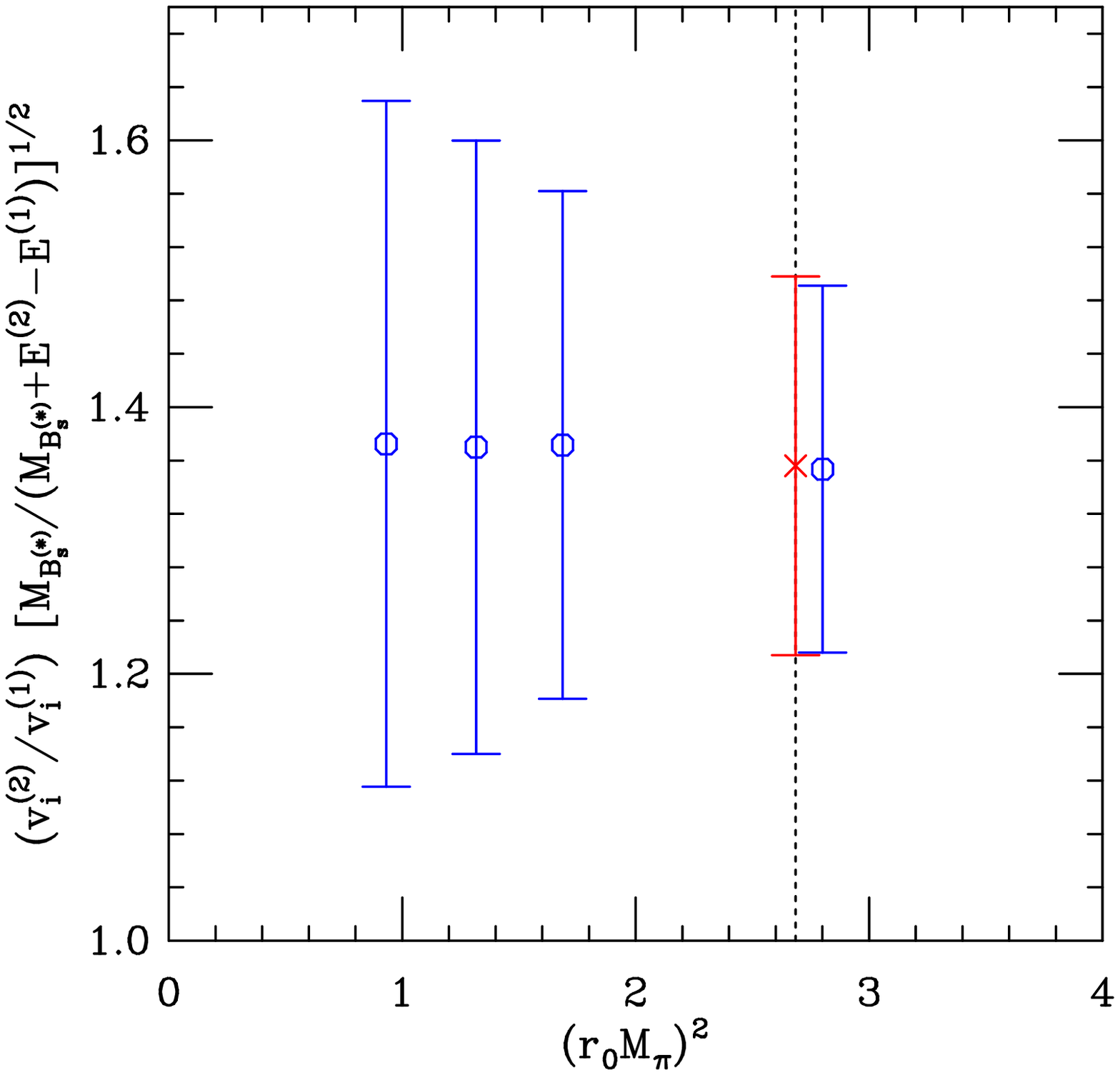}
\end{center}
\caption{
Sample of strange-mass (linear) interpolations of the quantity 
$f_{PS}^{(2)}/f_{PS}^{(1)}$ on the $\beta=7.90$ and $\beta=8.15$ quenched 
lattices and the $\beta=4.65$ dynamical lattice (from left to right, 
respectively).
}
\label{strange_fBp_over_fB}
\end{figure*}

\begin{figure}[!t]
\begin{center}
\includegraphics*[width=8cm]{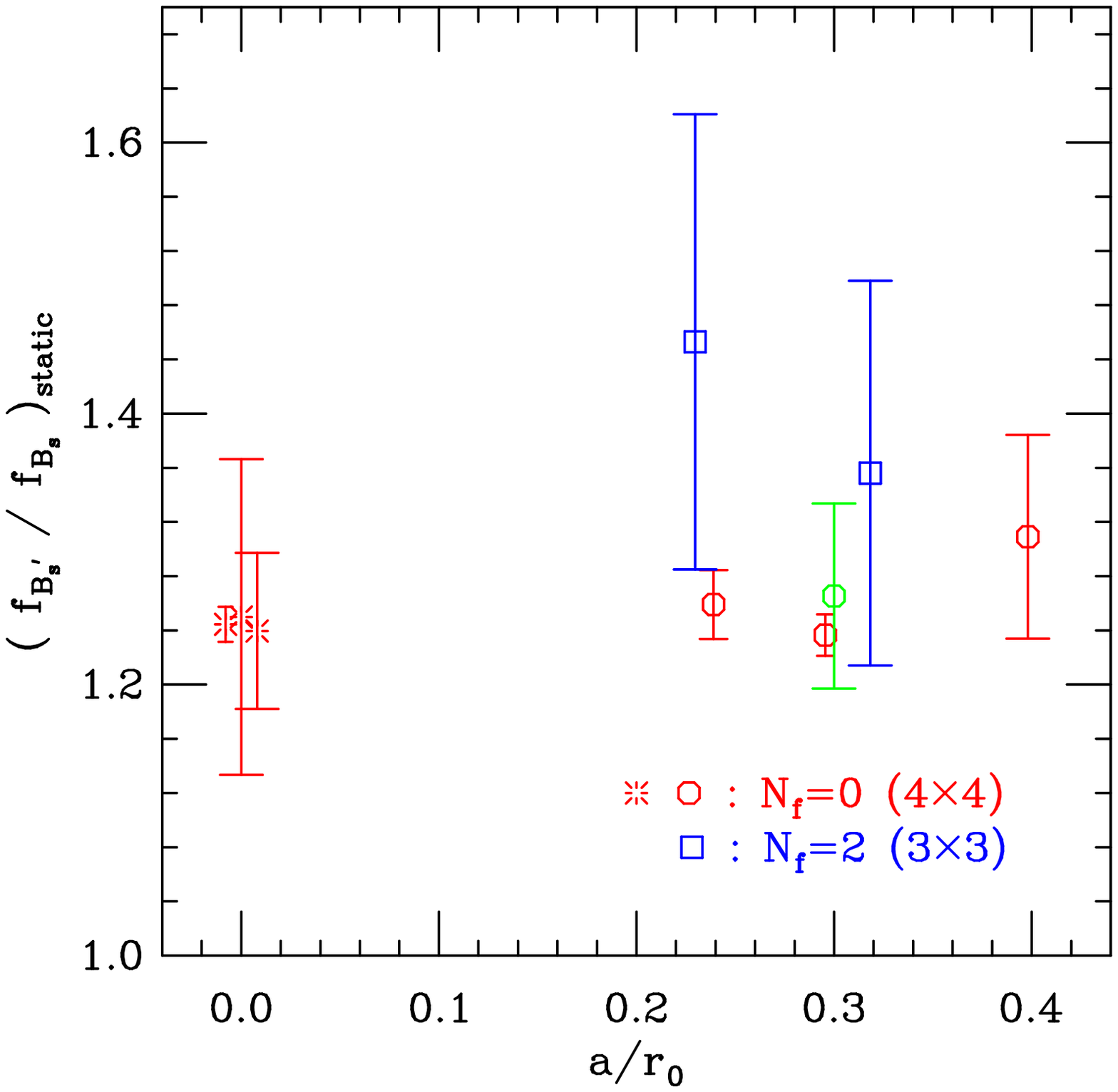}
\end{center}
\caption{
Ratio of meson decay constants $(f_{B_s'}/f_{B_s})_{static}$ as a function of 
lattice spacing. 
The symbols have the same meanings as in Fig.\ \ref{f_Bs_over_f_B_vs_a}.
}
\label{f_2S_over_f_1S_vs_a}
\end{figure}

Turning now to the first-excited $S$-waves, we form ratios of the form of 
Eq.\ (\ref{f2_over_f1}). 
In Fig.\ \ref{strange_fBp_over_fB} we plot these quantities from the two 
finer quenched lattices and the larger dynamical lattice as functions of the 
light-quark mass. 
The strange-mass interpolations then give the decay constant ratios 
$f_{B_s'}/f_{B_s}$, which we plot versus the lattice spacing in 
Fig.\ \ref{f_2S_over_f_1S_vs_a}.

The most striking feature here is that we obtain 
$(f_{B_s'}/f_{B_s})_{static} \gtrsim 1$. 
Perhaps naively, we would have expected this ratio to be $<1$ as the 
numerator involves the local current coupling to the excited state, which one 
could easily imagine to be less than that for the ground state (e.g., in a 
nonrelativistic potential-model view, the normalized wavefunction at the 
origin is smaller for the first-excited state than that for the ground state).

We therefore try to test the robustness of this rather strange result by 
looking for some systematic effects: 
We first tried reducing the size of the basis (to $3 \times 3$ for quenched) 
to see if the ratio is significantly affected by the fact that we do not have 
a truly orthogonal basis (the different overlaps can thus ``trade'' their 
relative contributions without sacrificing their usefulness in isolating a 
given state \cite{ExcAmp}). 
What we find is that, whenever we get the same energy for the first-excited 
state (from fits to $R(t)_i^{(2)}$) from the 3- and 4-dimensional bases, we 
achieve the same results for the $v_i^{(2)}$ and $v_i^{(1)}$ values, and hence 
$f_{B_s'}/f_{B_s}$, too. 
The smaller basis, however, can sometimes lead to higher results for 
$v_i^{(2)}$ (due to higher-order corrections) and we find up to $\sim 15$\% 
higher results for $f_{PS}^{(2)}/f_{PS}^{(1)}$ on the $\beta=7.90$ lattice. 
These come together with larger values for $E^{(2)}$, so we know in such 
cases that the larger basis better isolates the first-excited state, giving 
more reliable (and lower) results for this decay constant ratio. 
It therefore appears that, with the $4 \times 4$ basis, we have better 
isolated the $2S$ state for a determination of $f_{PS}^{(2)}$. 
However, without an even larger basis, we cannot strictly rule out that there 
are still higher-state corrections enhancing this quantity.

Another test we apply is to attempt the extraction of the same numbers via a 
different method: direct fitting of the local-local correlators. 
Since we want an accurate amplitude for the first-excited state from this, 
it is necessary to perform such a fit to a three-state ansatz (otherwise, 
higher-state contamination will enter strongly at the level we are after). 
This proves rather difficult with our data and we are only able to get said 
fits to converge on our finest quenched lattice ($\beta=8.15$). 
The results, however, are consistent with those from the $R(t)_i^{(n)}$. 
For example, at $am_q=0.064$ we arrive at the result 
$f_{PS}^{(2)}/f_{PS}^{(1)}=1.416(143)$ from a three-state fit to the 
local-local correlator over the range of $t/a=7-16$ ($\chi^2/dof=8.2/4$), 
with energies $E^{(1)}$ and $E^{(2)}$ consistent with those found from the 
variational analysis. 
Although this is not a very good fit and the window of meaningful fits is 
small (starts at $t/a=6$ and 8 result in a much larger $\chi^2/dof$ and 
degenerate $E^{(2)}$ and $E^{(3)}$ values, respectively), this is only 
$\approx 1.1\sigma$ above the corresponding result from jackknifed fits to 
the ratios $R(t)_i^{(1)}$ and $R(t)_i^{(2)}$: 
$f_{PS}^{(2)}/f_{PS}^{(1)}=1.263(20)$.

We also switch to unsmeared links for the static quark and check the results 
again. 
A sample of such a ratio on the $\beta=7.90$ lattice appears as the green 
circle at $a/r_0 \approx 0.3$ (shifted to the right for clarity) in 
Fig.\ \ref{f_2S_over_f_1S_vs_a}. 
This value is obviously consistent with the one from the Hyp-smeared static 
quark.

Further discussion of these tests, as well as that of the other decay 
constant ratios we determine, appears in Sec.\ \ref{subsect_Swaves}.

Results for the ratio $f_{B_s'}/f_{B'}$ are also presented in 
Table \ref{decay_constant_table}. 
All of these values are consistent with 1.

\subsection{Kinetic corrections}

\begin{figure}[!t]
\begin{center}
\includegraphics*[width=8cm]{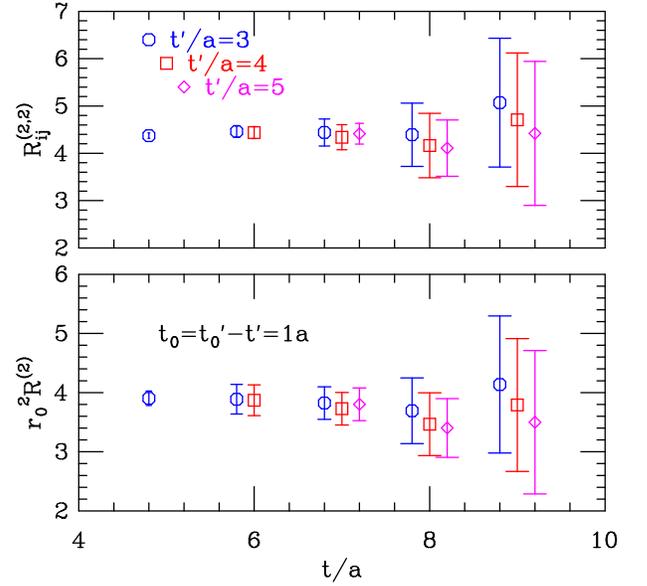}
\end{center}
\caption{
Three- to two-point correlator ratios ${\cal R}_{ij}^{(2,2)}$ (above) and 
$r_0^2{\cal R}^{(2)}$ (below) for the first-excited $S$-wave; see Eqs.\ 
(\ref{generalized_3pt}) and (\ref{simp_generalized_3pt}), respectively. 
The plateaus in $t$ and $t'$ show that the exponential terms from the 
first-excited static-light state have been successfully isolated in the 
projected three-point functions and canceled by the projected two-point 
functions in the denominator. 
Displayed is the case of $i=j$ for the local operator in the basis, but the 
results are the same (well within the errors) for other combinations of $i$ 
and $j$ in ${\cal R}_{ij}^{(2,2)}$. 
These results are from 80 of the $\beta=7.90$ quenched configurations at 
$am_q=0.08$, with only one of the two possible boundary crossings of the 
static-light correlators.
}
\label{prelim_ratio3_vs_t}
\end{figure}

Applying the techniques presented in Sec.\ \ref{subsect_kin_corr}, we create 
ratios of eigenstate projections of three- and two-point correlator matrices, 
where the eigenvectors from the two-point variational method are used (see 
Eqs.\ (\ref{3pt_function}), (\ref{2_2pt_functions}), (\ref{generalized_3pt}), 
and (\ref{simp_generalized_3pt})). 
These we check for independence from the $t$ and $t'$ variables and the 
corresponding, bare kinetic corrections $\varepsilon^{(n,n')}$ may then be 
found.

Figure \ref{prelim_ratio3_vs_t} shows our first results for the ratios 
${\cal R}_{ij}^{(2,2)}$ and $r_0^2{\cal R}^{(2)}$ (versus $t$ and $t'$) from 80 
of the quenched $\beta=7.90$ configurations at $am_q=0.08$. 
The first ratio should approach the quantity 
$\varepsilon^{(2,2)}/(v_i^{(2)*}v_j^{(2)})$, the second 
$r_0^2\varepsilon^{(2,2)}$. 
Each of these shows plateaus (within the errors) in $t$ and $t'$, indicating 
the proper cancellation of the exponentials associated with the first-excited 
state. 
We then perform correlated fits to obtain the bare kinetic corrections to the 
$2S$ static-light state.

\begin{figure}[!t]
\begin{center}
\includegraphics*[width=8cm]{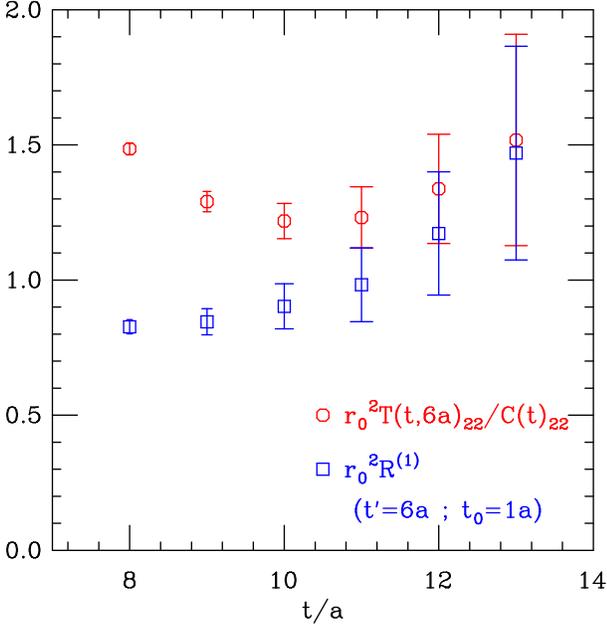}
\end{center}
\caption{
Comparison of the ``traditional'' ($T(t,t')_{22}/C(t)_{22}$) and the 
variational-method (${\cal R}^{(1)}$) ratios used for extracting the kinetic 
correction to the ground state. 
The results agree at $t-t'=6a$, consistent with the starting value of our 
chosen fit range for the ground-state eigenvalue (see Table 
\ref{fittable_b790}). 
These results are from 80 of the $\beta=7.90$ quenched configurations at 
$am_q=0.08$, with only one of the two possible boundary crossings of the 
static-light correlators.
}
\label{prelim_simpratio3compar_vs_t}
\end{figure}

One should be careful in applying this method by looking for plateaus in 
the time range suggested by the two-point variational problem. 
For example, Fig.\ \ref{prelim_simpratio3compar_vs_t} shows a comparison 
of the variational-method ratio ${\cal R}^{(1)}$ with the usual way of 
extracting the correction to the ground state: $T(t,t')_{22}/C(t)_{22}$, 
where the subscript 2 refers to the smeared operator. 
Since we fit the two-point eigenvalues for $E^{(1)}$ starting at $t/a=6$, 
we use $t'/a=6$ for the time of the $\bar Q \vec D^2 Q$ interaction. 
One can clearly see in the plot that the two methods only agree when 
$t-t'$ also becomes $6a$, even though one might be tempted to choose a 
seemingly earlier plateau for ${\cal R}^{(1)}$. 
Also, at any given time slice, there is no apparent statistical advantage 
to the variational method for the ground state; in fact, the 
``traditional'' approach reaches an earlier, consistent plateau and for 
this reason we use this method to obtain the $\varepsilon^{(1,1)}$ 
values. 
The results for $n=2,3$, however, involve time ranges consistent with what 
would be chosen for the energy and coupling determinations (starting at 
$t_0+1a$).

\begin{figure}[!t]
\begin{center}
\includegraphics*[width=8cm]{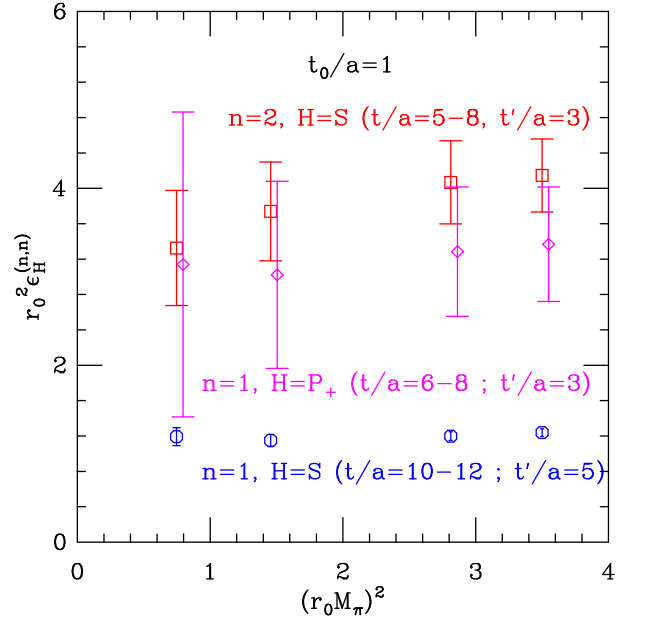}
\end{center}
\caption{
Unsubtracted, bare kinetic-energy corrections to the ground and first-excited 
$S$-waves (80 configurations, one boundary crossing) and the ground $P_+$ (50 
configurations, one boundary) from the $\beta=7.90$ lattice, plotted versus 
the light quark mass.
}
\label{prelim_r02eps_vs_r0Mpi2}
\end{figure}

Fits to the relevant ratios for the $3S$ and $1P_+$ states are also made and 
we plot some of the results versus the light-quark mass in 
Fig.\ \ref{prelim_r02eps_vs_r0Mpi2}. 
Linear chiral extrapolations and strange interpolations are then possible and 
differences of the results are taken to cancel the $\varepsilon_0^{}$ value 
(due to divergent $\bar Q \vec D^2 Q$ operator; see 
Eq.\ (\ref{kin_corr_mass})) and arrive at more meaningful shifts to the 
static-light mass splittings: see Table \ref{eps_nn_table}. 
Of course, these are still bare quantities in that the $\bar Q \vec D^2 Q$ 
operator, although tadpole-improved, requires a renormalization factor $Z$.

Assuming $Z \approx 1$ for the moment, we can form some rough estimates of the 
energy shifts to the static-light mass splittings. 
Working with a value of $m_b^{} \approx 4.2$ GeV \cite{Burch:2004aa}, we find 
that the $2S-1S$ mass differences should increase by $\approx 40(13)$ and 
$\approx 52(8)$ MeV at $m_q^{}=m_{ud}^{}$ and $m_s^{}$, respectively. 
For $3S-1S$, these work out to $\approx 120(80)$ and $\approx 120(50)$ MeV, 
while for $1P_+-1S$ we find $\approx 33(30)$ and $\approx 38(14)$ MeV. 
A systematic discrepancy in the $n=1$ values becomes obvious when we 
calculate the ${\cal O}(1/m_b^{})$ adjustment to $E_s^{(1)}-E_{ud}^{(1)}$ in 
going from the static limit to the $B$-meson system: 
$(\varepsilon_s^{(1,1)}-\varepsilon_{ud}^{(1,1)})/2m_b^{} \approx 2.4(1.2)$ MeV. 
Using experimental, spin-averaged, heavy-light mass differences to set the 
strange-quark mass in Sec.\ \ref{subsect_phys_ms}, we found 9.9 MeV for this 
shift. 
We need to boost our statistics here (and luckily, we can and will do so) and 
run at different lattice spacings to see if such a systematic difference 
persists.

For the ${\cal O}(1/m_b^{})$ mixing of the $1S$ and $2S$ static-strange 
states we find $\varepsilon_s^{(2,1)} \approx 0.26(5)$ GeV$^2$. 
If we have the sign correct and we assume $\varepsilon_0^{} \approx 0$ and 
$Z \approx 1$, this would contribute to an $\approx 6$\% decrease in $f_{B_s}$ 
and an $\approx 4$\% increase in $f_{B_s'}$, increasing $f_{B_s'}/f_{B_s}$ by 
$\approx 10$\% and exacerbating the ``dilemma'' we face in expecting 
$f_{B_s'}/f_{B_s} < 1$. 
However, without an actual value for $\varepsilon_0^{}$ and without the 
mixings with states of $n'>2$, we cannot accurately obtain the overall shifts 
in the couplings $\delta v_i^{(n)}$ (see Eq.\ (\ref{kin_corr_coupling})).

Simply put, we have much more work to do here in order to go beyond these 
rough estimations of the kinetic corrections to the static-light states. 
We are encouraged, however, by the ability of the variational method to 
isolate the excited-state contributions in these three-point function 
calculations. 
We point out again (see Sec.\ \ref{subsect_kin_corr}), though, that this 
method could exhibit instabilities. 
In the limit of large statistics or with a poor basis of operators, one will 
have to add the interaction (or current insertion), along with the subsequent 
transfer to the sink (and, separately, the source), to the basis since it is 
an additional, linearly independent operator.

\begin{table}[hb]
\caption{
Preliminary differences of bare heavy-quark kinetic energy corrections on the 
$\beta=7.90$ ($a=0.145(3)$ fm) quenched lattice.
}
\label{eps_nn_table}
\begin{center}
\begin{tabular}{cc}
\hline \hline
difference & GeV$^2$ \\
\hline
$\varepsilon_s^{(1,1)}-\varepsilon_{ud}^{(1,1)}$ & 0.02(1)(?) \\
$\varepsilon_s^{(2,2)}-\varepsilon_{ud}^{(2,2)}$ & 0.12(7) \\
$\varepsilon_{P+,s}^{(1,1)}-\varepsilon_{P+,ud}^{(1,1)}$ & 0.06(12) \\ \hline
$\varepsilon_{ud}^{(2,2)}-\varepsilon_{ud}^{(1,1)}$ & 0.34(11) \\
$\varepsilon_s^{(2,2)}-\varepsilon_s^{(1,1)}$ & 0.44(7) \\
$\varepsilon_{ud}^{(3,3)}-\varepsilon_{ud}^{(1,1)}$ & 1.0(7) \\
$\varepsilon_s^{(3,3)}-\varepsilon_s^{(1,1)}$ & 1.0(4) \\
$\varepsilon_{P+,ud}^{(1,1)}-\varepsilon_{ud}^{(1,1)}$ & 0.28(25) \\
$\varepsilon_{P+,s}^{(1,1)}-\varepsilon_s^{(1,1)}$ & 0.32(12) \\
\hline \hline
\end{tabular}
\end{center}
\end{table}

\section{Discussion}
\label{SectDiscussion}

We now discuss our results separately for the $S$-, $P$-, and $D$-waves, and 
the baryons. 
In Table \ref{summary_table} we provide a summary listing of the lower-lying 
static-light meson and baryon mass splittings, in which we include {\it all} 
lattice spacings for the quenched continuum extrapolations (see 
Sec.\ \ref{subsubsect_extraps} for details and caveats); all experimental 
values \cite{B**,Bs**,Sigmab*,Xib-,Omegab-,PDG} for spin-averaged singly $b$ 
hadrons are also listed. 
The decay constant ratios and preliminary kinetic corrections have already 
been well summarized in Tables \ref{decay_constant_table} and 
\ref{eps_nn_table}, respectively.

\begin{table}[!b]
\caption{
Summary of our static-light (-strange) mass splittings (in MeV), 
approximating the spin-averaged mass differences of lower-lying $b$-hadron 
states. 
Above appear results for which there are experimental values 
($\dagger$ only the $\Omega_b$ has been observed experimentally); 
below, those without. 
Our values are given using $r_0=0.49(1)$ fm \cite{r0_error} and {\it all} 
lattice spacings for the quenched continuum extrapolations 
(see Sec.\ \ref{subsubsect_extraps} and Tables \ref{chiral_contin_table} and 
\ref{strange_contin_table} for associated caveats). 
}
\label{summary_table}
\begin{center}
\begin{tabular}{cccc}
\hline \hline
difference & $N_f=0$ & $N_f=2$ & experiment \\
 & $a\rightarrow0$ & $a=0.156(3)$ fm & \cite{B**,Bs**,Sigmab*,Xib-,Omegab-,PDG} \\
 & & $M_{\pi,\mbox{sea}}=461(6)(9)$ & \\
\hline
$B_{1(2)}^{(*)}-B^{(*)}$ & 423(13)(9) & 446(17)(9) & 423(4) \\
$B_{s1(2)}^{(*)}-B_s^{(*)}$ & 400(8)(8) & 417(10)(9) & 436(1) \\
\hline
$\Lambda_b-B^{(*)}$ & 415(23)(8) & 358(55)(7) & 306(2) \\
$\Sigma_b^{(*)}-B^{(*)}$ & 604(16)(12) & 555(47)(11) & 512(4) \\
$\Xi_b-B^{(*)}$ & 466(17)(10) & 426(37)(9) & 476(5) \\
$\Omega_b^{(*)}-B_s^{(*)}$ & 683(9)(14) & 624(21)(13) & 765(16)$^\dagger$ \\
\hline
$\Sigma_b^{(*)}-\Lambda_b$ & 200(27)(4) & 195(72)(4) & 206(4) \\
$\Xi_b-\Lambda_b$ & 95(17)(2) & 111(37)(2) & 170(5) \\
$\Omega_b^{(*)}-\Lambda_b$ & 340(23)(7) & 342(55)(7) & 545(16)$^\dagger$ \\
\hline \hline
$B^{(*)'}-B^{(*)}$ & 612(31)(13) & 674(66)(14) & - \\
$B_s^{(*)'}-B_s^{(*)}$ & 604(26)(12) & 664(39)(13) & - \\
$B_{0,1}^*-B^{(*)}$ & 435(15)(9) & 454(19)(9) & - \\
$B_{s0,1}^*-B_s^{(*)}$ & 412(10)(8) & 421(12)(9) & - \\
\hline
$\Xi_b^{(',*)}-\Lambda_b$ & 272(23)(6) & 269(55)(5) & - \\
$\Xi_b^{(',*)}-\Xi_b$ & 173(20)(4) & 158(50)(3) & - \\
\hline \hline
\end{tabular}
\end{center}
\end{table}

\subsection{$S$-waves}
\label{subsect_Swaves}

The ground-state $S$-wave states are well determined on all ensembles. 
The lattice energies ($E^{(1)}$) we extract for these are very precise and 
provide good points from which the energies of all other states can be 
referenced. 
(A future implementation of the procedures of Ref.\ \cite{RenormsHQET} should 
allow us to obtain values for $E_0^{}$, $Z$, and $\varepsilon_0^{}$, and 
hence $m_b^{}$ and absolute hadron masses.) 
The differences among the ground-state energies also act as good measures of 
the relative light-quark masses and we use $E_s^{(1)}-E_{ud}^{(1)}$ to set the 
strange quark mass (see Sec.\ \ref{subsect_phys_ms}).

The local couplings to the ground states are also quite precise and we obtain 
values for the decay constant ratio $(f_{B_s}/f_B)_{static}$ with statistical 
errors of a few \% (see Table \ref{decay_constant_table}) with 
${\cal O}(100)$ configurations ($\lesssim 2$\% error for the bare $f_{B_s}$ 
on the $\beta=8.15$ lattice). 
Quenched results for this ratio \cite{Aoki:1998ji,AliKhan:2007tm} are 
typically lower than those from dynamical calculations 
\cite{Gray:2005ad,Bernard:2004kz,Bernard:2006zz}, by $\approx 5-10$\%. 
Our quenched value 
(static-light, $a^2 \rightarrow 0$ continuum limit: $f_{B_s}/f_B=1.087(31)$) 
is consistent with that in \cite{AliKhan:2007tm} and $\approx 2\sigma$ below 
that in \cite{Aoki:1998ji}. 
The result from our larger dynamical lattice is $\approx 2\sigma$ higher than 
the quenched result at similar lattice spacing. 
The finer dynamical lattice gives a value consistent with the quenched, but 
represents a rather small physical volume (1.35 fm); it may be that the local 
coupling is therefore enhanced more for the lighter quarks, reducing the 
ratio $f_{B_s}/f_B$ on such a small lattice. 
Results on fine dynamical configurations at larger volumes and lighter 
sea-quark masses are necessary to properly handle the systematics described 
above. 
A thorough treatment of the necessary renormalization factors 
\cite{DellaMorte:2007ij} would also permit us to go beyond decay constant 
ratios.

Without explicitly removing the divergences associated with the static quark 
($E_0$, $\varepsilon_0^{}$), absolute masses and kinetic corrections elude us 
for the moment. 
Up to a renormalization factor, differences of the kinetic corrections determine 
the ${\cal O}(1/m_b^{})$ shifts to the static-light mass splittings. 
Our first results for $\varepsilon_s^{(1,1)}-\varepsilon_{ud}^{(1,1)}$ 
underestimate the expected ${\cal O}(1/m_b^{})$ correction to the 
$B_s^{(*)}-B^{(*)}$ mass difference ($\approx 10$ MeV) and we need to do further 
running of the three-point functions $T(t,t')_{ij}$ to improve our statistics 
and go to finer lattice spacing. 
Our preliminary results for the $\varepsilon^{(2,1)}$ values are not yet precise 
enough for us to say anything about the corrections to the ratio $f_{B_s}/f_B$.

With our choice of operator bases, the variational method 
\cite{VarMeth1,VarMeth2,GBs,ExcAmp,Dudek:2007wv,SommerLat08} 
works quite well for finding excited static-light $S$-wave states. 
In particular, the first-excited $S$-wave ($2S$) energies are well determined. 
We find quenched, continuum values for the $B_s^{(*)'}-B_s^{(*)}$ and 
$B^{(*)'}-B^{(*)}$ mass splittings of around 600 MeV, the $a \approx 0.15$ fm 
dynamical results being about 10\% higher, yet consistent with the quenched 
values at similar lattice spacing. 
The few quenched energies which we find for the $3S$ state lead to 
estimates of the $B_{(s)}^{(*)''}-B_{(s)}^{(*)}$ mass differences of 
$\approx 1.1-1.2$ GeV (see Tables \ref{dubious_chiral_contin_table} and 
\ref{dubious_strange_contin_table}).

Hadronic decays of the $B^{(*)'}$ (and $B_s^{(*)'}$) should proceed through a 
two-pion $L=0$ emission, $B^{(*)}\pi\pi$, as the $\approx 600$ MeV puts 
these states well above this threshold. 
However, on our ensembles the sea quarks are too heavy (lightest 
$M_{\pi,\mbox{sea}} \approx 460$ MeV) for this channel to open. 
The same is true for the two-step decay through the broad $B_{0,1}^*$. 
(Single pion decays directly to the ground state are possible, but partially 
suppressed since these would be $L=1$.) 
The relative masses of our excited $S$-wave states therefore suffer 
systematic shifts due to the suppression of these decay modes.

The discussion of strong decays begs the question of how they should be 
handled in possible future studies and we therefore pause briefly here to 
comment on the subject. 
As the dynamical quark masses are reduced, multiple-hadron states with the 
same symmetries as the intended hadron may become energetically favored: 
e.g., in a lattice search for the $B'$ at near-physical pion masses and 
large volumes, the $B\pi\pi$, $L=0$ and $B\pi$, $L=1$ scattering states 
should more readily dominate the $S$-wave correlators than the $B'$ at 
large times. 
In (partially) quenched simulations it is known that the variational 
method can be used to isolate unphysical ``ghost'' states \cite{GBs}, 
and hence there is reason to believe that it may also work to separate 
hadronic scattering states as well. 
The inclusion of interpolators which better project the multiparticle 
states may also become necessary, along with a finite-volume study to 
identify the scattering states, or alternatively, an on-shell-mixing 
analysis to determine partial widths and mass shifts (see, e.g., 
Ref.\ \cite{McNeile:2004rf}). 
How much of this laundry list we will be able to cover in future 
simulations remains uncertain, but we shall certainly keep it in mind.

Using the methods outlined in Ref.\ \cite{ExcAmp} and 
Sec.\ \ref{subsect_couplings}, we are also able to find rather precise values 
for the local coupling to the first-excited state. 
Surprisingly, the resulting ratios, $(f_{B_s'}/f_{B_s})_{static}$, are greater 
than 1 (see Table \ref{decay_constant_table}). 
Results from the light-light \cite{McNeile:2006qy} and heavy-heavy 
\cite{Dudek:2006ej} sectors, as well as considering the heavy-light system in 
the context of nonrelativistic potential models, suggest that this should not 
be so. 
After some careful testing (see Sec.\ \ref{subsect_decay_const_ratios}), we 
are left with a few possible explanations for this seemingly odd result: 

{\it 1. Higher-state corrections:} 
We cannot completely rule out the possibility that there are still 
higher-state corrections in the eigenvectors. 
We check results from larger $t$ values and find noisier, yet consistent 
values for this ratio. 
However, we note that some results from a smaller basis give larger 
excited-state couplings. 
Whenever this happens, we also find significantly larger values for the 
energies, so we know that the larger basis better isolates this state. 
But without an even larger basis, we do not know for sure whether the 
measured first-excited energies and couplings would be reduced further.

{\it 2. The local operator has little effect in the basis:} 
It could be that the local operator is not having a large influence on the 
ability of the method to isolate the $2S$ state. 
If this were true, the associated eigenvector component would most likely 
have a value dictated by the other operators in the basis as their components 
did the work of separating the excited state. 
As far as we can tell, this does not seem to be happening. 
The decay constant ratios are quite similar on all ensembles, even though the 
smeared operators in the bases are not designed to be the same physical size. 
Results for the effective masses are similar whether we reduce the basis by 
removing the local operator or one of the others. 
Also, the few possible three-state fits that we are able to perform on the 
fine, quenched, local-local correlators give similar numbers for the ratio. 
However, such an effect as described above need only take place at about the 
$20-30$\% level (assuming an actual value of $f_{B_s'}/f_{B_s} \lesssim 1$) as 
we find $(f_{B_s'}/f_{B_s})_{static} \approx 1.2-1.3$. 
More extensive testing of this method against direct fitting in bottomonia 
systems \cite{ExcAmp} found no such systematic discrepancies. 
However, another group \cite{Dudek:2007wv}, using a slightly different 
approach, did see systematic enhancements of excited-state couplings in 
charmonia (e.g., $f_{\psi'}$).

{\it 3. The static approximation:} 
It may be that the values for $(f_{B_s'}/f_{B_s})_{static}$ are correct, but 
are $\gtrsim 1$ due to the use of the static approximation for the heavy 
quark. 
Our first step at relaxing this approximation (through 
$\varepsilon_s^{(2,1)}$ at $a \approx 0.15$ fm), however, seems to increase 
the ratio further, but we reiterate here that without a value for 
$\varepsilon_0^{}$ and without the static-light mixings involving the second- 
and higher-excited states, we cannot accurately say.

Of course, we are also left with the possibility that this effect is real for 
bottom-light systems. 
Indeed, some potential models which yield covariance display slightly higher 
decay constants for radially excited heavy-light states: 
in Ref.\ \cite{Morenas:1997rx} 
$f'\sqrt{M'}/f\sqrt{M} \approx 1.09$ is reported, whereas we find 
$f_{B_s'}\sqrt{M_{B_s'}}/f_{B_s}\sqrt{M_{B_s}}=1.31(6)$ in the quenched 
continuum limit and $1.44(15)$ on the larger dynamical lattice.

Fortunately, for the corrections to mass differences we are in a position to 
say something more. 
The $\varepsilon^{(2,2)}-\varepsilon^{(1,1)}$ differences thus far seem to 
indicate that the $B^{(*)'}-B^{(*)}$ and $B_s^{(*)'}-B_s^{(*)}$ mass 
splittings should be $\approx 40$ and $\approx 50$ MeV higher, respectively, 
than the static-light values found on the quenched $a \approx 0.15$ fm 
lattice.

\subsection{$P$-waves}

The ground-state $P_+$ states correspond to the spin-averages of the $B_1$ 
($B_{s1}$) and $B_2^*$ ($B_{s2}^*$) mesons recently observed at Fermilab 
\cite{B**,Bs**}. 
As can be seen in Table \ref{summary_table}, our results for the 
$B_{1(2)}^{(*)}-B^{(*)}$ mass splitting are in reasonable agreement with the 
experimental value. 
However, the true heavy-strange splitting, $B_{s1(2)}^{(*)}-B_s^{(*)}$, lies 
above our determination. 
Note that the experimental heavy-light splitting has a lower value than the 
heavy-strange one, as expected \cite{Becirevic:2004uv}. 
However, we see the wrong ordering here, due to the fact that the static 
results do not include the kinetic-energy correction for the heavy quark 
(using NRQCD up to such order \cite{Burch:2004aa}, the proper ordering was 
seen). 
We imagine that this will be resolved in the present approach when we have 
results for the kinetic corrections from our full statistics (these should be 
larger for the heavy-strange mesons). 
What may then be different, however, is the overall, relative proximity of 
our and the experimental $1P_+-1S$ splittings as our first results for the 
kinetic corrections (see Table \ref{eps_nn_table}) suggest that these 
static-strange (and -light) mass differences may increase by $\sim 40$ MeV 
(assuming $m_b^{} \approx 4.2$ GeV).

As single-pion decays of the $B_{1(2)}^{(*)}$ to the ground-state multiplet 
take place through $L=2$ emission, these are relatively narrow resonances 
and the associated systematic mass shifts should be small. 
The same is not the case for the $B_{0,1}^*$ (and $B_{s0,1}^*$). 
$L=0$, single-pion decays of these mesons are possible (see, e.g., 
Ref.\ \cite{McNeile:2004rf}), making them rather broad and therefore 
difficult to see in experiments. 
Our $1P_--1S$ results (e.g., $B_{0,1}^*-B^{(*)}$) should thus only be viewed 
as ``ballpark'' figures since they likely include strong systematic effects 
due to the suppression of these modes.

Calculation of the decay constants for the $P$-wave states is certainly 
possible \cite{Herdoiza:2006qv}, and in the works here.

Excited $P$-wave states are also observed, the mass splittings with the 
$1S$ states lying in the $800-900$ MeV range (see Tables 
\ref{dubious_chiral_contin_table} and \ref{dubious_strange_contin_table}). 
The results here are somewhat less certain as the effective-mass plateaus 
are usually quite short. 
Oddly enough, the $2P_-$ states lie slightly above the $2P_+$ ones. 
More statistics and finer lattices (along with cross correlations with decay 
products) are needed to better resolve these issues.

\subsection{$D$-waves}

We view the static-light $D_\pm$ meson as an approximation of the spin 
average of the $B^{*'}$, $B_2$, and $B_3^*$ states. 
The reason for the presence of the $B^{*'}$ in this average, rather than the 
$B^*$, is the fact that the heavy-quark spin interactions which normally 
allow the mixing of the (mostly orbitally excited) $1D_\pm$ configuration 
and the $1S$ one are not present. 
Of course, the $1D_\pm-2S$ configuration mixing is therefore also absent; 
but these orbital and radial excitations are certainly closer in energy. 
The results appear to support this averaging scheme since the $1D_\pm$ 
energies are slightly above the level of the $2S$ on all lattices.

All of the physical states associated with these operators are rather 
high-lying ones and should therefore experience a number of hadronic decays. 
Between our relatively heavy sea-quark masses and finite volumes (restricting 
the available momenta of decay products), we likely suppress these and we 
therefore have rough values for the mass differences here.

\subsection{Baryons}
\label{subsect_baryons}

With a four-operator basis on most ensembles and two-state fits to the 
largest eigenvalues of the variational method, we have a number of good 
results for the ground-state baryons. 
Differences between the resulting energies and those of the ground-state 
mesons are calculated, as well as those among the baryons (see Table 
\ref{summary_table}).

There are obvious discrepancies between our quenched, continuum results for 
the $\Lambda_b-B^{(*)}$ and $\Sigma_b^{(*)}-B^{(*)}$ mass splittings and the 
corresponding experimental values, which are $\approx 100$ MeV (or 
$\approx 5\sigma$) lower. 
Even considering the possible systematic increase in the fine lattice 
results (due to only having a $2 \times 2$ basis there) and leaving them out, 
the baryon masses only decrease by $\approx 60$ MeV for the $\Lambda_b$ and 
$\approx 30$ MeV for the $\Sigma_b^{(*)}$ (see Table 
\ref{chiral_contin_table}), while the errors also decrease, thereby 
maintaining a largely significant difference with experiment.

The large-volume, dynamical results for these baryon-meson mass differences 
are consistent with experiment, but with much larger errors, making them also 
consistent with the quenched results at similar lattice spacing.

The $\Sigma_b^{(*)}-\Lambda_b$ mass differences (both quenched continuum and 
finite-$a$ dynamical) agree with the experimental value. 
The $\Omega_b^{(*)}-\Lambda_b$ splitting also shows better agreement when 
comparing quenched continuum and dynamical results (but is below experiment; 
see below). 
This seems to support the notion that by taking differences between the 
baryon energies we are able to remove some of the possible higher-state 
corrections associated with the smaller basis on the fine quenched lattice 
(see Sec.\ \ref{subsubsect_extraps}). 
Taking the $\Sigma_b^{(*)}-\Lambda_b$ difference also partially removes the 
need for the heavy-quark relativistic corrections (see below) since the light 
valence content is the same for these two states.

Given the presence of two lighter valence quarks in the baryons, the heavy 
quark in these states most likely exhibits a larger kinetic energy than in 
the mesons. 
Accordingly, the kinetic corrections for our baryons may be quite large when 
compared to those for the ground-state mesons and the corresponding 
baryon-meson mass differences given here may become significantly greater. 
This could help to explain the much lower result we have for the 
$\Omega_b^{(*)}-B_s^{(*)}$ mass splitting: 
$\approx 80$ MeV ($\approx 9\sigma$) below the experimental 
$\Omega_b^--B_s^{(*)}$ for the quenched continuum and $\approx 140$ MeV 
($\approx 7\sigma$) lower for the dynamical $a=0.156(3)$ fm lattice. 
Also, since the $\Omega_b^{(*)}$ has two strange quarks, while the 
$\Lambda_b$ has two light ones, there is likely a sizeable amount of 
heavy-quark kinetic energy missing in our $\Omega_b^{(*)}-\Lambda_b$ 
difference; indeed, our values are $\approx 200$ MeV too low.

Hadronic decays and the associated mass shifts are also a more important 
consideration for the baryons and we have yet to consider any of these.

Until now, we have not said much about the $\Xi_b$ and $\Xi_b^{(',*)}$ states. 
We do not yet explicitly create the appropriate correlators for these as they 
require strange-light diquarks and we have only degenerate masses for ours. 
Nevertheless, we make the simplification of using our $\Lambda_Q$ results at 
$m_q=(m_{ud}^{}+m_s^{})/2$ to approximate the $\Xi_b$ and the corresponding 
$\Sigma_Q^{(*)}$ results for approximating the $\Xi_b^{(',*)}$. 
These lead to the mass differences in Table \ref{summary_table}. 
The only agreement with the experimental values comes from the quenched 
continuum limit of the $\Xi_b-B^{(*)}$ splitting. 
The other values are all lower than experiment; the $\Xi_b-\Lambda_b$ mass 
difference being $> 4\sigma$ away. 
This fits with our low $\Omega_b^{(*)}-\Lambda_b$ observations due to the 
missing extra heavy-quark kinetic energy for the baryons involving strange 
quarks. 
Given the possible over-simplification we use here for approximating the 
$\Xi_b$ and $\Xi_b^{(',*)}$ correlators, it becomes difficult to draw more 
meaning from these results.

A few effective-mass plateaus for a possible, first-excited state are seen in 
the second $\Lambda_Q$ eigenvalues on the fine quenched ($\beta=8.15$) 
lattice. 
The chiral extrapolation of these energies leads to an estimate of a 
$\Lambda_b'-\Lambda_b$ mass difference of $\approx 900$ MeV (see Table 
\ref{chiral_contin_Baryon_table}). 
This represents a large gap to the first excitation and, if real, would 
entail large systematic effects due to omitted kinetic corrections and 
decay channels. 
We are, however, rather skeptical of this ``state'' as it arises in the case 
where we have the smallest basis (2 equally smeared operators with different 
diquark spin structures; see Table \ref{operatortable}). 
We see similar, yet somewhat less certain, plateaus on the $\beta=7.90$ 
quenched lattice when we reduce the basis to the two-dimensional one similar 
to that above; these lead to a $\Lambda_b'-\Lambda_b$ splitting estimate of 
$\approx 970$ MeV. 
However, a telling result is that the evidence of this state vanishes when we 
then enlarge the basis on this ensemble. 
Perhaps the same will happen with a larger basis on the fine quenched 
lattice.

\section{Conclusions and Outlook}
\label{SectConclusion}

In this work we presented static-light approximations of $B$- and $B_s$-meson 
mass splittings and decay constant ratios in the quenched continuum limit and 
at two finite lattice spacings which include two flavors of improved sea 
quarks. 
Results were reported for a number of excitations: $2S$, $3S$, $1P$, $2P$, 
$1D$, $2D$. 
We also included our first results for kinetic corrections to the $1S$, $2S$, 
$3S$, and $1P_+$ mesons, indicating the approximate shifts to the associated 
mass differences and decay constants. 
In addition, static-light-light baryon correlators and the corresponding mass 
splittings were calculated.

The strengths of the current calculation include: the use of improved gauge 
(L\"uscher-Weisz) and fermion (Chirally Improved) actions for the light 
degrees of freedom; a good separation of the static-light states, achieved 
via the variational method and a robust basis of hadron operators; and a 
significant boost in the statistics due to the domain-decomposition 
improvement (half-to-half) of the light-quark propagator estimation.

Comparison of our results with experiment 
\cite{B**,Bs**,Sigmab*,Xib-,Omegab-,PDG} 
and with similar results from other lattice groups 
\cite{MVR,AliKhan:1999yb,Hein:2000qu,Lewis:2000sv,Wingate:2002fh,Green:2003zz,Foley:2007ui,Koponen:2007nx,Jansen:2008ht,Bowler:1996ws,Mathur:2002ce,Na:2007pv,Lewis:2008fu} 
is difficult for a number of reasons: mainly systematic effects inherent in 
our lattice formulation of HQET for the former, and those between the 
different formulations of HQET or QCD for the latter. 
However, we have pointed out the relevant literature above and have discussed 
our findings in detail. 
Overall we have shown: While our mass splittings from the ground state 
$B^{(*)}$ to the excited meson states are quite reasonable, those up to the 
baryons are in general too large when near the light-quark chiral limit and 
too small for bottom-strange baryons. 
We discuss possible sources for these discrepancies. 
Our decay constant for $B_s'$ turns out to be somewhat larger than that for 
$B_s$. 
As this was initially unexpected, we perform multiple tests of this result, 
but nevertheless find it to persist. 
We offer a few scenarios whereby this may still be due to systematic effects, 
which we plan to test further. 
However, we also entertain the possibility of this being a real effect for 
heavy-light hadronic systems (see also, e.g., Ref.\ \cite{Morenas:1997rx}).

Along the way, we have pointed out areas for further improvement. 
Higher statistics, finer lattices, larger volumes, lighter sea quarks, and 
more basis operators have all been mentioned and we shall endeavor to boost 
each of these. 
In the near future, we aim to remove the divergences associated with the 
static quark, make a determination of the bottom-quark mass, and calculate 
the kinetic corrections for all of the lower-lying hadrons on the full range 
of ensembles contained herein.

\begin{acknowledgments}
We would like to thank Christof Gattringer, Stephan D\"urr, Carleton DeTar, 
Alexander Lenz, Rainer Sommer, Georg von Hippel, Marc Wagner, and 
Benoit Blossier for helpful discussions. 
Simulations were performed at the LRZ in Munich. 
This work is supported by GSI. 
The work of T.B.\ is currently supported by the NSF (NSF-PHY-0555243) and 
the DOE (DE-FC02-01ER41183). 
The work of M.L.\ is supported by the Fonds zur F\"orderung der 
wissenschaftlichen Forschung in \"Osterreich (FWF DK W1203-N08).
\end{acknowledgments}


\begin{thebibliography}{1234567}
\bibitem{subscript_note}
  The quantities $E^{(n)}$, $\bar\Lambda^{(n)}$, $k^{(n)}$, and 
  $\varepsilon^{(n,n')}$ may have additional subscripts like $ud$, $s$, $P+$, 
  $\Lambda$, $\Sigma$, or $\Omega$ to denote that they correspond to the 
  $B^{(*)}$, $B_s^{(*)}$, $B_{1(2)}^{(*)}$, $\Lambda_b$, $\Sigma_b^{(*)}$, or 
  $\Omega_b^{(*)}$ systems, respectively.
\bibitem{StateNotation}
  We use a parenthetical notation to denote spin averages. For example, by 
  $B^{(*)}$ we mean the spin average of the pseudoscalar $B$ and vector $B^*$ 
  states: $M_{B^{(*)}}=(M_B+3M_{B^*})/4$. We use $B_{s1(2)}^{(*)}$ for the average 
  of $B_{s1}$ and $B_{s2}^*$: $M_{B_{s1(2)}^{(*)}}=(3M_{B_{s1}}+5M_{B_{s2}^*})/8$. 
  The bewildering $B_{2(3)}^{(*',*)}$ is meant to denote the spin average of 
  the $B^{*'}$, $B_2$, and $B_3^*$ states: 
  $M_{B_{2(3)}^{(*',*)}}=(3M_{B^{*'}}+5M_{B_2}+7M_{B_3^*})/15$. A similar notation is 
  used for the baryons. 
  Also, we leave out any charge designations (e.g., $0,+,-$) since we do not 
  include any EM effects. Wherever possible, we average experimental 
  quantities over these too.
\bibitem{r0_error}
  Wherever present, the second error estimate on quantities is due to the 
  $\pm\approx2$\% range we use for $r_0=0.49(1)$ fm (the only exceptions 
  being $a$ or $a^{-1}$, where it is the first error). 
  This systematic is not included in any extrapolations of heavy-light 
  states; there, it is only added in the end. 
  So, when using values in the tables, e.g., to construct new mass 
  differences, one should not propagate this error. 
  For the heavy-strange hadrons, the different choice of $r_0$ also affects 
  the physical $m_s$ point and this additional systematic is determined by 
  performing the strange-mass interpolations (along with any associated 
  continuum extrapolations) using $m_s$ values from $r_0=0.48$ and 0.50 fm. 
  This effect is found to be insignificant when compared to the previously 
  mentioned systematic and statistical errors.
\bibitem{new_results_note}
  The use of larger $t_0$ values and the corresponding changes in fit ranges 
  (since $t>t_0$) cause the greatest differences with our previous results. 
  For some ensembles, we also have greater statistics here than before, 
  including a second set of correlators where the static quark crosses the 
  other domain boundary.
\end{thebibliography}
\end{document}